\documentclass[bibyear]{aa}  

\usepackage{graphicx}
\usepackage{natbib}
\usepackage{float}
\usepackage[usenames,dvipsnames]{color} 
\usepackage{morefloats}
\usepackage{longtable}
\usepackage{lscape}
\usepackage{amsmath,upgreek}
\usepackage{txfonts}
\usepackage{textcomp}
\usepackage{gensymb}
\usepackage[]{hyperref}
\begin{document} 
\newcommand{\dmu}{pc/cm$^3$}
\newcommand{\su}{\sigma^\mathrm{unkn}_{\lg S}}

\title{A LOFAR census of non-recycled
pulsars: extending below 80 MHz}
\titlerunning{A LOFAR census of non-recycled pulsars, LBA}

\author{A.~V.~Bilous \inst{\ref{uva}}
        \and L.~Bondonneau\inst{\ref{lpc2e}}
        \and V.~I.~Kondratiev\inst{\ref{astron},\ref{asc}}
        \and J.-M.~Grie{\ss}meier\inst{\ref{lpc2e},\ref{nancay}}
        \and G.~Theureau\inst{\ref{lpc2e},\ref{nancay},\ref{paris}}
        \and J.~W.~T.~Hessels\inst{\ref{uva},\ref{astron}}
        \and M.~Kramer\inst{\ref{mpifr},\ref{jb}}
        \and J.~van~Leeuwen\inst{\ref{uva},\ref{astron}}
        \and C.~Sobey\inst{\ref{csiro}}
        \and B.~W.~Stappers\inst{\ref{jb}}
        \and S.~ter~Veen\inst{\ref{astron}}
        \and P.~Weltevrede\inst{\ref{jb}}
                }

\institute{
Anton Pannekoek Institute for Astronomy, University of Amsterdam, Science Park 904, 1098 XH Amsterdam, The Netherlands\\ \email{A.Bilous@uva.nl}\label{uva}
\and
LPC2E - Universit\'{e} d'Orl\'{e}ans / CNRS, 45071 Orl\'{e}ans cedex 2, France\label{lpc2e}
\and
ASTRON, the Netherlands Institute for Radio Astronomy, Postbus
2, 7990 AA Dwingeloo, The Netherlands\label{astron} 
\and
Astro Space Centre, Lebedev Physical Institute, Russian Academy of Sciences, Profsoyuznaya Str. 84/32, Moscow 117997, Russia\label{asc} 
\and
Station de Radioastronomie de Nan\c{c}ay, Observatoire de Paris, PSL Research University, CNRS, Univ. Orl\'{e}ans, OSUC, 18330 Nan\c{c}ay, France\label{nancay}
\and
Laboratoire Univers et Th\'{e}ories LUTh, Observatoire de Paris, CNRS/INSU, Université Paris Diderot, 5 place Jules Janssen, 92190 Meudon, France\label{paris}
\and
Max-Planck-Institut f\"ur Radioastronomie, Auf dem H\"ugel 69, 53121 Bonn, Germany\label{mpifr} 
\and
Jodrell Bank Centre for Astrophysics, School of Physics and Astronomy, University of Manchester, Manchester M13 9PL, UK\label{jb}
\and
CSIRO Astronomy and Space Science, PO Box 1130 Bentley, WA 6102, Australia\label{csiro}
}


\abstract{We present the results from the low-frequency (40--78\,MHz) extension of the first LOFAR pulsar census of
  non-recycled pulsars. We have used the Low-Band Antennas of the LOFAR core stations to observe 87 pulsars out of 
  158 that have been detected previously with the High-Band Antennas. Forty-three
  pulsars have been detected and we present here their flux densities and flux-calibrated profiles. Seventeen of these 
  pulsars have not been, to our knowledge, detected before at such low frequencies. 
  We re-calculate the spectral indices using the new low-frequency flux density measurements from the LOFAR census
  and discuss the prospects of studying pulsars at the very low frequencies with the current and upcoming facilities, 
  such as NenuFAR.}


   \maketitle
%

\section{Introduction}
\label{sec:intro}

Half a century ago, the work on interplanetary scintillation at the frequency of 81.5\,MHz 
led to the serendipitous discovery of pulsars \citep{Hewish1968}. However, until recently 
most pulsar observations were conducted at higher frequencies of 300--3000\,MHz because of 
various challenges. Properties of pulsar emission at radio frequencies below 200\,MHz 
remained relatively poorly explored for two reasons: the high level of the background 
Galactic emission, and the deleterious influence of the electron plasma in the interstellar 
medium (ISM) and Earth's ionosphere. 

The last decade brought rapid advances both in hardware and computing capabilities, for the 
first time allowing broadband sensitive observations of pulsars with precise compensation for 
dispersive delay at frequencies below 200\,MHz. These observations deepen our understanding of 
pulsars as astrophysical objects: e.g., changes in spectral shape of radio emission and the 
morphology of the average pulse shape provide information about the microphysics of pulsar 
radio emission and magnetospheric configurations. Also, because of their increased effects 
on the received signal, the ISM and the ionosphere can be more accurately studied at lower 
frequencies.

The new generation of low-frequency radio telescopes has already started charting the 
meter-wavelength pulsar sky. Several surveys of the known pulsar population have been conducted 
over the last few years. The newly upgraded second modification of the Ukrainian T-shaped radio 
telescope (UTR-2) was utilised to detect 40 pulsars at 10--30\,MHz, the lowest radio frequencies 
visible from Earth \citep{Zakharenko2013}. The first station of the Long Wavelength Array 
(LWA1) was used to measure the flux densities of 44 pulsars at 30--88 MHz \citep{Stovall2015}.
At 185\,MHz, the Murchison Widefield Array (MWA) was used to detect 50 pulsars \citep[including 
six millisecond pulsars,][]{Xue2017} and also to measure flux densities from continuum images 
\citep{Murphy2017}.

In 2014, we undertook a large campaign of observing almost all known non-recycled radio pulsars
with declination, ($\mathrm{Dec}$) $\mathrm{Dec}>8\degree$ and Galactic latitude, ($\mathrm{Gb}$)  
$|\mathrm{Gb}|>3\degree$. The observations were performed with the High-Band Antennas (HBA) of the 
LOFAR telescope at frequencies 110--188\,MHz \citep{vanHaarlem2013}. The census (hereafter HBA 
census) encompassed 194 such sources and resulted in 158 detections, updating DMs and measuring 
flux density values \citep[][hereafter B16]{Bilous2016}. Based on the measurements at 110--188\,MHz 
and the previously published flux densities, broadband spectra were constructed and the spectral 
indices were measured with a single or broken power-law model. It appeared that the spectra of 
most pulsars are, in fact, not known very well and regular flux density measurements are needed, 
as flux densities can typically vary up to an order of magnitude due to diffractive and refractive 
interstellar scintillation, and/or due to intrinsic variability. With the exception of a handful 
of bright pulsars with hundreds of flux density measurements, the choice of the model and the frequency
of the spectral turnover depends greatly on the poorly explored low-frequency end of the spectrum.

To investigate the shape of the pulsar spectra further, we undertook an LBA extension of the HBA 
census (hereafter, LBA census), encompassing 87 out of 158 pulsars that have been detected in the 
HBA census. This paper presents the average profiles, DMs and flux density measurements for the 
pulsars that were detected. The results presented will be also made available through the European 
Pulsar Network (EPN) Database for Pulsar Profiles\footnote{\url{http://www.epta.eu.org/epndb}}.

\section{Source selection}
For the HBA census, we have selected pulsars from version 1.51 of the ATNF Pulsar 
Catalogue\footnote{\texttt{http://www.atnf.csiro.au/people/pulsar/psrcat/} (\citealt{Manchester2005})}
which satisfied the following criteria: (a) $\mathrm{Dec}>8\degree$;
(b) $|\mathrm{Gb}|>3\degree$; (c) surface magnetic field strength, $B_\mathrm{surf}>10^{10}$\,G;
(d) positional uncertainty within half of LOFAR's beam width at the upper edge of the HBA band 
\citep[$130\arcsec$, ][]{vanHaarlem2013}; (e) not in a globular cluster. For a more detailed 
discussion of the selection criteria we refer the reader to B16. 

Ideally, the LBA extension of the HBA census would include all pulsars that had been detected with 
the high-band antennas, except, perhaps the pulsars with considerable scattering and without any 
prospects of detecting very strong single pulses. In practice, when this project started, the HBA 
census was not yet processed and completed and only preliminary detection estimates were available.

Originally, observations using the LBAs were planned to be conducted with an incoherent dedispersion 
scheme. Under this scheme the observing band is split into many narrow channels and interstellar 
dispersion is only compensated for between channels, but not within the channels themselves. 
The proposed source sample therefore only included pulsars with sufficiently small intra-channel 
smearing at 30\,MHz, with the exact smearing threshold depending on the preliminary S/N estimates 
from the HBA census, made without proper radio frequency interference (RFI) excision and without 
updated ephemerides. We did not exclude sources with considerable scattering in the LBA band, 
hoping to detect strong single pulses. 

\begin{figure}
   \centering
  \includegraphics[scale=1.0]{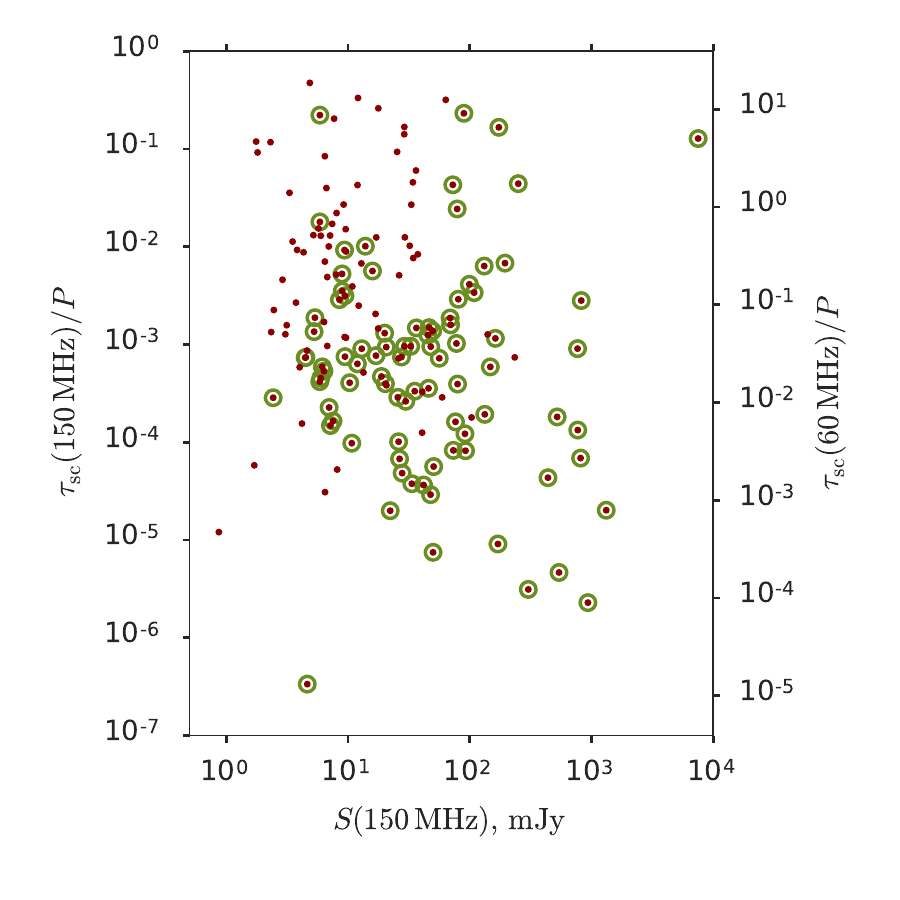}
  \caption{Band-integrated fluxes and the ratio of scattering time in the middle of the HBA 
  (left y-axis) and LBA (right y-axis) bands to the pulsar spin period for all sources detected 
  in the HBA census (red dots). Green circles mark the pulsars selected for the follow-up with LBAs. 
  The scattering time was estimated with the Galactic electron density model from \citet{Yao2017} and scaled 
  to respective frequencies with an exponent, $\alpha = -4.0$ in $\tau_\mathrm{sc} \sim\nu^{\alpha}$. 
  }
  \label{fig:src}   
\end{figure}

Before the start of observations with the LBAs, the incoherent dedispersion observing scheme was 
replaced with a coherent observing scheme, which made the intra-channel smearing criterion 
obsolete. However, the initial target list for the LBA follow-up remained unchanged. At present, 
with all HBA observations being processed and analysed (leading to substantial changes in 
some of the S/N estimates), we can regard the LBA census source sample as being an
arbitrary subsample of pulsars detected in the HBA census, with some preference towards 
closer and/or brighter sources (see Fig.~\ref{fig:src}).

\begin{figure*}
   \centering
 \label{fig:RFI_mitigation}   
 \includegraphics[scale=0.8]{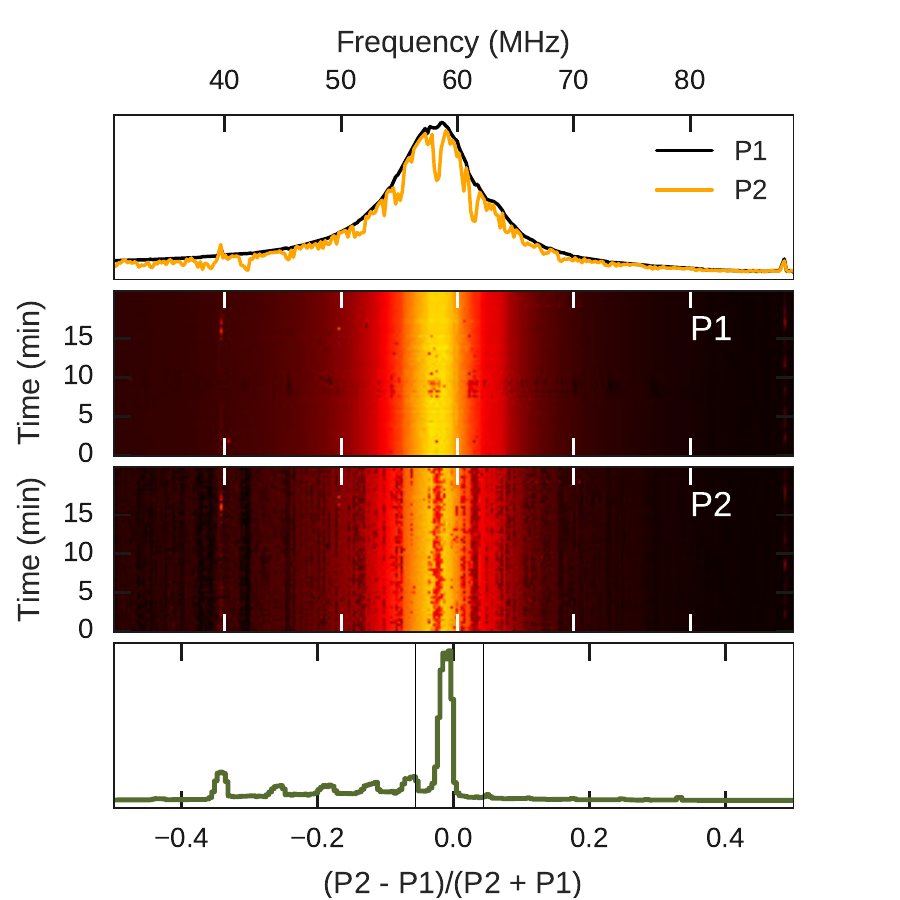}\includegraphics[scale=0.8]{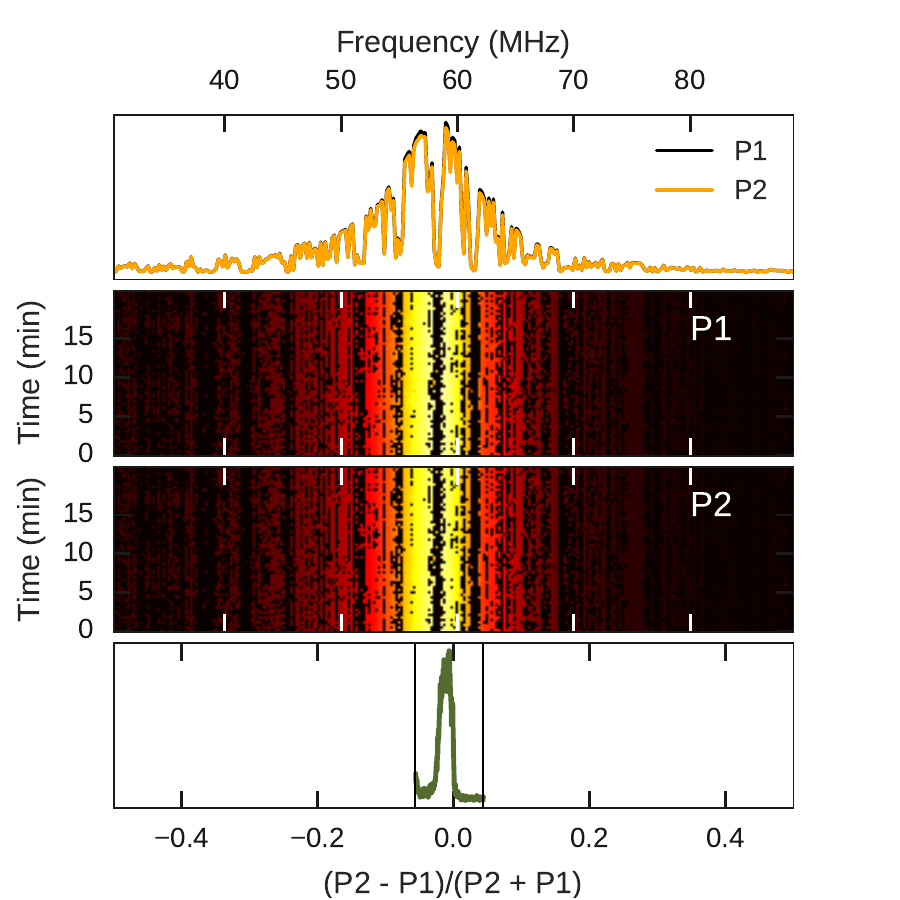}
 \includegraphics[scale=0.8]{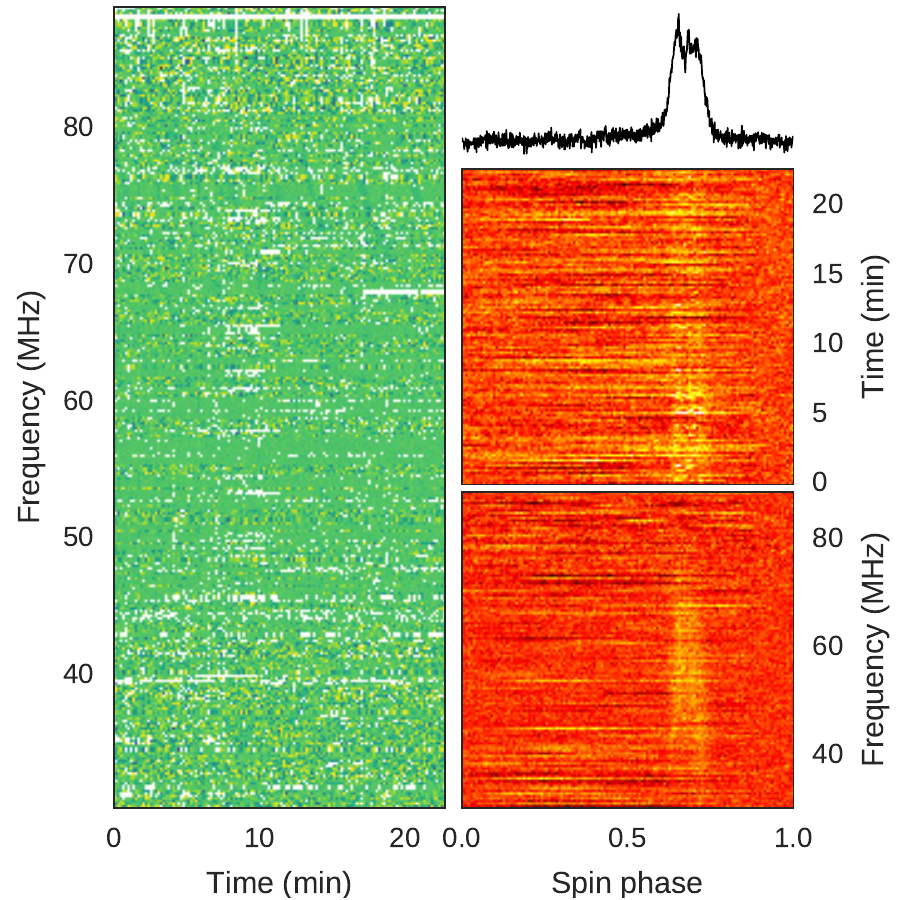}\includegraphics[scale=0.8]{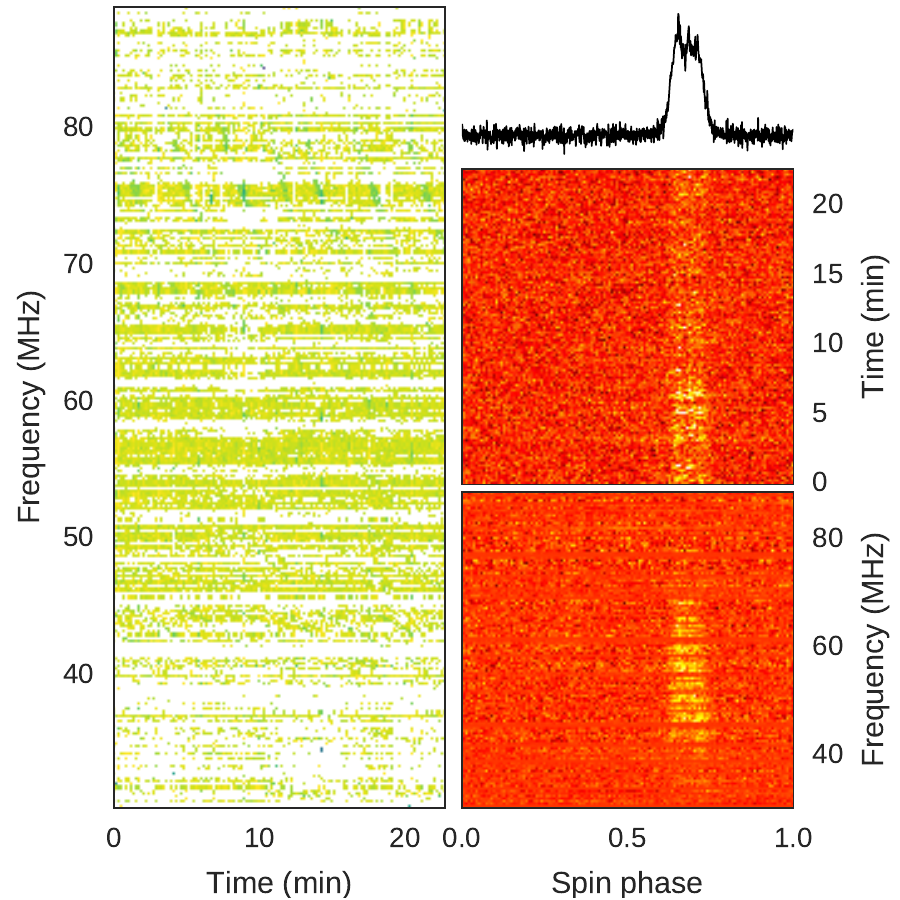}
   \caption{An example of diagnostic plots without (left) and with (right) dropped packet 
   cleaning applied    for one observation of the bright pulsar B0809+74. The upper row of plots 
   shows the statistics of the two polarisations, the lower plots show the dynamic 
   and folded spectra, waterfall diagram, and the average profile.
 }
\end{figure*}

\section{Observations and data reduction}

Similarly to the HBA census, each pulsar was observed during one session for either at 
least 1000 rotational periods, or at least 20\,min. Pulsars were observed in June 2014 
-- May 2015 using the LBAs of the LOFAR core stations in the frequency range of 30--89\,MHz. 
In order to compensate for the refraction in the ionosphere, seven beams were formed 
around each source (beam 0 on the target and beams 1--6 in a hexagonal grid around beam 0 
on the nominal position of the target) at a distance of about $210\arcsec$, approximately 
half of the telescope resolution at 60\,MHz \citep[$412.5\arcsec$,][]{vanHaarlem2013}.
The coordinates of the sources were taken from the ATNF pulsar catalogue, or from the timing 
observations conducted with the Lovell telescope at Jodrell Bank and the 100-m Robert C. 
Byrd Green Bank Telescope.

For each beam, the coherently summed complex-voltage signal from individual stations 
was coherently dedispersed. Raw data were stored in the LOFAR Long-Term 
Archive\footnote{\url{http://lofar.target.rug.nl/}}. For a more detailed description of 
LOFAR and its pulsar observing modes, we refer the reader to  \citet{vanHaarlem2013} and 
\citet{Stappers2011}.

Observations were pre-processed with the standard LOFAR pulsar pipeline \citep{Stappers2011}, 
which uses the PSRCHIVE software package \citep{Hotan2004,vanStraten2010}. Raw data were 
converted to full-Stokes samples which were recorded in PSRFITS format \citep{Hotan2004}, 
with time resolution of 5.12\,$\upmu$s and 300 channels of 195\,kHz. Folding produced 5-s 
sub-integrations with 1024 phase bins. In this paper we focus only on total intensity data.
Table~\ref{table:obssum} gives the basic observation summary for all pulsars in the LBA sample.

In most cases the raw data were folded using the same ephemerides that were used for 
folding the HBA census data. Analysis of the HBA census data revealed that in many cases the 
DM as derived from higher-frequency observations was substantially different from the one 
obtained from census data. Thus, dedispersing and folding LBA data using incorrect DMs caused 
substantial pulse smearing within one frequency channel. To mitigate that, we re-dedispersed 
(coherently) and re-folded 25 pulsars that were affected the most using the DM value obtained 
in the HBA census. For the remaining pulsars, the smearing was less than one phase bin at 
60\,MHz for the downsampled number of bins used in the analysis.

After the observations took place it was found that a substantial fraction of data packets 
was dropped,resulting in numerous data gaps\footnote{The observations were carrying out 
during the timewhen new \textit{Cobalt} correlator was put online. But, unfortunately, one of 
the network switches was misconfigured that resulted in somewhat lower network throughput 
for the used observing setup that was preliminary tested with the old BG/P beamformer.}.
These gaps appeared independently in two polarisations because of how the data is recorded 
to disk and led to significant decrease of overall signal-to-noise ratio (Fig.~\ref{fig:RFI_mitigation}, 
left). In order to mitigate this adverse effect, we performed an additional step during 
the RFI cleaning procedure on all data archives. Working with 5-s, 300-channel archives
with two polarisations ($P_1=XX^{*}$ and $P_2=YY^{*}$), we computed the histogram of the relative 
signal strength difference, $dP = (P_2-P_1)/(P_2+P_1)$ for each 5-s/195-kHz data cell.  
We then assigned zero weights to the cells with $dP$ deviating more than by 0.05 from the 
peak of the histogram (Fig.~\ref{fig:RFI_mitigation}, right).

Since the bandpass in the LBA band is not uniform and has a large peak in sensitivity in 
the middle of the band, it is necessary to flatten the bandpass before cleaning RFI. Thus, 
we have divided the dynamic spectrum by an "ideal bandpass", obtained from interpolating 
the median bandpass from all observations. To remove RFI from the flattened data we used 
the \texttt{clean.py} tool from the \texttt{CoastGuard} package \citep{Lazarus2016}.

Archives that were automatically excised of RFI were also visually inspected for residual 
RFI. In many cases the cleaning procedure was not entirely sufficient, resulting in some 
relatively faint RFI biasing the baseline estimates for flux calibration. For only three 
pulsars (namely, PSRs B0105+68, B0643+80, B0656+14), the RFI prevented useful analysis, 
hence they were excluded from our sample.

Overall, the fraction of band that has been zapped due to dropped packets or RFI is quite 
substantial, ranging from a few percent to almost the entire band (Table~\ref{table:obssum}). 
Zapped fraction varies considerably from beam to beam and is present in most observing 
runs, not showing a clear dependence on the observing date. While the data used here may 
not use LOFAR to its full capabilities, and future and ongoing low-frequency observations 
may reach higher signal-to-noise, the results we present here still provide useful 
information about low-frequency end of the pulsar spectra (see Section~\ref{subsec:results}).

\subsection{Detection and ephemerides update}

We adjusted the folding period $P$ and the intra-channel dispersive delay with the PSRCHIVE 
program \texttt{pdmp}, maximising the integrated S/N of the frequency- and time-averaged 
profile. Initially, the entire band was used and the diagnostic output from $\texttt{pdmp}$ 
was visually inspected for a pulsar-like signal. For those non-detected in this manner, or 
the ones with spectra not being visually present across the whole band, we additionally 
zapped the edges of the band where the sensitivity is low and repeated the search for 
frequencies between 41 and 78\,MHz. To facilitate visual inspection of the average profiles, 
we downsampled the initial number of phase bins by a factor of 2, 4 or 8.

It is worth mentioning that our DM measurements, based on maximising S/N of the 
frequency-integrated profile did not take into account any profile evolution, which usually 
becomes rapid in the LBA band. Thus, the reported DM values may be subject to a bias 
depending on the assumed profile evolution model.

\begin{figure}
   \centering
  \includegraphics[scale=0.9]{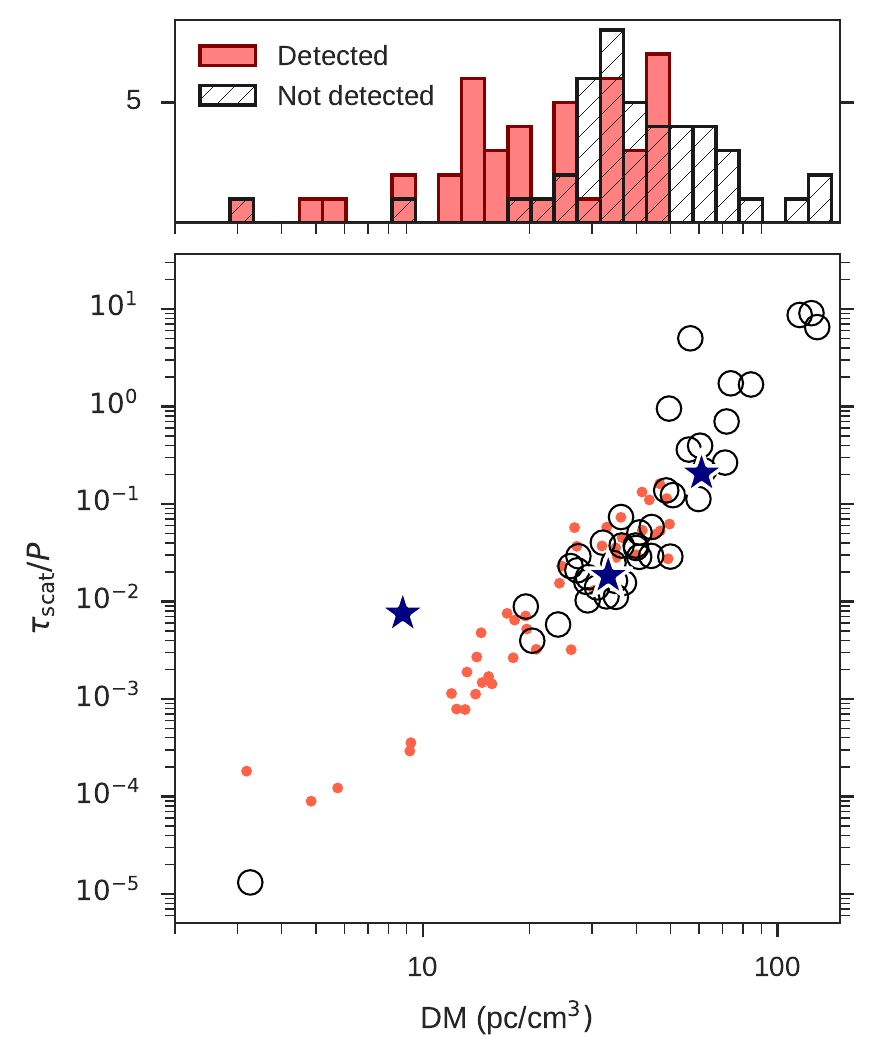}
  \caption{Detected pulsars (red dots) and the 
  non-detected ones (black circles) versus DM and estimated scattering time at 60\,MHz divided 
  by the pulsar period. Scattering times at 1\,GHz were taken from the \citet{Yao2017} electron 
  density model and scaled to 60\,MHz with an exponent, $\alpha = -4.0$ in 
  $\tau_\mathrm{sc} \sim\nu^{\alpha}$. Blue stars mark pulsars discarded due to an excess of RFI.
 }
 \label{fig:det_DM_scat}   
\end{figure}

Figure~\ref{fig:det_DM_scat} shows the correlation between DM and the estimated scattering 
time over pulsar period for the detected and non-detected pulsars. The same information is 
also available in Table~\ref{table:obssum}. Our detections do not extend beyond a DM of 
$\sim60$\,\dmu\ and an estimated scattering time fraction of $\sim20\%$ of the pulse period. 

Interestingly enough, one of the pulsars closest to Earth in our sample, J1503+2111, has 
not been detected. This pulsar had an ostensible error in DM measurements and the HBA census 
found it at $\mathrm{DM} =  3.260 \pm 0.004$\,\dmu\ instead of the previously published 
$\mathrm{DM}= 11.75 \pm 0.06$\,\dmu~\citep{ChampionA2005}. The pulsar was subsequently detected 
in HBAs with the LOFAR French station FR606 at the DM of the HBA census and this DM was used 
for folding in the current work. Since scattering is unlikely to be at play at this low DM, 
it is reasonable to assume that in our LBA observations the pulsar has not been detected either 
because it is intermittent or because its flux density is too low. The upper limit on the 
band-integrated flux density is $\sim35$\,mJy (Table~\ref{table:obssum}), which is comparable 
to the predicted flux density from the HBA census ($\sim20$\,mJy), so there is no clear 
indication of the spectral turnover. Note that both the upper limit and the predicted flux 
density value are subject to large, poorly constrained uncertainties. 

\subsection{Flux calibration}

The folded data files were calibrated in the same way as in the HBA census, thus we refer the 
reader for the details to B16 and \citet{Kondratiev2016}. In short, we have established the 
flux density scale using the radiometer equation \citep{Dicke1946}, which expresses the noise 
power through frequency-dependent antenna and sky temperatures, frequency- and direction-dependent
telescope gain, observing bandwidth, integration time and the number of polarisation summed. 
The instrument temperature was derived from  the measurements of \citet{Wijnholds2011}. 
The background sky temperature was scaled down to LBA frequencies from 408-MHz maps of \citet{Haslam1982}
with the spectral index of $-2.55$ \citep{Lawson1987}. For the antenna gain, we used the Hamaker 
model of a station beam \citep{Hamaker2006} calculated using the 
\texttt{mscorpol}\footnote{\url{https://github.com/2baOrNot2ba/mscorpol}} package by Tobia Carozzi.
A coherence factor of 0.85 was used  to scale the antenna gain with the actual number of 
stations involved in a given observation \citep{Kondratiev2016}.

For the Crab pulsar, the sky temperature was complemented with the contribution from the nebula. 
The latter was estimated with the relation $S_\mathrm{Jy} \approx 955 \nu_\mathrm{GHz}^{-0.27}$ 
\citep{Bietenholz1997,Cordes2004}. At 75\,MHz, the solid angle occupied by the nebula 
\citep[radius of $240\arcsec$,][]{Bietenholz1997} is smaller than the full-width at the half-maximum
of the LOFAR LBA beam \citep[$412.5\arcsec$][]{vanHaarlem2013}, thus the entire nebula
is contributing to the system temperature. We must note though that the Crab pulsar is scattered 
for more than a pulse period in the LBA band, thus the upper limit on the flux density is purely 
nominal, and, in fact, much smaller than the pulsar point source flux density measurements (B16).

The on- and off-pulse windows for calculating the mean S/N of the pulse profile were selected 
manually for each pulsar. Calibration was performed in each of 5-s subintegrations and 50 subbands 
of 11.7\,MHz. Zero-weighted sub-bands and/or sub-integrations were excluded from calculation of 
total band width / observing time. 

The nominal error on the flux density measurement, $\epsilon_{S\mathrm{nom}}$, set by the noise 
in the off-pulse window does not fully reflect the true flux density measurement uncertainty, 
since the latter is also influenced by the uncertainties in telescope parameters and scintillation. 
Since we did not observe any calibrator sources and were limited to only one session per pulsar, 
we can not estimate those errors separately. The only way to verify our measurements is to compare 
obtained flux density values to the ones from the literature. For those pulsars for which multiple 
spectra have been published (e.g. PSRs B0809+74, B1133+16, B1508+55, and others), the LBA 
measurements are consistent with the reported fluxes, which vary by a factor of a few.
More rigorous comparison performed by B16 for the HBA data, based on multiple observing sessions 
and more numerous literature measurements in the HBA frequency range, suggested adopting 50\% of 
measured flux density as the uncertainty on telescope parameters and scintillation. In this work 
we extend this uncertainty estimate to the LBA frequency range, deferring thorough study of 
telescope performance to future work. For our observing setup and pulsar sample, the 
scintillation-induced flux modulation index, calculated with the basic theory of diffractive and 
refractive scintillation (Appendix~\ref{app:scint}) is on the order of few percent for the majority 
of pulsars in the sample (but can be as large as 21\%, e.g. for the low-DM pulsar J1503+2111). 
The total flux density error was calculated by adding the nominal and the 50\% errors in
quadrature: $\epsilon_S = \sqrt{(0.5S)^2+\epsilon_{S\mathrm{nom}}^2}$. 

The mean band-integrated flux densities and their respective uncertainties are listed in 
Table~\ref{table:obssum}. For non-detected pulsars we adopted three times the nominal error as 
the upper limit, although this does not take into account possible signal smearing due to scattering.

Our observing setup involved six beams in a circle surrounding the central beam formed towards 
the pulsars' coordinates. Interestingly, 19 out of 44 detected pulsars were detected with 
the highest S/N in a side beam. This is indicative of refraction in the Earth’s ionosphere 
due to differential total electron content (TEC) between the lines-of-sight of different LBA stations.
We use the formula for the angular shift due to ionospheric refraction from \citet{Loi2015}:
\begin{equation}
|\Delta \theta|=\frac{40.3}{\nu^{2}} |\nabla_{\perp}\mathrm{TEC}|.
\label{eq:refr}
\end{equation}
Here the numerical coefficient stems from the combination of fundamental physical constants, 
the angular shift $\Delta \theta$ is in radians, the transverse gradient of total electron 
content (along the line of sight in the ionosphere) $\nabla_{\perp}\mathrm{TEC}$ 
is in electrons m$^{-3}$ and $\nu$ is in Hz. For the rough estimate of possible values of TEC 
gradients we used differences in TEC measured by the core HBA LOFAR stations in 
\citet{vanWeeren2016}. For $\delta \mathrm{TEC} \sim 10^{14}$\,m$^{-2}$ over the core size of 
80\,km, $\nabla_{\perp}\mathrm{TEC}$ is on the order of $10^{11}$\,m$^{-2}$. At the centre of 
LBA band ($\nu=60$\,MHz)  this gives $|\Delta \theta|\sim230\arcsec$ which is comparable to 
the beam separation. Smaller TEC differences, also reported by \citet{vanWeeren2016} will 
result in smaller refraction angles.

Appendix B shows average profiles, band-integrated or in 2--4 subbands
in case of strong pulsars or pulsars with interesting profile evolution.

\section{Flux density spectra}

\subsection{Fitting}
\label{subsec:fit}

For detected pulsars, flux density values were combined with the published measurements 
(see Table~\ref{table:obssum} for the full list of references) and these broadband spectra 
were fitted with a collection of power law (PL) functions. Similarly to B16, a Bayesian 
fit was performed in $\lg S-\lg\nu$ space. Each $\lg S$ was modelled as a normally distributed 
random variable. The mean of the normal distribution was defined by the assumed PL dependence 
and the standard deviation reflected any kind of  intrinsic variability or measurement 
uncertainty. See B16 for the remarks on PL applicability in general and the choice of a $\lg S$ model 
in particular.

Depending on the number of flux density measurements, we have approximated $\lg S_\mathrm{PL}$ 
either as a single PL (hereafter ``1PL''):
 \begin{equation}
  \lg S_\mathrm{1PL} = \alpha \lg (\nu/\nu_0) + \lg S_0,
 \end{equation}
 a broken PL with one break (2PL):
 \begin{equation}
\lg S_\mathrm{2PL} = \begin{cases}
          \alpha_\mathrm{lo}\lg(\nu/\nu_0) + \lg S_0,  & \nu<\nu_\mathrm{br}\\
        \alpha_\mathrm{hi}\lg(\nu/\nu_\mathrm{br}) + \alpha_\mathrm{lo}\lg(\nu_\mathrm{br}/\nu_0) + \lg S_0, & \nu>\nu_\mathrm{br},
        \end{cases}
 \end{equation}
or a broken PL with two breaks (3PL):
\begin{equation}
\lg S_\mathrm{3PL} = \begin{cases}
          \alpha_\mathrm{lo}\lg(\nu/\nu_0) + \lg S_0,  & \nu<\nu^\mathrm{lo}_\mathrm{br}\\
        \alpha_\mathrm{mid}\lg(\nu/\nu^\mathrm{lo}_\mathrm{br}) + \alpha_\mathrm{lo}\lg(\nu^\mathrm{lo}_\mathrm{br}/\nu_0) + \lg S_0, & \nu^\mathrm{lo}_\mathrm{br}<\nu<\nu^\mathrm{hi}_\mathrm{br}\\
        \alpha_\mathrm{hi}\lg(\nu/\nu^\mathrm{hi}_\mathrm{br}) + \alpha_\mathrm{mid}\lg(\nu^\mathrm{hi}_\mathrm{br}/\nu_0) + \lg S_0, & \nu>\nu^\mathrm{hi}_\mathrm{br}.
        \end{cases}
\end{equation}
The reference frequency $\nu_0$ was taken to be close to the geometric average of the 
minimum and maximum frequencies in the spectrum. 

As in B16, for the small number of flux density measurements (treating all measurements 
within 10\% in frequency as a single group), we fixed the uncertainty at the level defined 
by the reported errors. For larger number of frequency groups, an additional fit parameter
was introduced,  the unknown error $\sigma^\mathrm{unkn}_{\lg S}$. A single error per source 
was fitted, representing intrinsic variability, or any kind of unaccounted propagation or 
instrumental error. The total flux density uncertainty of any $\lg S$ was then taken as 
the known and unknown errors added in quadrature. 

The posterior distribution of $\sigma^\mathrm{unkn}_{\lg S}$ was used to discriminate between 
models. 1PL was taken as a null hypothesis and rejected in favour of 2PL or 3PL if the latter 
gave statistically smaller $\sigma^\mathrm{unkn}_{\lg S}$. For the details we refer the reader 
to B16. If no $\sigma^\mathrm{unkn}_{\lg S}$ was fitted, we adopted 1PL as the single model. 
In a few cases, when the data showed a hint of a spectral break, but no 
$\sigma^\mathrm{unkn}_{\lg S}$ was fitted, we fitted 2PL with the break frequency fixed at the 
frequency of the largest flux density measurement. For such pulsars we give both 1PL and 2PL 
values of the fitted parameters. Upper limits on flux densities were not taken into account 
while fitting.

\subsection{Results}
\label{subsec:results}

Seventeen out of the 43 detected pulsars did not have published flux density measurements in the 
LBA frequency range. Some of them do not show signs of scattering, which indicates an opportunity 
to study these pulsars at even lower frequencies: for example PSRs B0011+47, B0226+70, and B2022+50 
did not exhibit any sign of low-frequency turnover or apparent scattering down to 40\,MHz (see Figures 
in the Appendix~\ref{app:plots}).

Overall, our new spectral indices are very similar to the ones published in B16. For pulsars 
with relatively well-measured spectra, the  LBA flux densities agree reasonably well with previous 
measurements in this frequency range. In some cases (e.g. PSRs B0450+55, B0655+64, B2217+47), 
LBA flux densities are lower by a factor of a few with respect to the measurements of 
\citet{Stovall2015}. However, this is not the case for all LBA census pulsars that overlap with their 
sample (e.g. PSR B1929+21) and there is at least one example where fluxes from \citet{Stovall2015}
are much higher than the bulk of other literature measurements in the same frequency region (PSR B1133+16).

For pulsars with fewer spectral points and no previous measurements below 100\,MHz, LBA fluxes 
did not span enough of the frequency range to have a large influence on the spectral index. For 
some sources the S/N was sufficient to break the LBA band into two or more subbands and the flux 
density values hint to a possible spectral break (e.g. PSR J0611+30), however large errors and 
close proximity in frequency between the new data points do not make the break statistically significant. 

Some of the sources had a different number of PL breaks than in B16 (e.g. PSRs B0450+55, B1133+16, 
B1811+40, B2217+47). This mostly stems from the influence of separate flux measurements on the 
fitted $\sigma^\mathrm{unkn}_{\lg S}$: since we fitted only one unknown error per source. Sometimes 
the low-frequency (or even high-frequency) breaks were substantially different than in B16 (e.g. 
PSRs B0823+26, B1530+27, B1737+13).

It is worth mentioning that parallel to this study, a similar census of the pulsar population 
visible below 100 MHz was undertaken by the LOFAR international station FR606 (Bondonneau et al, 
submitted). They observed a similar sample of pulsars (103 compared to the 88 pulsars of the 
present study). The reduced collecting area ($\sim$10\%) was compensated by long integrations 
(on average 3h per target). With this, the authors detected 64 pulsars, compared to the 43 
pulsars of the present study. For the detailed comparison of the results, we refer the reader to 
Bondonneau et al. (submitted).

\section{Discussion}

Despite its limitations, the LBA extension of the HBA census provides useful information, 
identifying more than a dozen pulsars suitable for the subsequent follow-up at the lowest 
frequencies observable from Earth. We provide reference average profiles and fluxes for the 
43 pulsars detected, 17 of them having no previously published flux densities at these 
frequencies.

Overall, the main concerns raised in B16 remain standing: despite being one of the basic 
characteristics of pulsars, their spectra remain, to a large extent, poorly constrained due 
to the lack of robust, systematic multifrequency measurements. The situation is even worse 
in the low-frequency range, where a spectral break is widely expected.

Proper quantification of the low-frequency spectral break is essential for studying the emission 
mechanism and propagation of radio waves in the magnetosphere. The existence of a low-frequency 
turnover has been attributed previously to a number of physical processes, for example 
synchrotron self-absorption \citep{Sieber1973},  refraction of the ordinary radio wave mode 
\citep{Beskin2018}, free-free absorption \citep{Malov1979}, or stimulated scattering 
\citep{Lyubarskii1996}. In particular, \citet{Beskin2018} proposes clear dependence of 
the turnover frequency on the pulsar spin period, something that may be readily verified 
using the future observations. However, such study would require a large number of well-measured 
spectra from pulsars of substantially different periods. Furthermore, the influence of the ISM 
on the observed flux densities (for example, decrease in apparent pulsed flux density due 
to scattering) should be carefully accounted for, for example measuring continuum flux densities 
at low radio frequencies \citep{Shimwell2019}.

Future observations of pulsars below 100\,MHz can provide more robust flux density measurements 
and better constrain spectral break(s). This will be achieved, in particular, by the NenuFAR 
radio telescope \citep{ZarkaNenufarSF2A,nenufar,ZarkaNenufarIEEE} and its pulsar instrumentation 
LUPPI \citep{BondonneauURSI}, with which we are currently conducting a systematic census of 
the pulsar population (Bondonneau et al. in prep).  Due to its sensitivity, its constant 
antenna response across its frequency band (10--85 MHz), and long integrations, it is 
expected that NenuFAR will detect a much larger number of pulsars compared to the LBA census 
and its companion study with FR606.

\begin{acknowledgements}
This works makes extensive use of \texttt{matplotlib}\footnote{\url{http://matplotlib.org/}} \citep{Hunter2007}, 
\texttt{seaborn}\footnote{\url{http://stanford.edu/~mwaskom/software/seaborn/}}
Python plotting libraries and NASA’s Astrophysics Data System. 

This paper is based on data obtained with the International LOFAR Telescope (ILT) under project code \url{LC2_025}. 
LOFAR \citep{vanHaarlem2013} is the Low Frequency Array designed and constructed by ASTRON. It has observing, data 
processing, and data storage facilities in several countries, that are owned by various parties (each with their 
own funding sources), and that are collectively operated by the ILT foundation under a joint scientific policy. 
The ILT resources have benefited from the following recent major funding sources: CNRS-INSU, Observatoire de Paris 
and Université d'Orléans, France; BMBF, MIWF-NRW, MPG, Germany; Science Foundation Ireland (SFI), Department of 
Business, Enterprise and Innovation (DBEI), Ireland; NWO, The Netherlands; The Science and Technology Facilities Council, UK.

\end{acknowledgements}


\appendix

\section{Scintillation}
\label{app:scint}
The variation of observed flux due to scintillation on the inhomogeneities in the interstellar medium 
was estimated with a  simple thin-screen Kolmogorov model \citep[see][for review]{Lorimer2005}. 
The scintillation bandwidth was taken to be $\Delta f = 1.16/(2\pi \tau_\mathrm{scat})\times($60\,MHz/1\,GHz$)^{4.0}$, 
where $\tau_\mathrm{scat}$ is scattering time at 1\,GHz from \citet{Yao2017}. 
For all census pulsars the scintillation bandwidth $\Delta f$ was smaller than a few kHz,
satisfying the conditions of strong scintillation regime  ($\sqrt{f/\Delta f}>1$).

Diffractive interstellar scintillation (DISS) did not have a large impact on the flux variation since many scintles were 
averaged in the frequency domain, resulting in modulation index $m_\mathrm{DISS}$ 
(rms of the flux density divided by its mean value) was on the order of a percent or less.
The refractive scintillation (RISS) was much stronger  with typical $m_\mathrm{RISS} \approx 0.05 - 0.1$. 
Table\,\ref{table:obssum} lists the expected values of total modulation index.

\section{Tables}
\label{app:fit}

Table~\ref{table:obssum} summarises observations.  The columns indicate: pulsar name; 
approximate spin period (s); observing epoch (MJD); duration of an observing session (min);
frequency range (MHz); best beam (usually the one with highest S/N); fraction of zapped data in the dynamic spectrum 
for the least and the most affected beams;
peak S/N of the average profile; DM from B16; 
measured census DM; expected \citet{Yao2017} scattering time at 60\,MHz divided by pulsar period; expected
modulation index due to scintillation in the ISM; 
mean flux density within specified frequency range (upper limit for the non-detected pulsars), 
and the literature references to previous flux density measurements. The values 
in parentheses indicate the errors on the last two significant digits.

Tables \ref{table:1pl}--\ref{table:3pl} contain best fitted parameters for the pulsars with the spectra
modelled with a single PL, a PL with one break and a PL with two breaks, respectively. The columns 
include pulsar name; spectral frequency span (MHz); number of data points in spectrum, $N_p$; 
the reference frequency, $\nu_0$ (MHz); flux density at the reference frequency, $S_0$ (mJy); 
spectral index (or indices in case of broken PLs), $\alpha$; and fitted flux density scatter, $\su$, 
if applicable (see Sect.~\ref{subsec:fit}).
Tables \ref{table:2pl} and \ref{table:3pl} also include break frequency, $\nu_\mathrm{br}$ (MHz), together with its 68\% 
uncertainty range.

\onecolumn

\setlength\LTcapwidth{7.2in}
\setlength{\tabcolsep}{2pt}
\renewcommand{\arraystretch}{1.05}
\begin{center}
\footnotesize
\begin{longtable}{|l|c|c|c|l|c|c|c|l|l|c|c|l|l|}
\caption{Observation summary, DM and flux density measurements. \label{table:obssum}}\\
\hline
\parbox[t][][t]{1.5cm}{\centering PSR} & 
\parbox[t][][t]{0.9cm}{\centering Period $P$\\ (s) } & 
\parbox[t][][t]{1.4cm}{\centering Observing \\ epoch (MJD)} &
\parbox[t][][t]{0.7cm}{\centering Obs. time \\ (min)} & 
\parbox[t][][t]{1.4cm}{\centering Freq. range \\ (MHz)} & 
\parbox[t][][t]{0.7cm}{\centering Best \\ beam } & 
\parbox[t][][t]{0.8cm}{\centering Zapped fraction  } & 
\parbox[t][][t]{0.8cm}{\centering Peak S/N} & 
\parbox[t][][t]{1.6cm}{\centering $\mathrm{DM_\mathrm{cen}}$ \\ (pc cm$^{-3}$)} & 
\parbox[t][][t]{1.0cm}{\centering Expec-\\ted $\tau_\mathrm{scat}/P$\\ } &
\parbox[t][][t]{1.0cm}{\centering Exp. mod. index } &
\parbox[t][][t]{1.0cm}{\centering Mean \\ flux \\(mJy) } &
\parbox[t][][t]{2.1cm}{\centering Literature flux references } 
\\  
\hline
\endfirsthead
\multicolumn{7}{c}%
{\tablename\ \thetable\ -- \textit{Continued from previous page}} \\
\hline
\parbox[t][][t]{1.5cm}{\centering PSR} & 
\parbox[t][][t]{0.9cm}{\centering Period $P$\\ (s) } & 
\parbox[t][][t]{1.4cm}{\centering Observing \\ epoch (MJD)} &
\parbox[t][][t]{0.7cm}{\centering Obs. time \\ (min)} & 
\parbox[t][][t]{1.4cm}{\centering Freq. range \\ (MHz)} & 
\parbox[t][][t]{0.7cm}{\centering Best \\ beam } & 
\parbox[t][][t]{0.8cm}{\centering Zapped fraction} & 
\parbox[t][][t]{0.8cm}{\centering Peak S/N} & 
\parbox[t][][t]{1.6cm}{\centering $\mathrm{DM_\mathrm{cen}}$ \\ (pc cm$^{-3}$)} & 
\parbox[t][][t]{1.0cm}{\centering Expec-\\ted $\tau_\mathrm{scat}/P$\\ } &
\parbox[t][][t]{1.0cm}{\centering Exp. mod. index } &
\parbox[t][][t]{1.0cm}{\centering Mean  \\ flux \\(mJy) } &
\parbox[t][][t]{2.1cm}{\centering Literature flux references } 
\\  
\hline
\endhead
\hline \multicolumn{7}{r}{\textit{Continued on next page}} \\
\endfoot 
\multicolumn{13}{l}{\parbox[t][][t]{7.2in}{\vspace{-4mm}{
\tablebib{ 
[1] \citet{Bridle1970}; [2] \citet{Barr2013}; [3] \citet{Bilous2016}; [4] \citet{Boyles2013}; [5] \citet{Bhat1999}; 
[6] \citet{Bartel1978}; [7] \citet{ChampionA2005}; [8] \citet{Camilo1995}; [9] \citet{CamiloB1996}; 
[10] \citet{Downs1979}; [11] \citet{Dembska2014}; [12] \citet{Davies1977}; [13] \citet{Deshpande1992}; [14] \citet{Downs1973}; 
[15] \citet{Damashek1978}; [16] \citet{Dewey1985}; [17] \citet{Fomalont1992}; [18] \citet{Gould1994}; [19] \citet{Han2009}; 
[20] \citet{HobbsA2004}; [21] \citet{Izvekova1979}; [22] \citet{Izvekova1981}; [23] \citet{Johnston2006}; [24] \citet{Jankowski2018}; 
[25] \citet{Kaplan1998}; [26] \citet{Kijak2007}; [27] \citet{Kramer1997}; [28] \citet{Kijak1998}; [29] \citet{Kuzmin1986}; 
[30] \citet{Krzeszowski2014}; [31] \citet{Kuzmin1978}; [32] \citet{Karuppusamy2011}; [33] \citet{Kramer1996}; [34] \citet{Lane2014}; 
[35] \citet{Loehmer2008}; [36] \citet{Lewandowski2004}; [37] \citet{LorimerA2005}; [38] \citet{Lorimer1995}; [39] \citet{Lommen2000}; 
[40] \citet{Manchester1971}; [41] \citet{Malofeev1993}; [42] \citet{Malofeev1999}; [43] \citet{Morris1981}; [44] \citet{Moffett1999}; 
[45] \citet{Murphy2017}; [46] \citet{Maron2000}; [47] \citet{Manchester1978}; [48] \citet{Malofeev1980}; [49] \citet{Malofeev2000}; 
[50] \citet{Manchester1981}; [51] \citet{Navarro2003}; [52] \citet{Rankin1981}; [53] \citet{Rankin1970}; [54] \citet{Ray1996}; 
[55] \citet{Slee1986}; [56] \citet{Stinebring1990}; [57] \citet{Seiradakis1995}; [58] \citet{Sayer1997}; [59] \citet{Stovall2015}; 
[60] \citet{Stokes1986}; [61] \citet{Shrauner1998}; [62] \citet{Sieber1987}; [63] \citet{Thorsett1993}; [64] \citet{Taylor1995}; 
[65] \citet{Vivekanand1983}; [66] \citet{Wielebinski1993}; [67] \citet{Wang2005}; [68] \citet{Xue2017}; [69] \citet{Zakharenko2013}; [70] \citet{Zhao2017}. 
}}}}\\
\multicolumn{13}{l}{\parbox[t][][t]{7.0in}{\textbf{Notes.}
\tablefoottext{$\sharp$} {PSRs B0105+68, B0643+80, B0656+14 were excluded from analysis because of RFI contamination.}}}
\endlastfoot
\hline 
B0011+47 & 1.241 & 56949.82 & 21 & 42.12--76.15 & 0 & 0.21--0.43 & \hphantom{000}6 & \hphantom{0}30.3048\,(65) & 1e-02 & 0.08 & \hphantom{00.}45 $\pm$ 25 & \parbox[t][][t]{2.5cm}{\tiny  3, 15, 19, 22, 25, 38, 42, 46, 57} \\ 
B0045+33 & 1.217 & 57112.46 & 21 & 42.35--77.09 & \hphantom{0}$\dots$ & 0.45--0.55 & \hphantom{0}$\dots$ & \hphantom{0}$\dots$ & 4e-02 & 0.06 & \hphantom{00000}<56.0 & \parbox[t][][t]{2.5cm}{\tiny  3, 16, 38, 46, 49, 57, 64} \\ 
B0052+51 & 2.115 & 57114.46 & 36 & 42.10--77.23 & \hphantom{0}$\dots$ & 0.16--0.17 & \hphantom{0}$\dots$ & \hphantom{0}$\dots$ & 3e-02 & 0.06 & \hphantom{00000}<22.6 & \parbox[t][][t]{2.5cm}{\tiny  3, 16, 19, 25, 38, 46, 49} \\ 
B0053+47 & 0.472 & 57027.69 & 20 & 30.37--77.24 & 5 & 0.05--0.06 & \hphantom{00}15 & \hphantom{0}18.0954\,(10) & 6e-03 & 0.10 & \hphantom{0.}110 $\pm$ 60 & \parbox[t][][t]{2.5cm}{\tiny  3, 16, 28, 38, 49, 64, 69} \\ 
B0105+68$^\sharp$ & 1.071 & 57148.40 & 20 & 42.05--77.24 & \hphantom{0}$\dots$ & 0.81--0.85 & \hphantom{0}$\dots$ & \hphantom{0}$\dots$ & 2e-01 & 0.05 & \hphantom{00000}<140.7 & \parbox[t][][t]{2.5cm}{\tiny  3, 16, 38, 49, 57, 64} \\ 
B0114+58 & 0.101 & 57112.48 & 20 & 42.41--77.19 & \hphantom{0}$\dots$ & 0.48--0.63 & \hphantom{0}$\dots$ & \hphantom{0}$\dots$ & 1e+00 & 0.06 & \hphantom{00000}<99.8 & \parbox[t][][t]{2.5cm}{\tiny  3, 19, 35, 38, 49, 60} \\ 
J0137+1654 & 0.415 & 56827.22 & 20 & 42.15--77.29 & \hphantom{0}$\dots$ & 0.17--0.35 & \hphantom{0}$\dots$ & \hphantom{0}$\dots$ & 2e-02 & 0.08 & \hphantom{00000}<42.8 & \parbox[t][][t]{2.5cm}{\tiny  3, 37, 69} \\ 
B0136+57 & 0.272 & 56827.28 & 20 & 42.15--77.19 & \hphantom{0}$\dots$ & 0.26--0.45 & \hphantom{0}$\dots$ & \hphantom{0}$\dots$ & 2e+00 & 0.04 & \hphantom{00000}<47.8 & \parbox[t][][t]{2.5cm}{\tiny  3, 15, 22, 25, 38, 46, 49, 56, 57} \\ 
B0153+39 & 1.811 & 56827.24 & 31 & 42.14--77.13 & \hphantom{0}$\dots$ & 0.32--0.59 & \hphantom{0}$\dots$ & \hphantom{0}$\dots$ & 1e-01 & 0.05 & \hphantom{00000}<52.6 & \parbox[t][][t]{2.5cm}{\tiny  3, 16, 19, 38, 49, 64} \\ 
B0226+70 & 1.467 & 56827.26 & 25 & 42.26--77.05 & 0 & 0.22--0.36 & \hphantom{000}7 & \hphantom{0}46.7394\,(66) & 5e-02 & 0.06 & \hphantom{00.}49 $\pm$ 29 & \parbox[t][][t]{2.5cm}{\tiny  3, 16, 38, 42, 46, 49, 57, 64} \\ 
B0301+19 & 1.388 & 56975.00 & 24 & 42.13--77.13 & 0 & 0.28--0.57 & \hphantom{00}10 & \hphantom{0}15.6568\,(99) & 1e-03 & 0.11 & \hphantom{00.}61 $\pm$ 33 & \parbox[t][][t]{2.5cm}{\tiny  3, 17, 21, 22, 25, 28, 38, 42, 46, 47, 48, 49, 56, 57, 59, 69} \\ 
B0320+39 & 3.032 & 56903.21 & 65 & 42.08--77.29 & 3 & 0.38--0.41 & \hphantom{00}29 & \hphantom{0}26.1698\,(20) & 3e-03 & 0.08 & \hphantom{00.}76 $\pm$ 39 & \parbox[t][][t]{2.5cm}{\tiny  3, 15, 17, 22, 28, 29, 38, 42, 46, 57, 59, 69} \\ 
J0324+5239 & 0.337 & 56903.25 & 20 & 42.16--77.18 & \hphantom{0}$\dots$ & 0.28--0.54 & \hphantom{0}$\dots$ & \hphantom{0}$\dots$ & 9e+00 & 0.03 & \hphantom{00000}<70.5 & \parbox[t][][t]{2.5cm}{\tiny  2, 3} \\ 
B0410+69 & 0.391 & 56975.03 & 20 & 42.18--77.29 & \hphantom{0}$\dots$ & 0.17--0.47 & \hphantom{0}$\dots$ & \hphantom{0}$\dots$ & 3e-02 & 0.08 & \hphantom{00000}<38.4 & \parbox[t][][t]{2.5cm}{\tiny  3, 16, 19, 28, 38, 49, 57, 69} \\ 
J0417+35 & 0.654 & 56975.02 & 20 & 42.08--77.26 & \hphantom{0}$\dots$ & 0.21--0.52 & \hphantom{0}$\dots$ & \hphantom{0}$\dots$ & 1e-01 & 0.06 & \hphantom{00000}<38.0 & \parbox[t][][t]{2.5cm}{\tiny  3, 9, 49} \\ 
B0450+55 & 0.341 & 56903.26 & 20 & 42.23--76.93 & 0 & 0.31--0.53 & \hphantom{00}12 & \hphantom{0}14.590\,(77) & 5e-03 & 0.11 & \hphantom{0.}110 $\pm$ 60 & \parbox[t][][t]{2.5cm}{\tiny  3, 25, 38, 46, 49, 50, 57, 59, 61, 69} \\ 
B0531+21 & 0.034 & 56903.27 & 20 & 41.89--77.28 & \hphantom{0}$\dots$ & 0.45--0.53 & \hphantom{0}$\dots$ & \hphantom{0}$\dots$ & 5e+00 & 0.05 & \hphantom{00000}<70.6 & \parbox[t][][t]{2.5cm}{\tiny  1, 3, 28, 38, 40, 42, 44, 48, 49, 50, 53, 57} \\ 
J0611+30 & 1.412 & 56975.08 & 24 & 42.13--77.22 & 0 & 0.19--0.49 & \hphantom{000}8 & \hphantom{0}45.2951\,(81) & 5e-02 & 0.06 & \hphantom{00.}89 $\pm$ 46 & \parbox[t][][t]{2.5cm}{\tiny  3, 9} \\ 
B0609+37 & 0.298 & 56975.07 & 20 & 42.10--77.22 & 4 & 0.15--0.41 & \hphantom{000}6 & \hphantom{0}27.175\,(49) & 4e-02 & 0.08 & \hphantom{00.}46 $\pm$ 25 & \parbox[t][][t]{2.5cm}{\tiny  3, 28, 38, 46, 49, 57, 60, 69} \\ 
B0626+24 & 0.476 & 56903.29 & 20 & 42.04--77.24 & \hphantom{0}$\dots$ & 0.56--0.64 & \hphantom{0}$\dots$ & \hphantom{0}$\dots$ & 2e+00 & 0.04 & \hphantom{00000}<51.5 & \parbox[t][][t]{2.5cm}{\tiny  3, 15, 17, 25, 28, 38, 46, 49, 50, 57} \\ 
B0643+80$^\sharp$ & 1.215 & 56903.34 & 21 & 42.17--77.09 & \hphantom{0}$\dots$ & 0.68--0.79 & \hphantom{0}$\dots$ & \hphantom{0}$\dots$ & 2e-02 & 0.07 & \hphantom{00000}<66.8 & \parbox[t][][t]{2.5cm}{\tiny  3, 28, 38, 42, 49, 57} \\ 
B0655+64 & 0.196 & 56903.32 & 20 & 42.29--77.23 & 0 & 0.60--0.67 & \hphantom{00}14 & \hphantom{00}8.7739\,(19) & 2e-03 & 0.14 & \hphantom{00.}86 $\pm$ 49 & \parbox[t][][t]{2.5cm}{\tiny  3, 15, 38, 41, 57, 59, 61, 69} \\ 
B0656+14$^\sharp$ & 0.385 & 56903.33 & 20 & 42.06--77.06 & \hphantom{0}$\dots$ & 0.82--0.98 & \hphantom{0}$\dots$ & \hphantom{0}$\dots$ & 4e-03 & 0.11 & \hphantom{00000}<94.8 & \parbox[t][][t]{2.5cm}{\tiny  3, 22, 23, 24, 38, 46, 47, 49, 57, 65, 69, 70} \\ 
B0751+32 & 1.442 & 56975.11 & 25 & 42.09--77.35 & 4 & 0.19--0.50 & \hphantom{000}4 & \hphantom{0}39.846\,(84) & 3e-02 & 0.06 & \hphantom{00.}21 $\pm$ 13 & \parbox[t][][t]{2.5cm}{\tiny  3, 15, 22, 28, 38, 49, 57} \\ 
B0809+74 & 1.292 & 56903.37 & 22 & 30.64--77.17 & 3 & 0.55--0.68 & \hphantom{00}34 & \hphantom{00}5.7707\,(84) & 1e-04 & 0.17 & \hphantom{.}1400 $\pm$ 700 & \parbox[t][][t]{2.5cm}{\tiny  3, 6, 13, 21, 22, 25, 28, 29, 31, 34, 38, 42, 43, 46, 48, 49, 50, 57, 61, 69} \\ 
B0823+26 & 0.531 & 56975.12 & 20 & 30.37--77.19 & 4 & 0.16--0.44 & \hphantom{0}107 & \hphantom{0}19.4763\,(35) & 7e-03 & 0.10 & \hphantom{0.}700 $\pm$ 350 & \parbox[t][][t]{2.5cm}{\tiny  3, 5, 6, 10, 14, 18, 21, 22, 25, 29, 33, 35, 37, 38, 42, 43, 45, 46, 48, 49, 50, 57, 59, 61, 66, 67, 69} \\ 
B0841+80 & 1.602 & 56975.14 & 27 & 42.33--77.18 & \hphantom{0}$\dots$ & 0.19--0.49 & \hphantom{0}$\dots$ & \hphantom{0}$\dots$ & 2e-02 & 0.07 & \hphantom{00000}<24.5 & \parbox[t][][t]{2.5cm}{\tiny  3, 16, 19, 49, 57, 64} \\ 
B0917+63 & 1.568 & 56975.16 & 27 & 42.09--77.34 & 0 & 0.21--0.51 & \hphantom{00}13 & \hphantom{0}13.1542\,(42) & 8e-04 & 0.12 & \hphantom{00.}41 $\pm$ 22 & \parbox[t][][t]{2.5cm}{\tiny  3, 16, 19, 38, 49, 64, 69} \\ 
B0940+16 & 1.087 & 56903.40 & 20 & 42.01--77.19 & \hphantom{0}$\dots$ & 0.64--0.74 & \hphantom{0}$\dots$ & \hphantom{0}$\dots$ & 4e-03 & 0.10 & \hphantom{00000}<61.2 & \parbox[t][][t]{2.5cm}{\tiny  3, 17, 22, 24, 37, 38, 47, 57, 65, 69} \\ 
J0943+22 & 0.533 & 56903.39 & 20 & 42.08--77.53 & \hphantom{0}$\dots$ & 0.52--0.64 & \hphantom{0}$\dots$ & \hphantom{0}$\dots$ & 2e-02 & 0.08 & \hphantom{00000}<39.2 & \parbox[t][][t]{2.5cm}{\tiny  3, 49, 63} \\ 
B0943+10 & 1.098 & 56826.71 & 20 & 30.40--77.17 & 0 & 0.16--0.36 & \hphantom{00}50 & \hphantom{0}15.3585\,(72) & 2e-03 & 0.11 & \hphantom{0.}400 $\pm$ 200 & \parbox[t][][t]{2.5cm}{\tiny  3, 13, 22, 38, 42, 48, 49, 50, 52, 57, 61, 69} \\ 
J0947+27 & 0.851 & 57109.86 & 20 & 42.11--77.23 & \hphantom{0}$\dots$ & 0.07--0.09 & \hphantom{0}$\dots$ & \hphantom{0}$\dots$ & 2e-02 & 0.08 & \hphantom{00000}<19.7 & \parbox[t][][t]{2.5cm}{\tiny  3, 54, 69} \\ 
B1112+50 & 1.656 & 57028.15 & 28 & 30.37--77.24 & 0 & 0.07--0.10 & \hphantom{00}26 & \hphantom{00}9.1863\,(11) & 3e-04 & 0.14 & \hphantom{00.}43 $\pm$ 22 & \parbox[t][][t]{2.5cm}{\tiny  3, 21, 22, 32, 38, 42, 43, 46, 48, 49, 50, 57, 59, 69} \\ 
B1133+16 & 1.188 & 56826.74 & 20 & 30.39--77.21 & 4 & 0.17--0.27 & \hphantom{0}135 & \hphantom{00}4.8407\,(78) & 9e-05 & 0.18 & \hphantom{0.}880 $\pm$ 440 & \parbox[t][][t]{2.5cm}{\tiny  3, 5, 6, 10, 13, 14, 18, 21, 22, 23, 24, 25, 29, 31, 30, 32, 33, 34, 35, 38, 42, 43, 46, 47, 48, 49, 50, 57, 59, 61, 62, 66, 68, 69, 70} \\ 
J1238+21 & 1.119 & 56826.80 & 20 & 42.11--77.26 & 4 & 0.14--0.17 & \hphantom{00}18 & \hphantom{0}17.9706\,(79) & 3e-03 & 0.10 & \hphantom{00.}37 $\pm$ 20 & \parbox[t][][t]{2.5cm}{\tiny  3, 49, 54, 69} \\ 
B1237+25 & 1.383 & 56826.76 & 24 & 42.11--77.25 & 4 & 0.08--0.12 & \hphantom{00}61 & \hphantom{00}9.2716\,(90) & 4e-04 & 0.14 & \hphantom{0.}150 $\pm$ 80 & \parbox[t][][t]{2.5cm}{\tiny  3, 5, 6, 10, 17, 18, 21, 22, 25, 29, 31, 38, 42, 43, 46, 48, 49, 50, 57, 59, 61, 62, 69, 70} \\ 
J1313+0931 & 0.849 & 56826.79 & 20 & 42.25--76.00 & 4 & 0.19--0.39 & \hphantom{000}6 & \hphantom{0}12.0406\,(15) & 1e-03 & 0.12 & \hphantom{00.}24 $\pm$ 19 & \parbox[t][][t]{2.5cm}{\tiny  3, 39, 69} \\ 
B1322+83 & 0.670 & 57007.33 & 20 & 42.19--77.17 & 3 & 0.20--0.22 & \hphantom{000}5 & \hphantom{0}13.2962\,(30) & 2e-03 & 0.12 & \hphantom{00.}20 $\pm$ 13 & \parbox[t][][t]{2.5cm}{\tiny  3, 19, 28, 38, 42, 49, 50, 64, 69} \\ 
J1503+2111 & 3.314 & 56826.82 & 81 & 42.18--77.27 & \hphantom{0}$\dots$ & 0.54--0.67 & \hphantom{0}$\dots$ & \hphantom{0}$\dots$ & 1e-05 & 0.21 & \hphantom{00000}<35.9 & \parbox[t][][t]{2.5cm}{\tiny  3, 7, 19, 69} \\ 
B1508+55 & 0.740 & 56826.86 & 20 & 30.26--77.34 & 4 & 0.21--0.45 & \hphantom{00}82 & \hphantom{0}19.6189\,(48) & 5e-03 & 0.10 & \hphantom{0.}390 $\pm$ 190 & \parbox[t][][t]{2.5cm}{\tiny  3, 5, 18, 21, 22, 25, 29, 31, 34, 38, 42, 43, 46, 48, 49, 56, 57, 59, 61, 69} \\ 
B1530+27 & 1.125 & 57007.39 & 20 & 42.13--77.22 & 0 & 0.20--0.22 & \hphantom{00}24 & \hphantom{0}14.711\,(28) & 1e-03 & 0.11 & \hphantom{00.}78 $\pm$ 40 & \parbox[t][][t]{2.5cm}{\tiny  3, 15, 37, 38, 41, 42, 46, 49, 57, 69} \\ 
B1541+09 & 0.748 & 56826.89 & 20 & 42.20--77.15 & 4 & 0.18--0.38 & \hphantom{000}8 & \hphantom{0}34.9958\,(46) & 4e-02 & 0.07 & \hphantom{0.}310 $\pm$ 160 & \parbox[t][][t]{2.5cm}{\tiny  3, 21, 22, 28, 38, 42, 43, 46, 47, 48, 49, 50, 56, 57, 59, 61} \\ 
J1549+2113 & 1.263 & 56913.72 & 22 & 42.29--77.10 & \hphantom{0}$\dots$ & 0.55--0.73 & \hphantom{0}$\dots$ & \hphantom{0}$\dots$ & 6e-03 & 0.09 & \hphantom{00000}<91.8 & \parbox[t][][t]{2.5cm}{\tiny  3, 19, 37, 49, 69} \\ 
J1612+2008 & 0.427 & 56826.95 & 20 & 42.16--77.23 & \hphantom{0}$\dots$ & 0.17--0.34 & \hphantom{0}$\dots$ & \hphantom{0}$\dots$ & 9e-03 & 0.10 & \hphantom{00000}<55.1 & \parbox[t][][t]{2.5cm}{\tiny  3, 4} \\ 
J1627+1419 & 0.491 & 56826.91 & 20 & 42.11--77.21 & \hphantom{0}$\dots$ & 0.12--0.22 & \hphantom{0}$\dots$ & \hphantom{0}$\dots$ & 4e-02 & 0.07 & \hphantom{00000}<68.0 & \parbox[t][][t]{2.5cm}{\tiny  3, 19, 36, 49} \\ 
B1633+24 & 0.491 & 56949.66 & 20 & 42.20--77.43 & 5 & 0.41--0.62 & \hphantom{000}6 & \hphantom{0}24.2471\,(24) & 2e-02 & 0.09 & \hphantom{00.}72 $\pm$ 41 & \parbox[t][][t]{2.5cm}{\tiny  3, 38, 49, 50, 57, 64, 65, 69} \\ 
J1645+1012 & 0.411 & 56826.93 & 20 & 42.17--77.19 & 3 & 0.15--0.34 & \hphantom{000}4 & \hphantom{0}36.171\,(16) & 7e-02 & 0.07 & \hphantom{00.}64 $\pm$ 39 & \parbox[t][][t]{2.5cm}{\tiny  3, 36, 49} \\ 
J1649+2533 & 1.015 & 56949.68 & 20 & 41.97--77.42 & \hphantom{0}$\dots$ & 0.47--0.71 & \hphantom{0}$\dots$ & \hphantom{0}$\dots$ & 2e-02 & 0.07 & \hphantom{00000}<68.8 & \parbox[t][][t]{2.5cm}{\tiny  3, 19, 36, 49} \\ 
J1652+2651 & 0.916 & 57107.21 & 20 & 42.10--77.23 & \hphantom{0}$\dots$ & 0.05--0.07 & \hphantom{0}$\dots$ & \hphantom{0}$\dots$ & 5e-02 & 0.06 & \hphantom{00000}<31.9 & \parbox[t][][t]{2.5cm}{\tiny  3, 19, 36, 37, 49, 58} \\ 
J1720+2150 & 1.616 & 56913.74 & 27 & 42.57--77.12 & \hphantom{0}$\dots$ & 0.41--0.60 & \hphantom{0}$\dots$ & \hphantom{0}$\dots$ & 3e-02 & 0.06 & \hphantom{00000}<57.6 & \parbox[t][][t]{2.5cm}{\tiny  3, 19, 36, 49} \\ 
B1737+13 & 0.803 & 56826.96 & 20 & 42.06--77.23 & 2 & 0.19--0.44 & \hphantom{00}19 & \hphantom{0}48.6682\,(11) & 1e-01 & 0.06 & \hphantom{00.}87 $\pm$ 47 & \parbox[t][][t]{2.5cm}{\tiny  3, 25, 28, 38, 46, 47, 49, 50, 56, 57} \\ 
J1741+2758 & 1.361 & 56949.69 & 23 & 43.16--77.19 & \hphantom{0}$\dots$ & 0.50--0.77 & \hphantom{0}$\dots$ & \hphantom{0}$\dots$ & 1e-02 & 0.08 & \hphantom{00000}<56.3 & \parbox[t][][t]{2.5cm}{\tiny  3, 19, 36, 49, 69} \\ 
J1746+2245 & 3.465 & 57125.14 & 69 & 42.09--77.24 & \hphantom{0}$\dots$ & 0.29--0.29 & \hphantom{0}$\dots$ & \hphantom{0}$\dots$ & 3e-02 & 0.06 & \hphantom{00000}<21.6 & \parbox[t][][t]{2.5cm}{\tiny  3, 7, 19} \\ 
J1752+2359 & 0.409 & 56949.71 & 20 & 42.21--77.31 & \hphantom{0}$\dots$ & 0.29--0.46 & \hphantom{0}$\dots$ & \hphantom{0}$\dots$ & 7e-02 & 0.07 & \hphantom{00000}<51.3 & \parbox[t][][t]{2.5cm}{\tiny  3, 36, 49} \\ 
B1753+52 & 2.391 & 57107.23 & 40 & 42.11--77.23 & \hphantom{0}$\dots$ & 0.09--0.11 & \hphantom{0}$\dots$ & \hphantom{0}$\dots$ & 1e-02 & 0.07 & \hphantom{00000}<19.1 & \parbox[t][][t]{2.5cm}{\tiny  3, 16, 19, 28, 38, 49} \\ 
J1758+3030 & 0.947 & 56949.72 & 20 & 50.20--77.32 & 5 & 0.36--0.65 & \hphantom{000}5 & \hphantom{0}35.1074\,(28) & 3e-02 & 0.07 & \hphantom{00.}44 $\pm$ 31 & \parbox[t][][t]{2.5cm}{\tiny  3, 9, 19, 49, 58} \\ 
B1811+40 & 0.931 & 56899.88 & 20 & 42.17--77.17 & 0 & 0.28--0.45 & \hphantom{000}6 & \hphantom{0}41.5766\,(52) & 5e-02 & 0.06 & \hphantom{00.}36 $\pm$ 22 & \parbox[t][][t]{2.5cm}{\tiny  3, 11, 15, 22, 28, 38, 41, 50, 57} \\ 
J1838+1650 & 1.902 & 56949.74 & 32 & 41.89--77.38 & \hphantom{0}$\dots$ & 0.52--0.84 & \hphantom{0}$\dots$ & \hphantom{0}$\dots$ & 1e-02 & 0.07 & \hphantom{00000}<92.5 & \parbox[t][][t]{2.5cm}{\tiny  3, 19, 37} \\ 
B1839+09 & 0.381 & 56826.98 & 20 & 42.21--77.26 & 0 & 0.17--0.34 & \hphantom{000}5 & \hphantom{0}49.1779\,(54) & 2e-01 & 0.06 & \hphantom{0.}190 $\pm$ 100 & \parbox[t][][t]{2.5cm}{\tiny  3, 25, 28, 38, 41, 46, 47, 50, 57} \\ 
B1839+56 & 1.653 & 56826.99 & 28 & 30.37--77.24 & 0 & 0.07--0.08 & \hphantom{0}166 & \hphantom{0}26.7916\,(11) & 6e-03 & 0.08 & \hphantom{0.}440 $\pm$ 220 & \parbox[t][][t]{2.5cm}{\tiny  3, 22, 25, 34, 38, 42, 46, 49, 50, 57, 59, 61, 69} \\ 
B1842+14 & 0.375 & 56827.01 & 20 & 42.20--77.20 & 0 & 0.26--0.46 & \hphantom{00}32 & \hphantom{0}41.5056\,(46) & 1e-01 & 0.06 & \hphantom{0.}830 $\pm$ 420 & \parbox[t][][t]{2.5cm}{\tiny  3, 22, 38, 42, 46, 47, 50, 55, 57, 59} \\ 
J1900+30 & 0.602 & 56899.89 & 20 & 42.26--77.46 & \hphantom{0}$\dots$ & 0.19--0.37 & \hphantom{0}$\dots$ & \hphantom{0}$\dots$ & 7e-01 & 0.04 & \hphantom{00000}<48.7 & \parbox[t][][t]{2.5cm}{\tiny  3, 9} \\ 
B1905+39 & 1.236 & 57006.53 & 21 & 42.22--77.26 & \hphantom{0}$\dots$ & 0.27--0.36 & \hphantom{0}$\dots$ & \hphantom{0}$\dots$ & 1e-02 & 0.08 & \hphantom{00000}<35.3 & \parbox[t][][t]{2.5cm}{\tiny  3, 15, 22, 28, 38, 46, 57} \\ 
B1919+21 & 1.337 & 56827.03 & 23 & 30.38--77.23 & 0 & 0.12--0.14 & \hphantom{0}453 & \hphantom{0}12.444\,(87) & 8e-04 & 0.12 & \hphantom{.}4600 $\pm$ 2300 & \parbox[t][][t]{2.5cm}{\tiny  3, 5, 12, 13, 17, 18, 21, 22, 31, 34, 38, 42, 43, 46, 48, 49, 55, 57, 59, 61, 68, 69} \\ 
B1929+10 & 0.227 & 56827.04 & 20 & 30.40--77.12 & 0 & 0.17--0.34 & \hphantom{00}28 & \hphantom{00}3.1832\,(34) & 2e-04 & 0.22 & \hphantom{0.}950 $\pm$ 480 & \parbox[t][][t]{2.5cm}{\tiny  3, 5, 6, 10, 12, 14, 18, 19, 20, 22, 25, 27, 29, 33, 35, 37, 38, 42, 43, 46, 47, 48, 49, 50, 55, 57, 62, 67, 68, 69, 70} \\ 
B1946+35 & 0.717 & 56827.06 & 20 & 42.18--77.23 & \hphantom{0}$\dots$ & 0.12--0.23 & \hphantom{0}$\dots$ & \hphantom{0}$\dots$ & 7e+00 & 0.03 & \hphantom{00000}<79.5 & \parbox[t][][t]{2.5cm}{\tiny  3, 12, 17, 25, 38, 42, 43, 46, 49, 52, 55, 57, 70} \\ 
B1953+50 & 0.519 & 57006.52 & 20 & 42.05--77.27 & 0 & 0.19--0.26 & \hphantom{000}4 & \hphantom{0}31.9827\,(53) & 4e-02 & 0.07 & \hphantom{00.}22 $\pm$ 17 & \parbox[t][][t]{2.5cm}{\tiny  3, 15, 22, 38, 41, 46, 49, 57} \\ 
J2017+2043 & 0.537 & 57126.25 & 20 & 42.10--77.25 & \hphantom{0}$\dots$ & 0.06--0.10 & \hphantom{0}$\dots$ & \hphantom{0}$\dots$ & 4e-01 & 0.05 & \hphantom{00000}<43.1 & \parbox[t][][t]{2.5cm}{\tiny  3, 19, 51} \\ 
B2016+28 & 0.558 & 56827.10 & 20 & 30.38--77.25 & 0 & 0.12--0.23 & \hphantom{00}38 & \hphantom{0}14.2239\,(36) & 3e-03 & 0.11 & \hphantom{0.}490 $\pm$ 250 & \parbox[t][][t]{2.5cm}{\tiny  3, 5, 6, 10, 12, 21, 22, 28, 29, 38, 43, 46, 48, 49, 50, 55, 56, 57, 59, 61, 68, 69} \\ 
B2020+28 & 0.343 & 56827.08 & 20 & 42.09--77.21 & 0 & 0.25--0.49 & \hphantom{00}16 & \hphantom{0}24.6311\,(40) & 2e-02 & 0.09 & \hphantom{0.}120 $\pm$ 60 & \parbox[t][][t]{2.5cm}{\tiny  3, 5, 6, 12, 21, 22, 29, 33, 35, 38, 43, 46, 48, 49, 50, 55, 56, 57, 59, 62, 66, 67, 68, 69} \\ 
B2022+50 & 0.373 & 57006.66 & 20 & 42.09--77.24 & 5 & 0.08--0.10 & \hphantom{000}6 & \hphantom{0}33.0282\,(50) & 6e-02 & 0.07 & \hphantom{00.}69 $\pm$ 38 & \parbox[t][][t]{2.5cm}{\tiny  3, 16, 19, 35, 38, 46, 49, 57} \\ 
B2034+19 & 2.074 & 57126.27 & 35 & 42.12--77.25 & \hphantom{0}$\dots$ & 0.07--0.12 & \hphantom{0}$\dots$ & \hphantom{0}$\dots$ & 2e-02 & 0.07 & \hphantom{00000}<25.7 & \parbox[t][][t]{2.5cm}{\tiny  3, 60} \\ 
J2040+1657 & 0.866 & 57006.60 & 20 & 42.04--77.33 & \hphantom{0}$\dots$ & 0.12--0.17 & \hphantom{0}$\dots$ & \hphantom{0}$\dots$ & 1e-01 & 0.06 & \hphantom{00000}<36.7 & \parbox[t][][t]{2.5cm}{\tiny  3, 37} \\ 
B2044+15 & 1.138 & 56949.76 & 20 & 42.27--77.11 & \hphantom{0}$\dots$ & 0.30--0.45 & \hphantom{0}$\dots$ & \hphantom{0}$\dots$ & 4e-02 & 0.06 & \hphantom{00000}<48.1 & \parbox[t][][t]{2.5cm}{\tiny  3, 22, 28, 38, 42, 46, 47} \\ 
B2053+21 & 0.815 & 57126.29 & 20 & 42.10--77.24 & \hphantom{0}$\dots$ & 0.04--0.06 & \hphantom{0}$\dots$ & \hphantom{0}$\dots$ & 4e-02 & 0.07 & \hphantom{00000}<34.2 & \parbox[t][][t]{2.5cm}{\tiny  3, 38, 42, 46, 60} \\ 
B2113+14 & 0.440 & 56827.12 & 20 & 42.18--77.19 & \hphantom{0}$\dots$ & 0.15--0.34 & \hphantom{0}$\dots$ & \hphantom{0}$\dots$ & 4e-01 & 0.05 & \hphantom{00000}<46.0 & \parbox[t][][t]{2.5cm}{\tiny  3, 38, 41, 42, 47, 50, 57, 65} \\ 
J2139+2242 & 1.083 & 57126.30 & 20 & 42.10--77.25 & \hphantom{0}$\dots$ & 0.06--0.11 & \hphantom{0}$\dots$ & \hphantom{0}$\dots$ & 6e-02 & 0.06 & \hphantom{00000}<39.0 & \parbox[t][][t]{2.5cm}{\tiny  3, 19, 49, 58} \\ 
B2154+40 & 1.525 & 56827.14 & 26 & 42.09--77.28 & \hphantom{0}$\dots$ & 0.25--0.50 & \hphantom{0}$\dots$ & \hphantom{0}$\dots$ & 3e-01 & 0.04 & \hphantom{00000}<45.2 & \parbox[t][][t]{2.5cm}{\tiny  3, 21, 22, 25, 28, 29, 38, 43, 46, 49, 50, 56, 57} \\ 
B2217+47 & 0.538 & 56827.16 & 20 & 30.32--77.24 & 0 & 0.22--0.47 & \hphantom{00}62 & \hphantom{0}43.5062\,(35) & 1e-01 & 0.06 & \hphantom{.}1300 $\pm$ 600 & \parbox[t][][t]{2.5cm}{\tiny  3, 10, 12, 17, 21, 22, 25, 34, 38, 43, 48, 49, 56, 57, 59, 61, 64} \\ 
B2224+65 & 0.683 & 56949.78 & 20 & 42.02--77.37 & 6 & 0.37--0.59 & \hphantom{00}25 & \hphantom{0}36.5036\,(17) & 5e-02 & 0.07 & \hphantom{0.}370 $\pm$ 180 & \parbox[t][][t]{2.5cm}{\tiny  3, 12, 21, 22, 25, 29, 38, 42, 43, 46, 50, 57, 61} \\ 
B2227+61 & 0.443 & 56949.79 & 20 & 42.01--77.23 & \hphantom{0}$\dots$ & 0.22--0.39 & \hphantom{0}$\dots$ & \hphantom{0}$\dots$ & 9e+00 & 0.03 & \hphantom{00000}<60.5 & \parbox[t][][t]{2.5cm}{\tiny  3, 38, 49, 50, 57} \\ 
J2253+1516 & 0.792 & 56899.91 & 20 & 42.26--77.25 & \hphantom{0}$\dots$ & 0.16--0.30 & \hphantom{0}$\dots$ & \hphantom{0}$\dots$ & 2e-02 & 0.08 & \hphantom{00000}<42.5 & \parbox[t][][t]{2.5cm}{\tiny  3, 8, 19, 49, 69} \\ 
B2303+30 & 1.576 & 56899.94 & 27 & 42.22--77.32 & 0 & 0.30--0.56 & \hphantom{000}5 & \hphantom{0}49.6445\,(50) & 6e-02 & 0.06 & \hphantom{00.}27 $\pm$ 21 & \parbox[t][][t]{2.5cm}{\tiny  3, 21, 22, 28, 37, 38, 43, 50, 55, 57, 59} \\ 
B2303+46 & 1.066 & 56949.81 & 20 & 42.23--77.15 & \hphantom{0}$\dots$ & 0.29--0.49 & \hphantom{0}$\dots$ & \hphantom{0}$\dots$ & 2e-01 & 0.05 & \hphantom{00000}<44.5 & \parbox[t][][t]{2.5cm}{\tiny  3, 16, 19, 26, 38, 49} \\ 
B2306+55 & 0.475 & 56899.93 & 20 & 42.14--77.17 & 0 & 0.11--0.16 & \hphantom{000}7 & \hphantom{0}46.559\,(71) & 2e-01 & 0.06 & \hphantom{0.}180 $\pm$ 90 & \parbox[t][][t]{2.5cm}{\tiny  3, 12, 17, 22, 38, 46, 50, 57} \\ 
B2310+42 & 0.349 & 56827.19 & 20 & 42.12--77.12 & 0 & 0.12--0.25 & \hphantom{000}8 & \hphantom{0}17.2969\,(19) & 8e-03 & 0.10 & \hphantom{00.}66 $\pm$ 35 & \parbox[t][][t]{2.5cm}{\tiny  3, 5, 15, 17, 18, 22, 28, 38, 42, 46, 49, 50, 57, 69} \\ 
B2315+21 & 1.445 & 56827.20 & 25 & 30.37--77.06 & 0 & 0.25--0.44 & \hphantom{00}18 & \hphantom{0}20.8896\,(94) & 3e-03 & 0.09 & \hphantom{00.}33 $\pm$ 19 & \parbox[t][][t]{2.5cm}{\tiny  3, 15, 22, 28, 37, 38, 46, 49, 50, 57, 69} \\ 

\hline                                   
\end{longtable}
\end{center}

\tiny
\setlength{\tabcolsep}{2pt}
\renewcommand{\arraystretch}{1.01}
\begin{center}
\begin{longtable}{|l|l|c|c|c|c|c||l|l|c|c|c|c|c|}
\caption{Fit results for pulsars with a single PL spectrum.  
} \label{table:1pl} \\ 

\hline 
\parbox[t][][t]{1.2cm}{\centering PSR } & 
\parbox[t][][t]{1.4cm}{\centering Frequency \\ span \\ (MHz)} & 
\parbox[t][][t]{0.9cm}{\centering \# of points,\\ $N_p$ } & 
\parbox[t][][t]{0.8cm}{\centering Ref. \\ freq., \\ $\nu_0$ (MHz)} & 
\parbox[t][][t]{0.8cm}{\centering Ref. \\ flux, \\ $S_0$ (mJy)} & 
\parbox[t][][t]{1.5cm}{\centering Spectral \\ index, \\ $\alpha$ } & 
\parbox[t][][t]{0.8cm}{\centering Fitted flux \\ scatter, \\ $\su$ } &
\parbox[t][][t]{1.5cm}{\centering PSR } & 
\parbox[t][][t]{1.4cm}{\centering Frequency \\ span \\ (MHz)} & 
\parbox[t][][t]{0.9cm}{\centering \# of points, \\ $N_p$} & 
\parbox[t][][t]{0.8cm}{\centering Ref. \\ freq., \\ $\nu_0$ (MHz)} & 
\parbox[t][][t]{0.8cm}{\centering Ref. \\ flux, \\ $S_0$ (mJy)} & 
\parbox[t][][t]{1.5cm}{\centering Spectral \\ index, \\ $\alpha$ } & 
\parbox[t][][t]{0.8cm}{\centering Fitted flux \\ scatter, \\ $\su$ } \\ \hline 
\endfirsthead
\hline \hline
\multicolumn{14}{l}{\parbox[t][][t]{17.5cm}{\textbf{Notes.}
\tablefoottext{$\sharp$} {These pulsars have also broken PL fit, with break frequency fixed at the frequency of the largest
 measured flux density. See Table~\ref{table:2pl} for the values of fitted parameters.}}}
\endlastfoot
B0011+47 & \hphantom{0}59 -- 4850\hphantom{00} & 19 & \hphantom{0}500 & 11.0 & $-0.9 \pm 0.1$ & 0.11 & J1238+21$^{\sharp}$ & \hphantom{0}25 -- 430\hphantom{000} & 6\hphantom{0} & \hphantom{0}100 & 22.0 & $-0.8 \pm 0.3$ & \dots\\ 
B0053+47 & \hphantom{0}20 -- 4850\hphantom{00} & 10 & \hphantom{0}300 & \hphantom{0}4.7 & $-1.3 \pm 0.2$ & 0.43 & J1313+0931$^{\sharp}$ & \hphantom{0}59 -- 1400\hphantom{00} & 4\hphantom{0} & \hphantom{0}300 & \hphantom{0}6.2 & $-2.3 \pm 0.3$ & \dots\\ 
B0226+70 & \hphantom{0}59 -- 1420\hphantom{00} & 10 & \hphantom{0}300 & \hphantom{0}3.8 & $-1.6 \pm 0.2$ & 0.08 & J1645+1012$^{\sharp}$ & \hphantom{0}59 -- 430\hphantom{000} & 4\hphantom{0} & \hphantom{0}200 & 14.0 & $-2.1 \pm 0.4$ & \dots\\ 
J0611+30$^{\sharp}$ & \hphantom{0}45 -- 430\hphantom{000} & 4\hphantom{0} & \hphantom{0}100 & 38.0 & $-1.9 \pm 0.4$ & \dots & J1758+3030 & \hphantom{0}59 -- 800\hphantom{000} & 9\hphantom{0} & \hphantom{0}200 & 20.0 & $-1.4 \pm 0.3$ & 0.11\\ 
B0609+37 & \hphantom{0}59 -- 4850\hphantom{00} & 13 & \hphantom{0}500 & \hphantom{0}5.2 & $-1.4 \pm 0.2$ & 0.30 & B1842+14 & \hphantom{0}47 -- 4850\hphantom{00} & 21 & \hphantom{0}500 & 14.0 & $-1.95 \pm 0.09$ & 0.05\\ 
B0655+64 & \hphantom{0}45 -- 1408\hphantom{00} & 15 & \hphantom{0}300 & 14.0 & $-2.0 \pm 0.2$ & 0.34 & B1953+50 & \hphantom{0}59 -- 4850\hphantom{00} & 13 & \hphantom{0}500 & 17.0 & $-1.2 \pm 0.1$ & 0.08\\ 
B0751+32 & \hphantom{0}59 -- 4850\hphantom{00} & 10 & \hphantom{0}500 & \hphantom{0}4.4 & $-1.4 \pm 0.2$ & 0.13 & B2022+50 & \hphantom{0}59 -- 32000 & 15 & 1400 & \hphantom{0}1.9 & $-1.10 \pm 0.08$ & 0.07\\ 
B0917+63 & \hphantom{0}45 -- 1408\hphantom{00} & 10 & \hphantom{0}300 & \hphantom{0}6.3 & $-1.4 \pm 0.2$ & 0.07 & B2224+65 & \hphantom{0}45 -- 10700 & 26 & \hphantom{0}700 & \hphantom{0}7.7 & $-1.62 \pm 0.07$ & 0.07\\

\end{longtable}
\end{center}

\tiny
\setlength{\tabcolsep}{2pt}
\renewcommand{\arraystretch}{1.05}
\begin{center}
\begin{longtable}{|l|l|c|c|c|c|c|c|c|c|}
\caption{Fit results for pulsars where the spectrum was modelled with a broken PL.} \label{table:2pl} \\ 

\hline 
\parbox[t][][t]{1.2cm}{\centering PSR } & 
\parbox[t][][t]{1.5cm}{\centering Frequency \\ span \\ (MHz)} & 
\parbox[t][][t]{1.0cm}{\centering \# of points, \\ $N_p$} & 
\parbox[t][][t]{1.0cm}{\centering Ref. \\ freq., \\ $\nu_0$ (MHz)} & 
\parbox[t][][t]{1.0cm}{\centering Ref. \\ flux, \\ $S_0$ (mJy)} & 
\parbox[t][][t]{1.5cm}{\centering Spectral \\ index, \\ $\alpha_\mathrm{lo}$ } & 
\parbox[t][][t]{1.0cm}{\centering Break freq., \\ $\nu_\mathrm{br}$ \\ (MHz)} &
\parbox[t][][t]{2.0cm}{\centering Uncertainty \\ range for \\ $\nu_\mathrm{br}$ \\ (MHz)  } & 
\parbox[t][][t]{1.5cm}{\centering Spectral \\ index, \\ $\alpha_\mathrm{hi}$ } & 
\parbox[t][][t]{1.0cm}{\centering Fitted flux \\ scatter, \\ $\su$ }

\\ \hline 
\endfirsthead

\hline \hline
\endlastfoot
B0301+19 & \hphantom{0}59 -- 4850\hphantom{00} & 35 & \hphantom{0}500 & \hphantom{0}24.0 & $-0.5 \pm 0.2$ & \hphantom{0}500 & 354 -- 761\hphantom{0} & $-1.9 \pm 0.3$ &  0.11 \\ 
B0320+39 & \hphantom{0}25 -- 4850\hphantom{00} & 26 & \hphantom{0}300 & \hphantom{0}52.0 & \hphantom{$-$}$0.9 \pm 0.5$ & \hphantom{0}157 & 133 -- 195\hphantom{0} & $-2.4 \pm 0.2$ &  0.10 \\ 
J0611+30 & \hphantom{0}45 -- 430\hphantom{000} & \hphantom{0}4 & \hphantom{0}100 & \hphantom{0}70.0 & \hphantom{$-$}$1.4 \pm 1.5$ & \hphantom{00}74 & \hphantom{000} \dots \hphantom{0000} & $-2.5 \pm 0.5$ &  \dots \\ 
B0809+74 & \hphantom{0}12 -- 14800 & 69 & \hphantom{0}400 & 110.0 & \hphantom{$-$}$0.8 \pm 0.3$ & \hphantom{00}66 & \hphantom{0}58 -- 73\hphantom{0}\hphantom{0} & $-1.66 \pm 0.06$ &  0.14 \\ 
B0943+10 & \hphantom{0}20 -- 1400\hphantom{00} & 35 & \hphantom{0}200 & \hphantom{0}83.0 & \hphantom{$-$}$0.2 \pm 0.6$ & \hphantom{0}114 & \hphantom{0}87 -- 167\hphantom{0} & $-2.8 \pm 0.7$ &  0.31 \\ 
B1112+50 & \hphantom{0}20 -- 4900\hphantom{00} & 37 & \hphantom{0}300 & \hphantom{0}45.0 & \hphantom{$-$}$1.1 \pm 0.4$ & \hphantom{0}148 & 133 -- 172\hphantom{0} & $-2.3 \pm 0.2$ &  0.36 \\ 
B1133+16 & \hphantom{0}16 -- 32000 & 130 & \hphantom{0}700 & 100.0 & \hphantom{$-$}$0.1 \pm 0.1$ & \hphantom{0}232 & 212 -- 257\hphantom{0} & $-1.93 \pm 0.06$ &  0.17 \\ 
J1238+21 & \hphantom{0}25 -- 430\hphantom{000} & \hphantom{0}6 & \hphantom{0}100 & \hphantom{0}64.0 & \hphantom{$-$}$0.8 \pm 0.5$ & \hphantom{0}102 & \hphantom{000} \dots \hphantom{0000} & $-2.4 \pm 0.4$ &  \dots \\ 
J1313+0931 & \hphantom{0}59 -- 1400\hphantom{00} & \hphantom{0}4 & \hphantom{0}300 & \hphantom{00}8.6 & \hphantom{$-$}$0.9 \pm 1.3$ & \hphantom{0}149 & \hphantom{000} \dots \hphantom{0000} & $-2.6 \pm 0.3$ &  \dots \\ 
B1322+83 & \hphantom{0}25 -- 1408\hphantom{00} & 12 & \hphantom{0}200 & \hphantom{0}59.0 & \hphantom{$-$}$0.8 \pm 0.6$ & \hphantom{0}214 & 173 -- 306\hphantom{0} & $-2.5 \pm 0.6$ &  0.18 \\ 
B1508+55 & \hphantom{0}20 -- 10750 & 59 & \hphantom{0}500 & \hphantom{0}52.0 & \hphantom{$-$}$2.4 \pm 0.4$ & \hphantom{00}88 & \hphantom{0}82 -- 97\hphantom{0}\hphantom{0} & $-2.04 \pm 0.08$ &  0.14 \\ 
B1530+27 & \hphantom{0}25 -- 4850\hphantom{00} & 22 & \hphantom{0}300 & \hphantom{0}17.0 & \hphantom{$-$}$1.3 \pm 0.5$ & \hphantom{00}92 & \hphantom{0}74 -- 100\hphantom{0} & $-1.6 \pm 0.1$ &  0.05 \\ 
B1541+09 & \hphantom{0}39 -- 10550 & 49 & \hphantom{0}600 & \hphantom{0}29.0 & \hphantom{$-$}$0.7 \pm 0.4$ & \hphantom{0}144 & 133 -- 156\hphantom{0} & $-2.15 \pm 0.09$ &  0.09 \\ 
B1633+24 & \hphantom{0}25 -- 1400\hphantom{00} & 12 & \hphantom{0}200 & \hphantom{0}44.0 & \hphantom{$-$}$0.6 \pm 0.5$ & \hphantom{0}155 & 131 -- 191\hphantom{0} & $-2.3 \pm 0.3$ &  0.06 \\ 
J1645+1012 & \hphantom{0}59 -- 430\hphantom{000} & \hphantom{0}4 & \hphantom{0}200 & \hphantom{0}21.0 & \hphantom{$-$}$1.5 \pm 1.7$ & \hphantom{0}102 & \hphantom{000} \dots \hphantom{0000} & $-2.9 \pm 0.5$ &  \dots \\ 
B1737+13 & \hphantom{0}45 -- 4850\hphantom{00} & 18 & \hphantom{0}500 & \hphantom{0}20.0 & $-0.6 \pm 0.5$ & \hphantom{0}330 & 153 -- 556\hphantom{0} & $-1.7 \pm 0.2$ &  0.04 \\ 
B1839+09 & \hphantom{0}59 -- 4850\hphantom{00} & 13 & \hphantom{0}500 & \hphantom{0}14.0 & $-0.6 \pm 0.8$ & \hphantom{0}229 & 135 -- 294\hphantom{0} & $-1.9 \pm 0.2$ &  0.06 \\ 
B1839+56 & \hphantom{0}20 -- 4850\hphantom{00} & 30 & \hphantom{0}300 & \hphantom{0}32.0 & \hphantom{$-$}$4.8 \pm 1.4$ & \hphantom{00}39 & \hphantom{0}35 -- 41\hphantom{0}\hphantom{0} & $-1.6 \pm 0.1$ &  0.23 \\ 
B1919+21 & \hphantom{0}16 -- 4850\hphantom{00} & 92 & \hphantom{0}300 & 250.0 & \hphantom{$-$}$0.4 \pm 0.2$ & \hphantom{0}135 & 120 -- 149\hphantom{0} & $-2.7 \pm 0.1$ &  0.21 \\ 
B1929+10 & \hphantom{0}20 -- 43000 & 98 & \hphantom{0}900 & 110.0 & \hphantom{$-$}$0.3 \pm 0.4$ & \hphantom{0}342 & 240 -- 511\hphantom{0} & $-1.74 \pm 0.09$ &  0.23 \\ 
B2016+28 & \hphantom{0}25 -- 10700 & 55 & \hphantom{0}500 & 150.0 & \hphantom{$-$}$0.1 \pm 0.2$ & \hphantom{0}331 & 293 -- 379\hphantom{0} & $-2.27 \pm 0.09$ &  0.13 \\ 
B2020+28 & \hphantom{0}45 -- 32000 & 47 & 1200 & \hphantom{0}24.0 & \hphantom{$-$}$0.6 \pm 0.6$ & \hphantom{0}307 & 225 -- 430\hphantom{0} & $-1.6 \pm 0.1$ &  0.21 \\ 
B2306+55 & \hphantom{0}59 -- 4850\hphantom{00} & 13 & \hphantom{0}500 & \hphantom{0}15.0 & $-0.7 \pm 0.5$ & \hphantom{0}357 & 178 -- 532\hphantom{0} & $-2.0 \pm 0.2$ &  0.07 \\ 
B2310+42 & \hphantom{0}25 -- 10700 & 34 & \hphantom{0}500 & 130.0 & \hphantom{$-$}$0.1 \pm 0.3$ & \hphantom{0}645 & 498 -- 909\hphantom{0} & $-2.1 \pm 0.3$ &  0.18 \\ 
B2315+21 & \hphantom{0}25 -- 4850\hphantom{00} & 16 & \hphantom{0}400 & \hphantom{0}17.0 & \hphantom{$-$}$0.6 \pm 0.5$ & \hphantom{0}186 & 167 -- 231\hphantom{0} & $-2.1 \pm 0.2$ &  0.06 \\ 

\end{longtable}
\end{center}

\tiny
\setlength{\tabcolsep}{2pt}
\renewcommand{\arraystretch}{1.10}
\begin{center}
\begin{longtable}{|l|l|c|c|c|c|c|c|c|c|c|c|c|}
\caption{Fit results for pulsars where the spectrum was modelled by a PL with two breaks. 
} \label{table:3pl} \\ 

\hline 
\parbox[t][][t]{1.2cm}{\centering PSR } & 
\parbox[t][][t]{0.8cm}{\centering Frequency \\ span \\ (MHz)} & 
\parbox[t][][t]{0.8cm}{\centering \# of points, \\ $N_p$} & 
\parbox[t][][t]{0.8cm}{\centering Ref. \\ freq., \\ $\nu_0$ (MHz)} & 
\parbox[t][][t]{0.9cm}{\centering Ref. \\ flux, \\ $S_0$ (mJy)} & 
\parbox[t][][t]{1.4cm}{\centering Spectral \\ index, \\ $\alpha_\mathrm{lo}$ } & 
\parbox[t][][t]{1.2cm}{\centering Lower \\ break freq., \\ $\nu^\mathrm{lo}_\mathrm{br}$ \\ (MHz)} &
\parbox[t][][t]{1.3cm}{\centering Uncertainty \\ range for \\ $\nu^\mathrm{lo}_\mathrm{br}$ \\ (MHz)  } & 
\parbox[t][][t]{1.4cm}{\centering Spectral \\ index \\ $\alpha_\mathrm{mid}$ } & 
\parbox[t][][t]{1.2cm}{\centering Higher \\ break freq., \\ $\nu^\mathrm{hi}_\mathrm{br}$ \\ (MHz)} &
\parbox[t][][t]{1.5cm}{\centering Uncertainty \\ range for \\ $\nu^\mathrm{hi}_\mathrm{br}$ \\ (MHz)  } &
\parbox[t][][t]{1.7cm}{\centering Spectral \\ index, \\ $\alpha_\mathrm{hi}$ } &
\parbox[t][][t]{1.0cm}{\centering Fitted flux \\ scatter, \\ $\su$ }
\\ \hline 
\endfirsthead
\hline \hline
\endlastfoot
B0450+55 & \hphantom{0}25 -- 14600 & 24 & \hphantom{0}600 & \hphantom{0}30.0 & \hphantom{$-$}$0.5 \pm 1.4$ & \hphantom{00}94 & \hphantom{0}45 -- 243\hphantom{0} & $-1.3 \pm 0.5$ &  1914 & 708 -- 4676 & $-1.8 \pm 0.7$ & 0.31 \\ 
B0823+26 & \hphantom{0}20 -- 32000 & 86 & \hphantom{0}800 & \hphantom{0}39.0 & \hphantom{$-$}$2.0 \pm 0.8$ & \hphantom{00}54 & \hphantom{0}45 -- 65\hphantom{0}\hphantom{0} & $-1.25 \pm 0.08$ &  2808 & 1199 -- 4182 & $-2.2 \pm 0.3$ & 0.08 \\ 
B1237+25 & \hphantom{0}20 -- 24620 & 82 & \hphantom{0}700 & \hphantom{0}48.0 & \hphantom{$-$}$2.6 \pm 1.4$ & \hphantom{00}55 & \hphantom{0}36 -- 69\hphantom{0}\hphantom{0} & $-0.9 \pm 0.2$ &  \hphantom{0}843 & 709 -- 917\hphantom{0} & $-2.2 \pm 0.1$ & 0.13 \\ 
B1811+40 & \hphantom{0}59 -- 2600\hphantom{00} & 12 & \hphantom{0}400 & \hphantom{0}10.0 & $-0.9 \pm 0.6$ & \hphantom{0}260 & 130 -- 413\hphantom{0} & $-1.4 \pm 0.6$ &  \hphantom{0}956 & 711 -- 1060 & $-2.8 \pm 0.6$ & 0.06 \\ 
B2217+47 & \hphantom{0}35 -- 4900\hphantom{00} & 39 & \hphantom{0}400 & 110.0 & $-1.0 \pm 0.4$ & \hphantom{0}241 & 147 -- 357\hphantom{0} & $-2.7 \pm 0.6$ &  1257 & 711 -- 1574 & $-1.8 \pm 0.8$ & 0.20 \\ 
B2303+30 & \hphantom{0}49 -- 4850\hphantom{00} & 19 & \hphantom{0}500 & \hphantom{0}14.0 & $-0.5 \pm 0.4$ & \hphantom{0}337 & 236 -- 418\hphantom{0} & $-2.7 \pm 0.6$ &  \hphantom{0}928 & 726 -- 1057 & $-1.2 \pm 0.4$ & 0.06 \\ 

\end{longtable}
\end{center}
\begingroup

\twocolumn

\section{Profiles and spectra}
\label{app:plots}

\begin{figure*}
\centering
\includegraphics[scale=0.43]{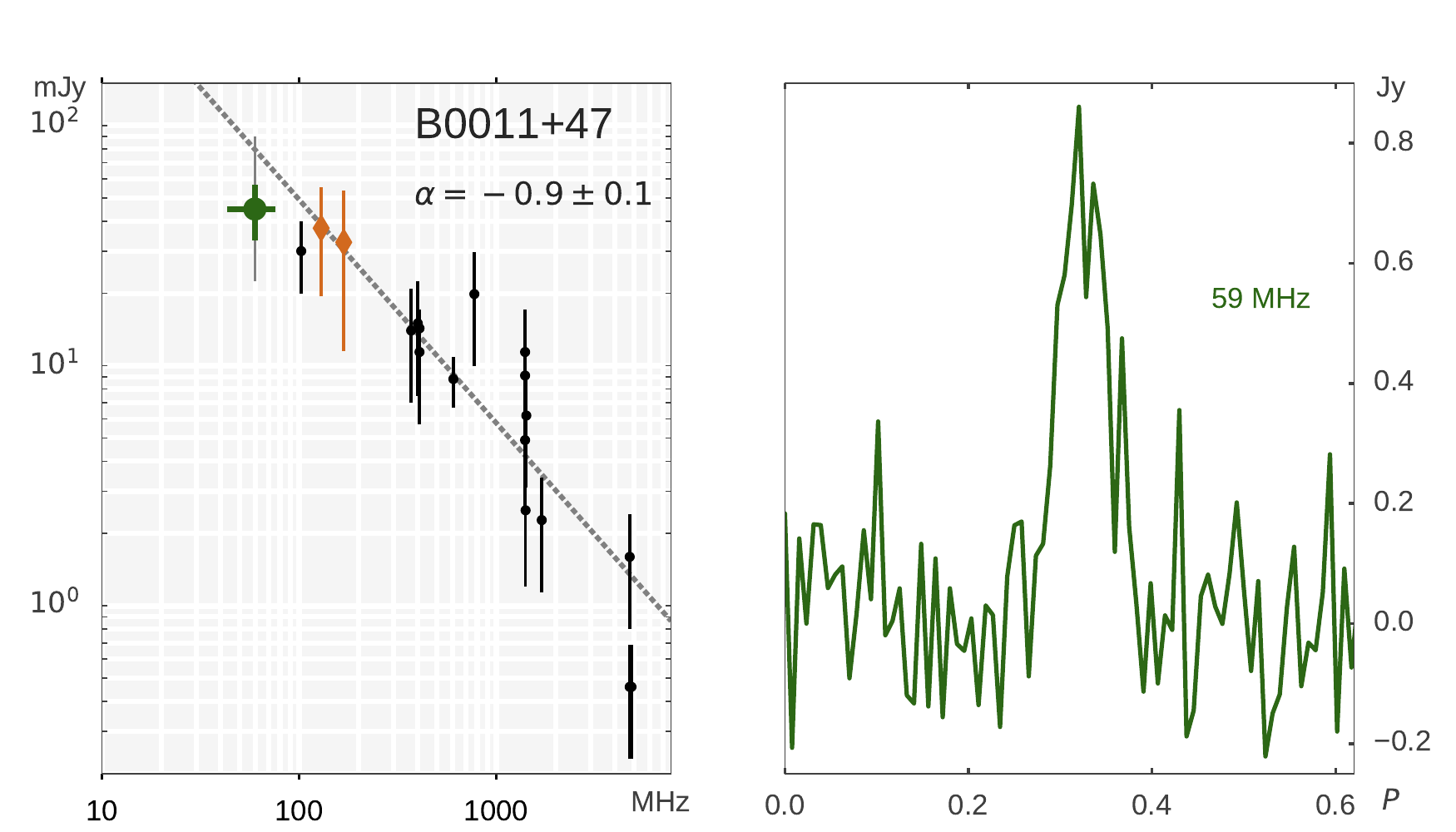}\includegraphics[scale=0.43]{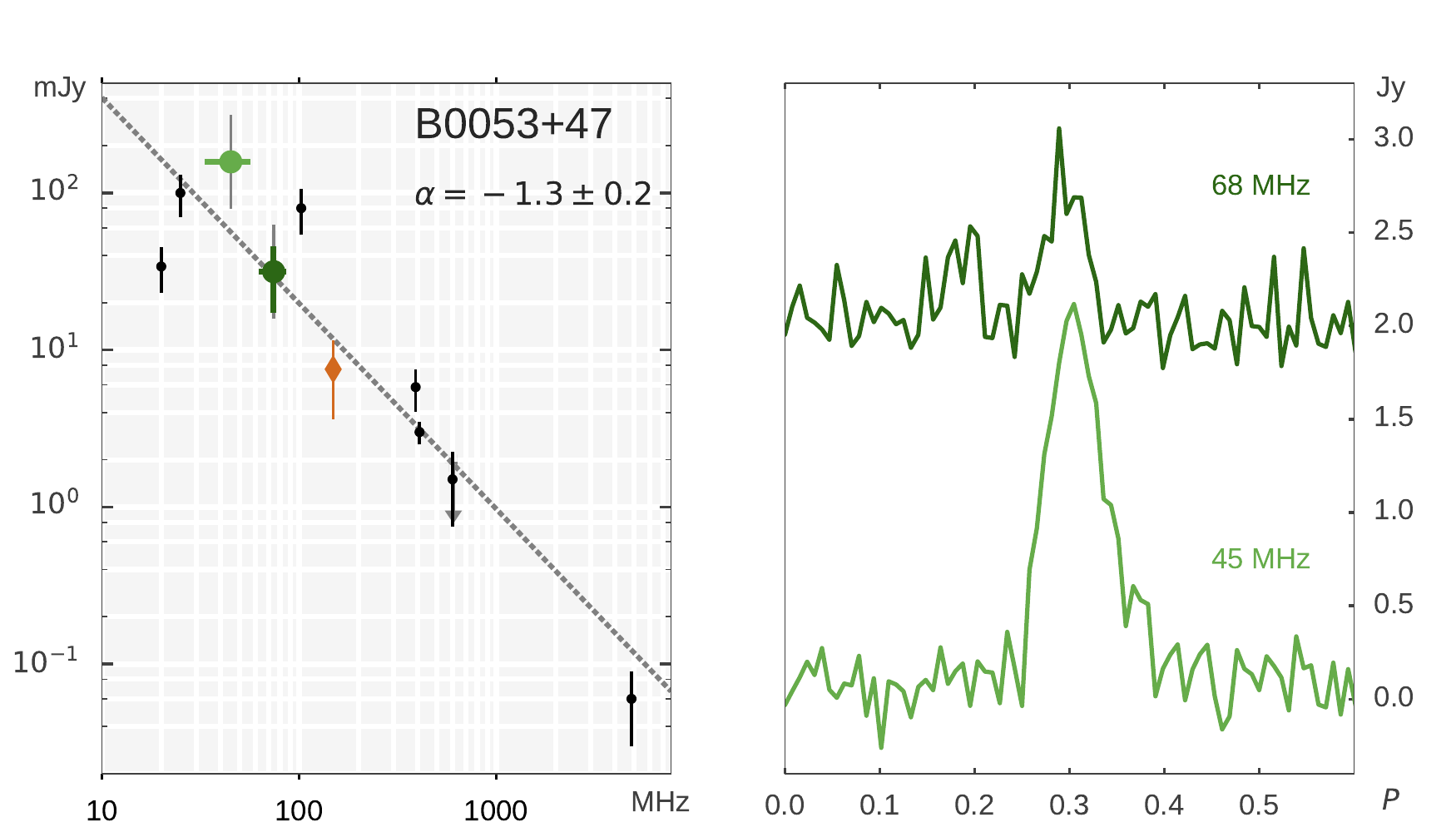}
\includegraphics[scale=0.43]{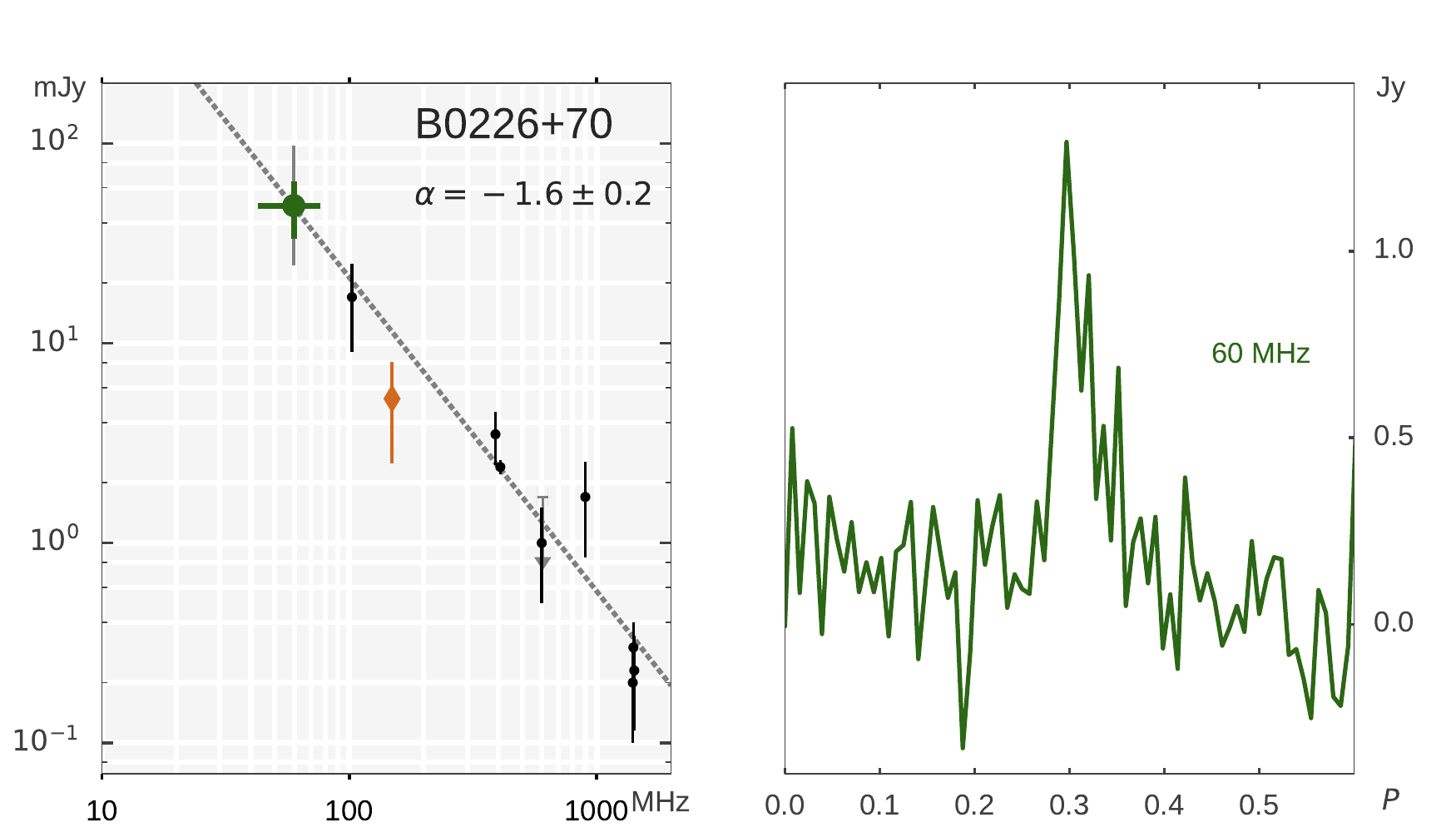}\includegraphics[scale=0.43]{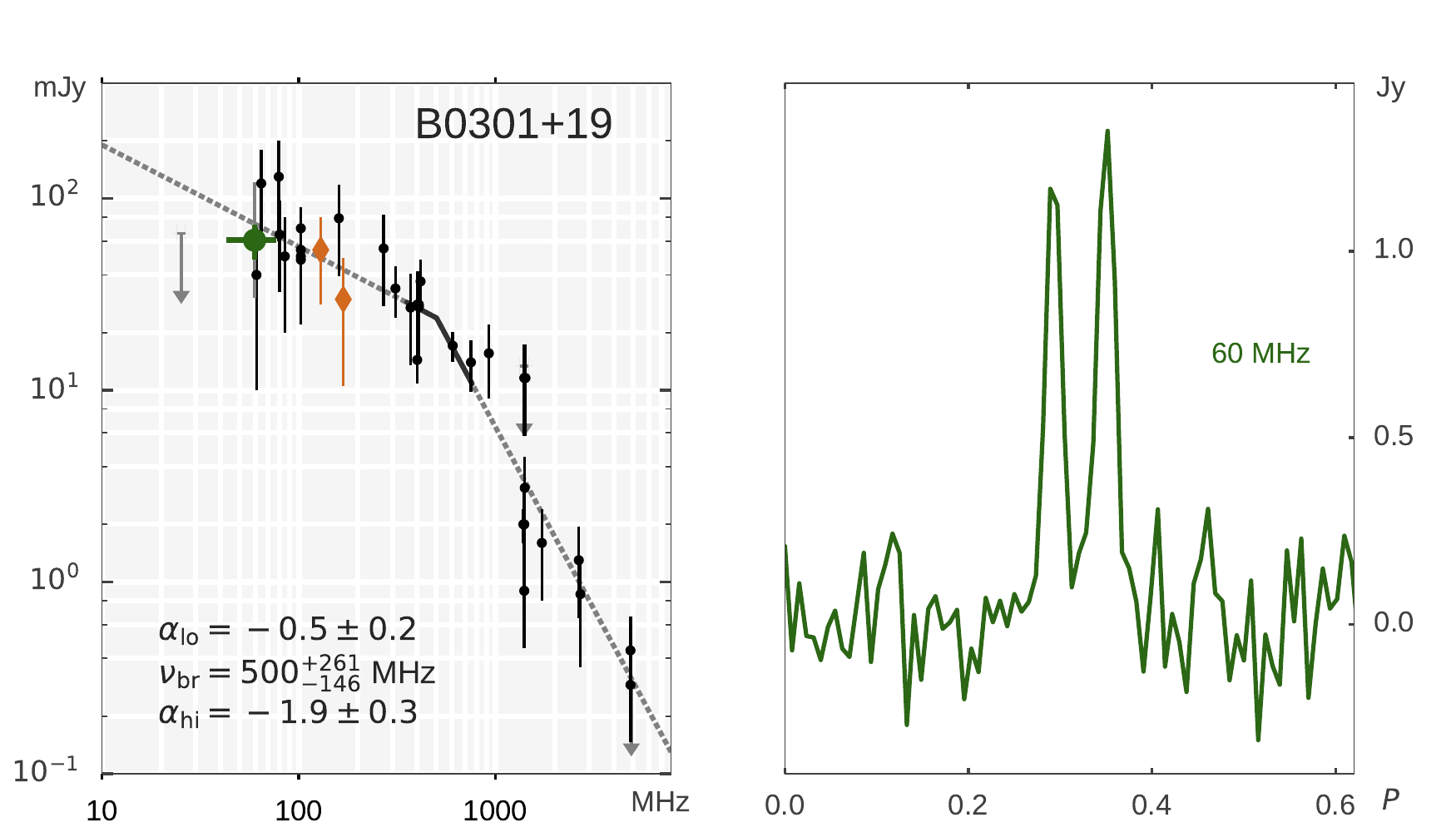}
\includegraphics[scale=0.43]{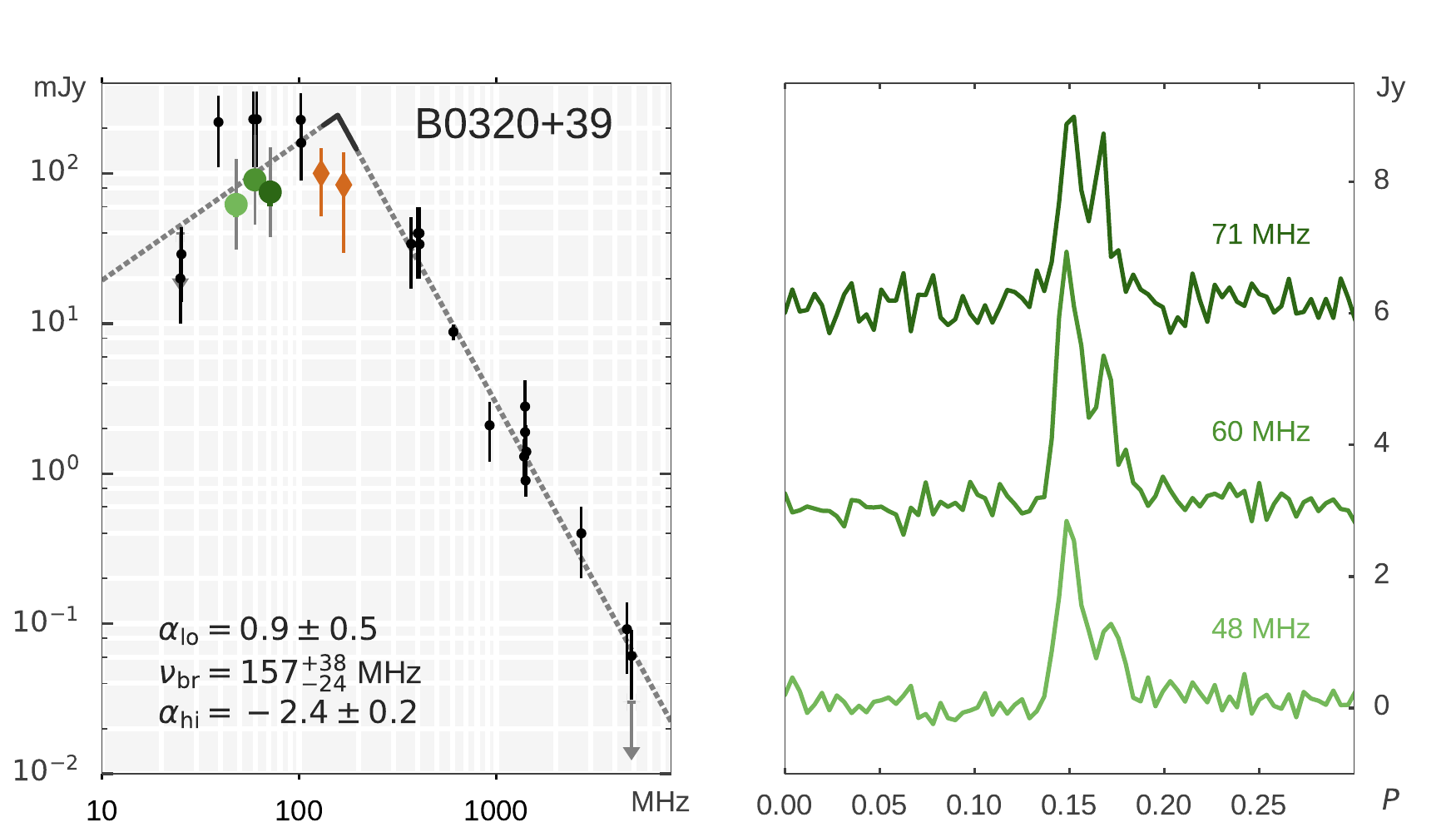}\includegraphics[scale=0.43]{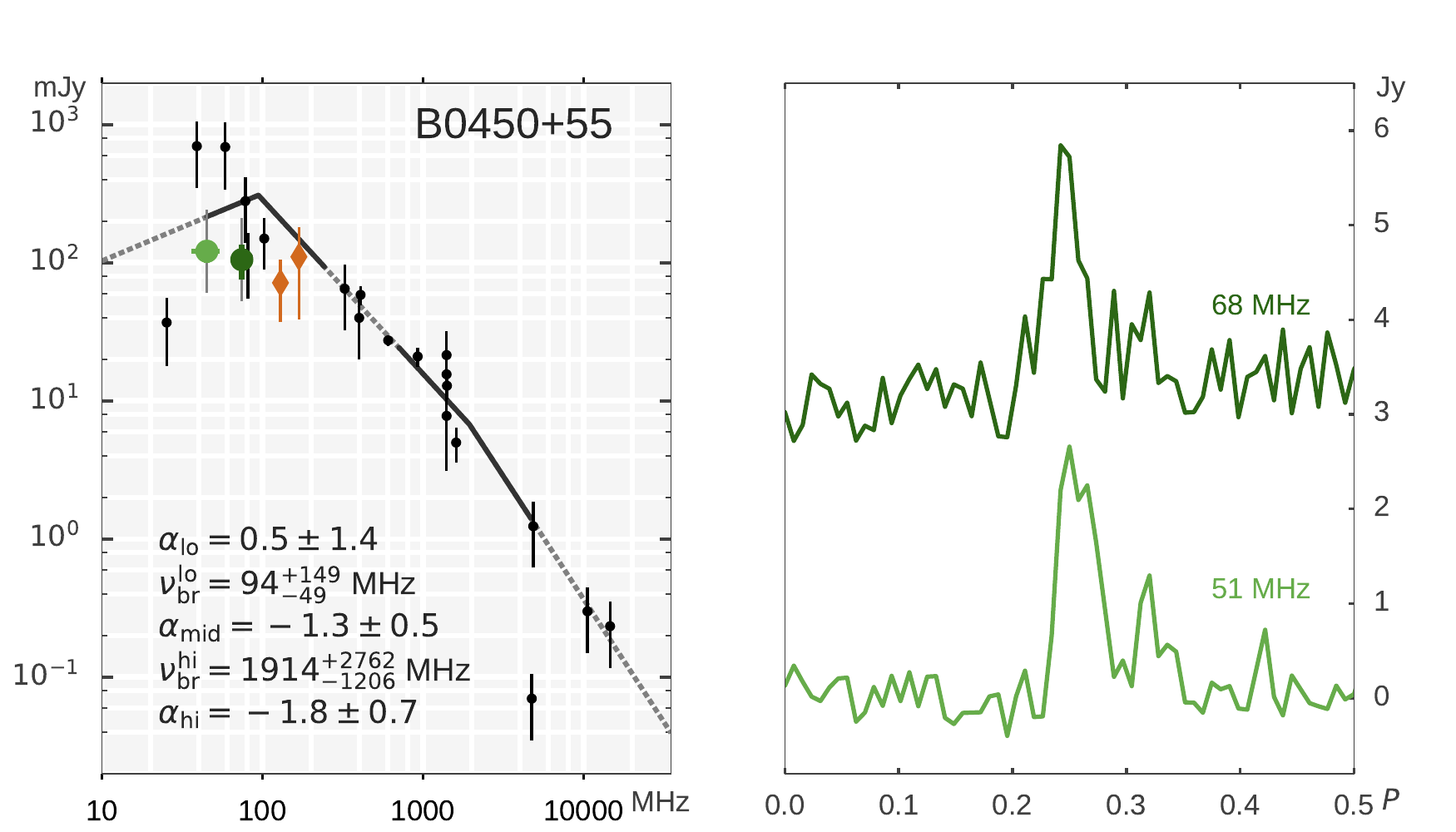}
\includegraphics[scale=0.43]{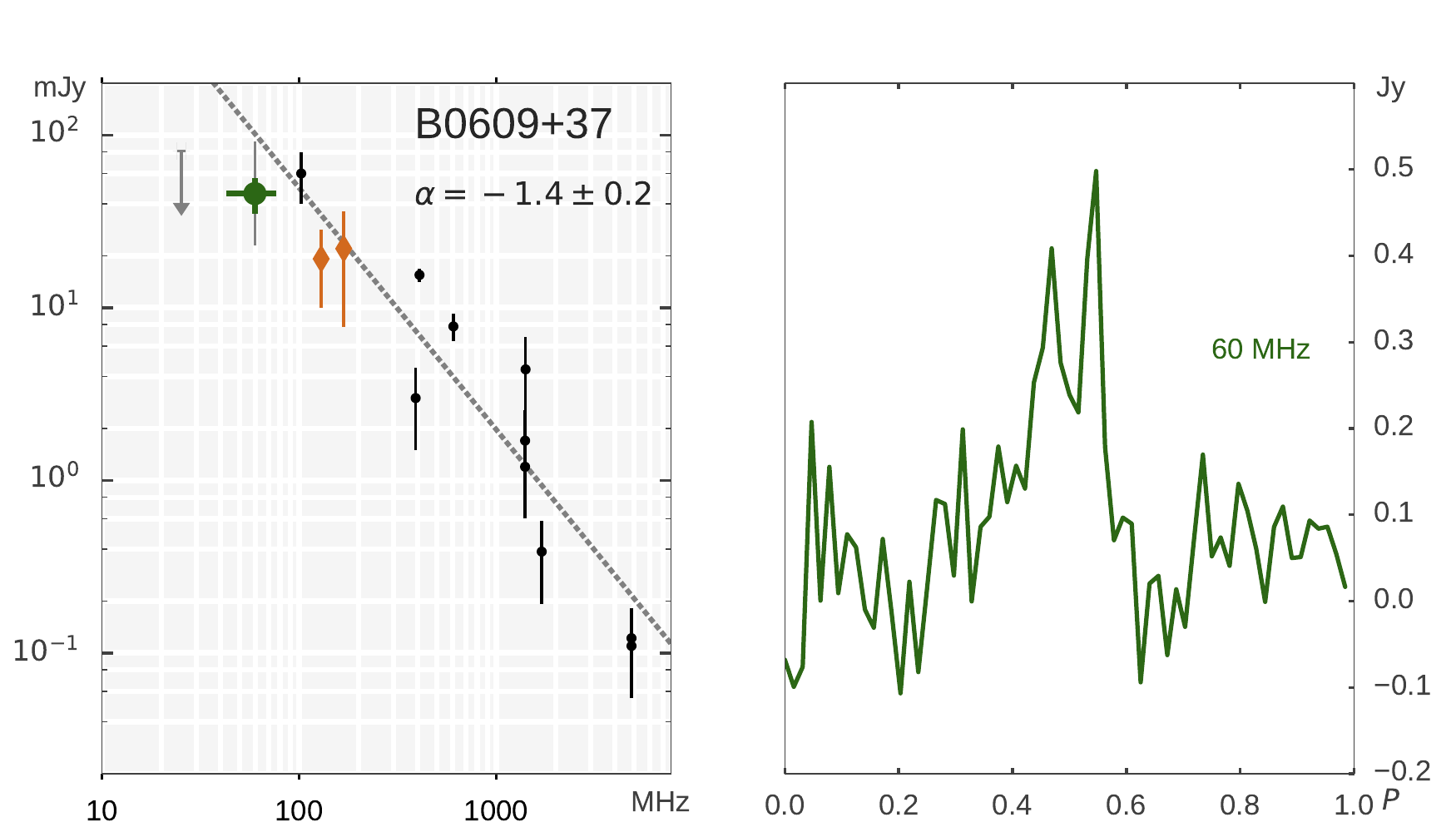}\includegraphics[scale=0.43]{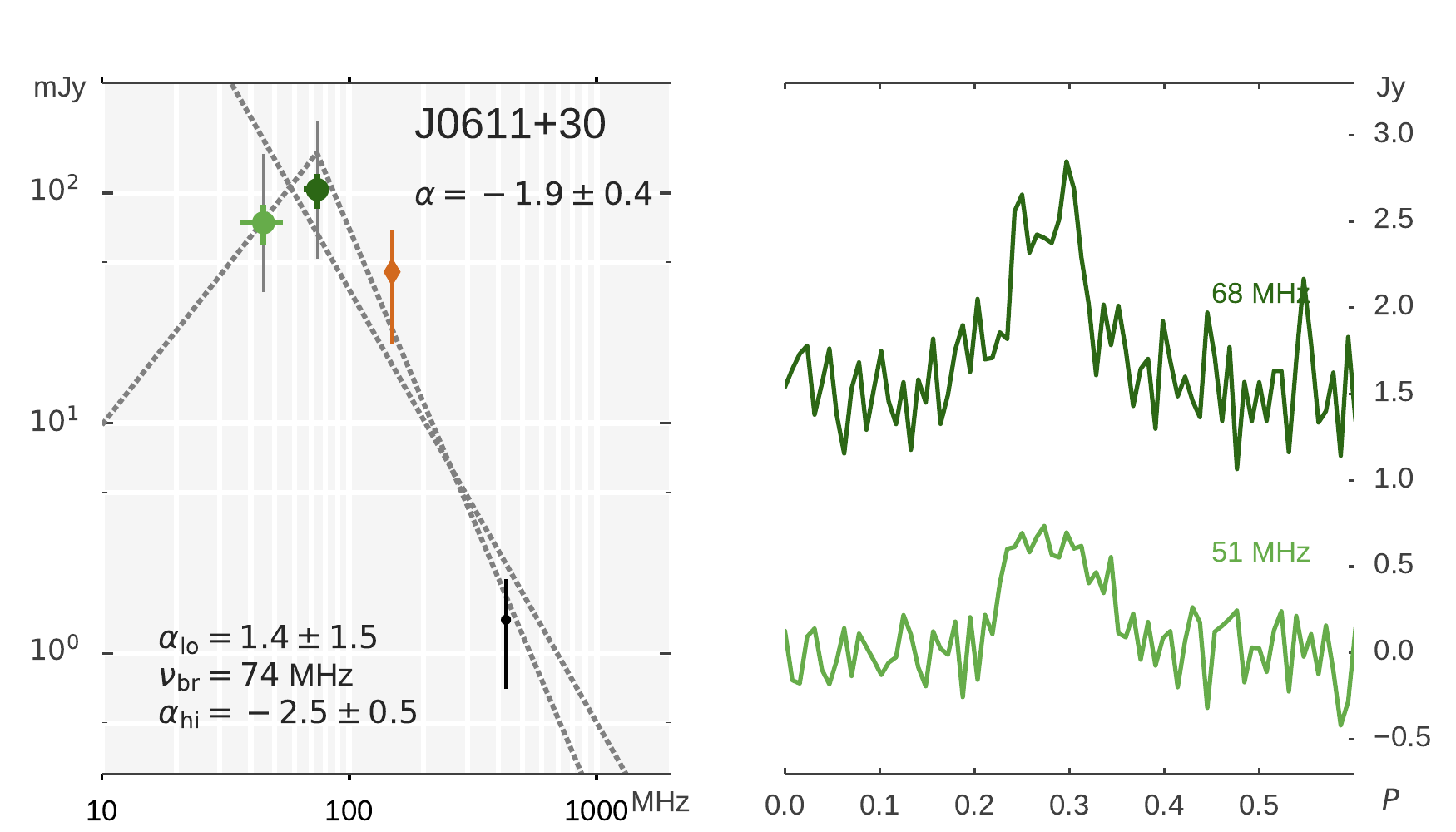}
\includegraphics[scale=0.43]{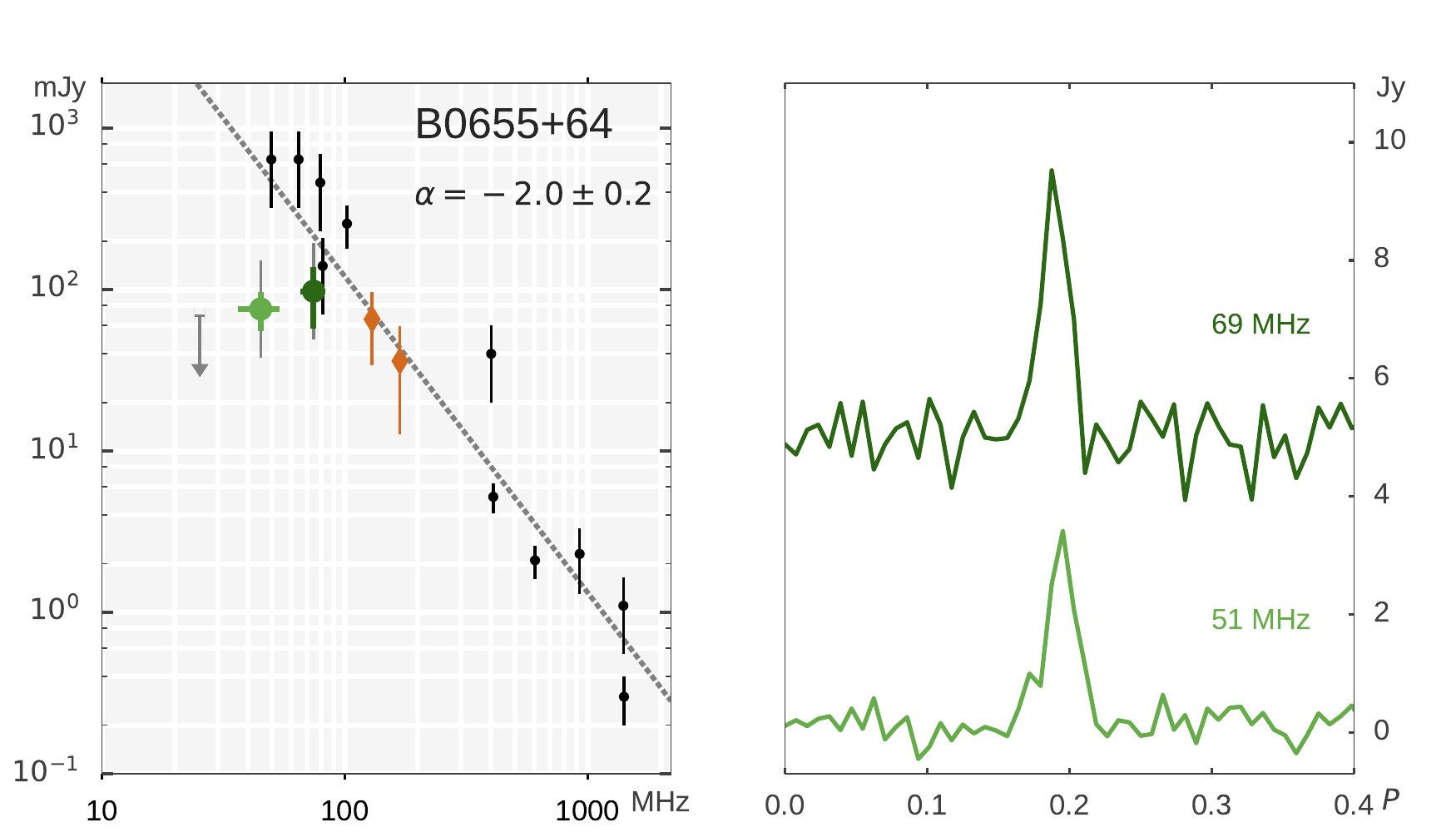}\includegraphics[scale=0.43]{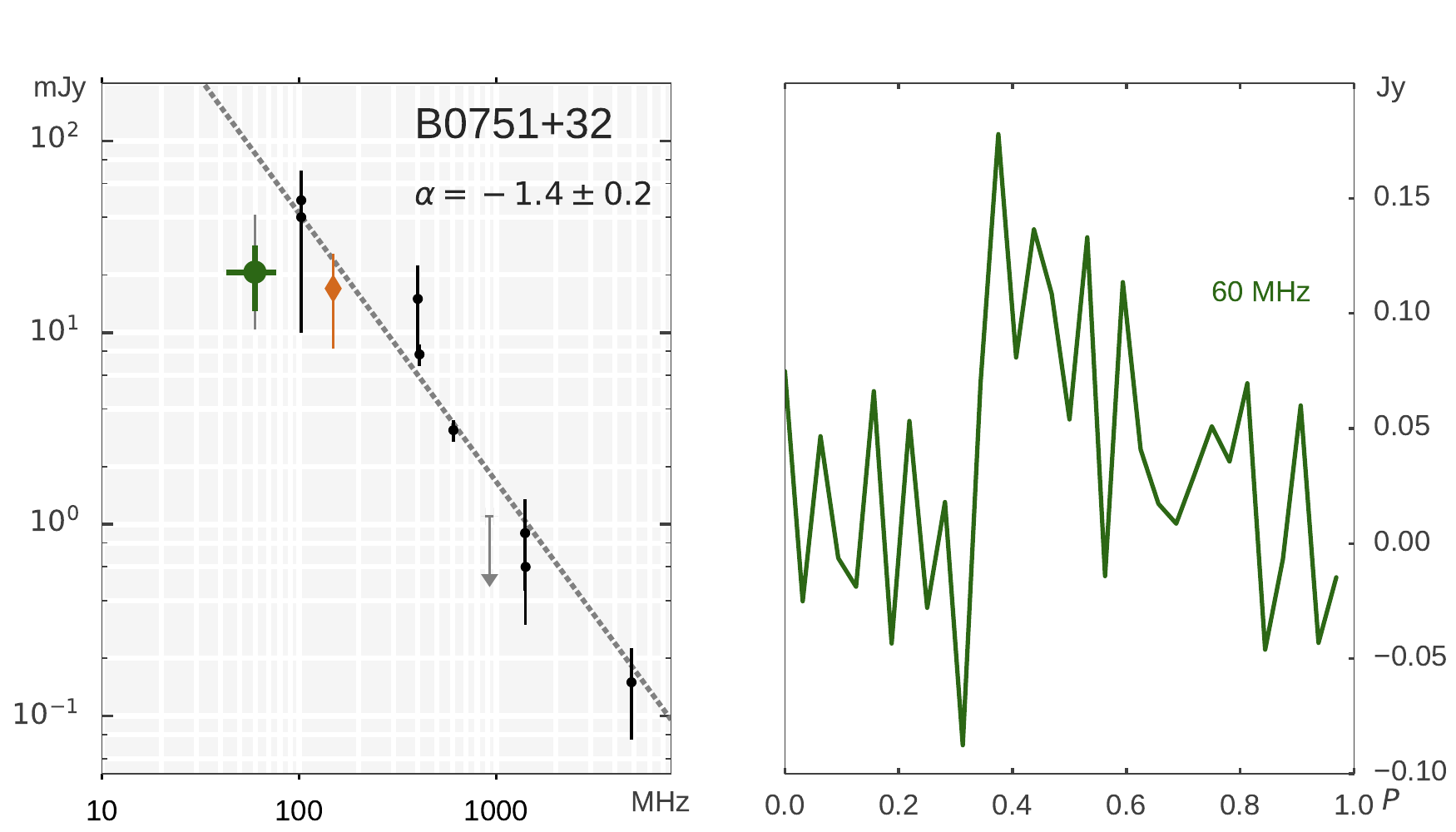}

\caption{For each pair of plots, \textit{Left:} Spectra of radio emission for pulsars detected in the census. 
Smaller black points with error bars mark literature flux densities, the larger green dots show the LOFAR 
LBA census measurements at various frequencies (with the horizontal errorbars indicating the frequency span 
of a given census measurement), brown diamonds show flux densities from HBA census (B16), and the arrows 
show upper limits. For the LBA census measurements, thin grey errorbars show $\pm50\%$ flux uncertainty 
and thicker green ones show uncertainty due to limited S/N. See text for both 
census and literature flux density errors and upper limit discussion. In the case of a multiple-PL fit, 
the uncertainty on the break frequency is marked with a broken black line. \textit{Right:} Flux-calibrated 
average profiles for LOFAR LBA census observations. Multiple profiles per band are shown with a constant 
flux offset between separate sub-bands. The choice of the number of sub-bands was defined by the peak S/N 
ratio of the average profile, the presence of profile evolution within the observing band and the number 
of previously published flux density values.}
\label{fig:prof_sp_1}
\end{figure*}

\begin{figure*}
\includegraphics[scale=0.48]{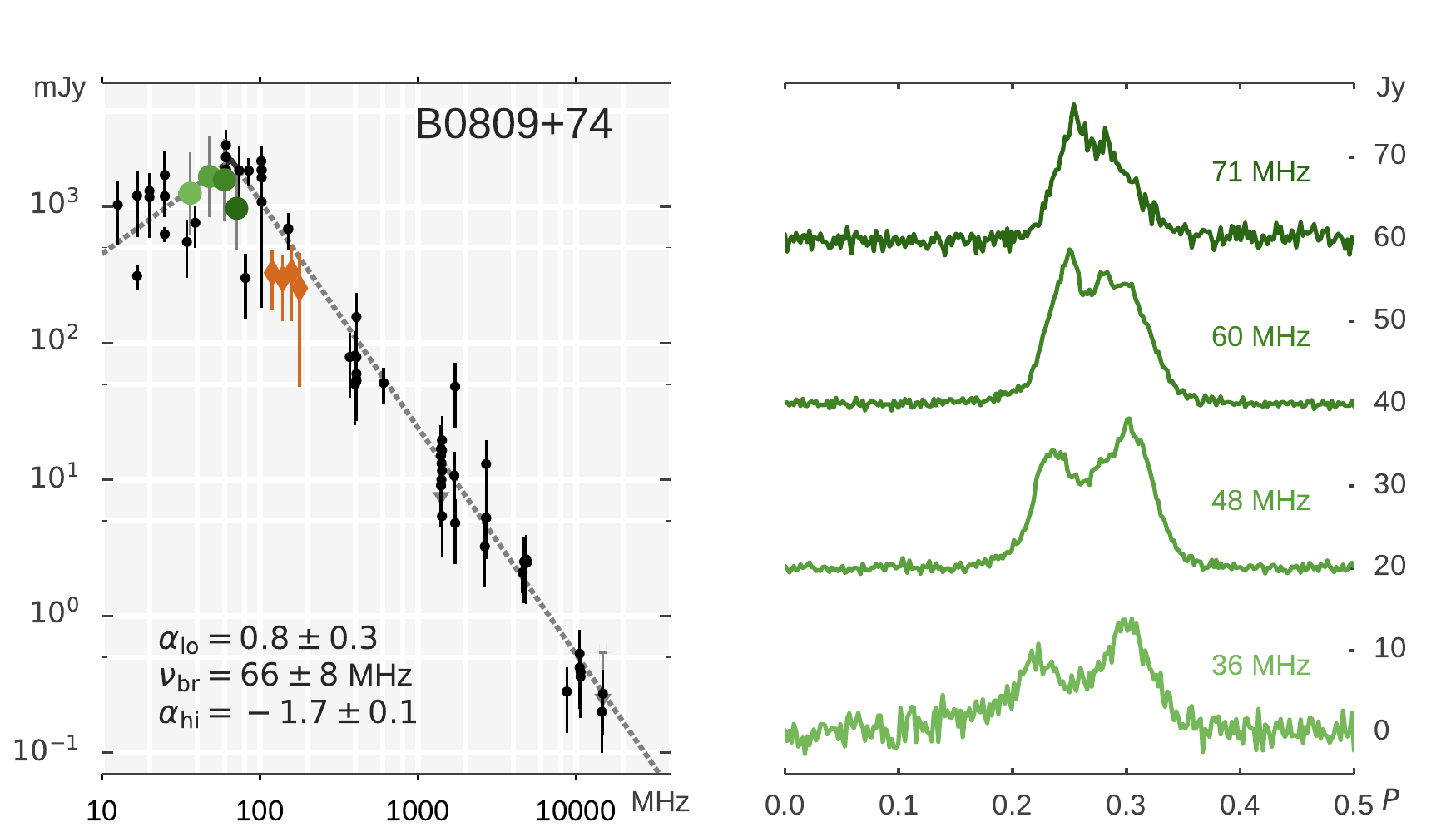}\includegraphics[scale=0.48]{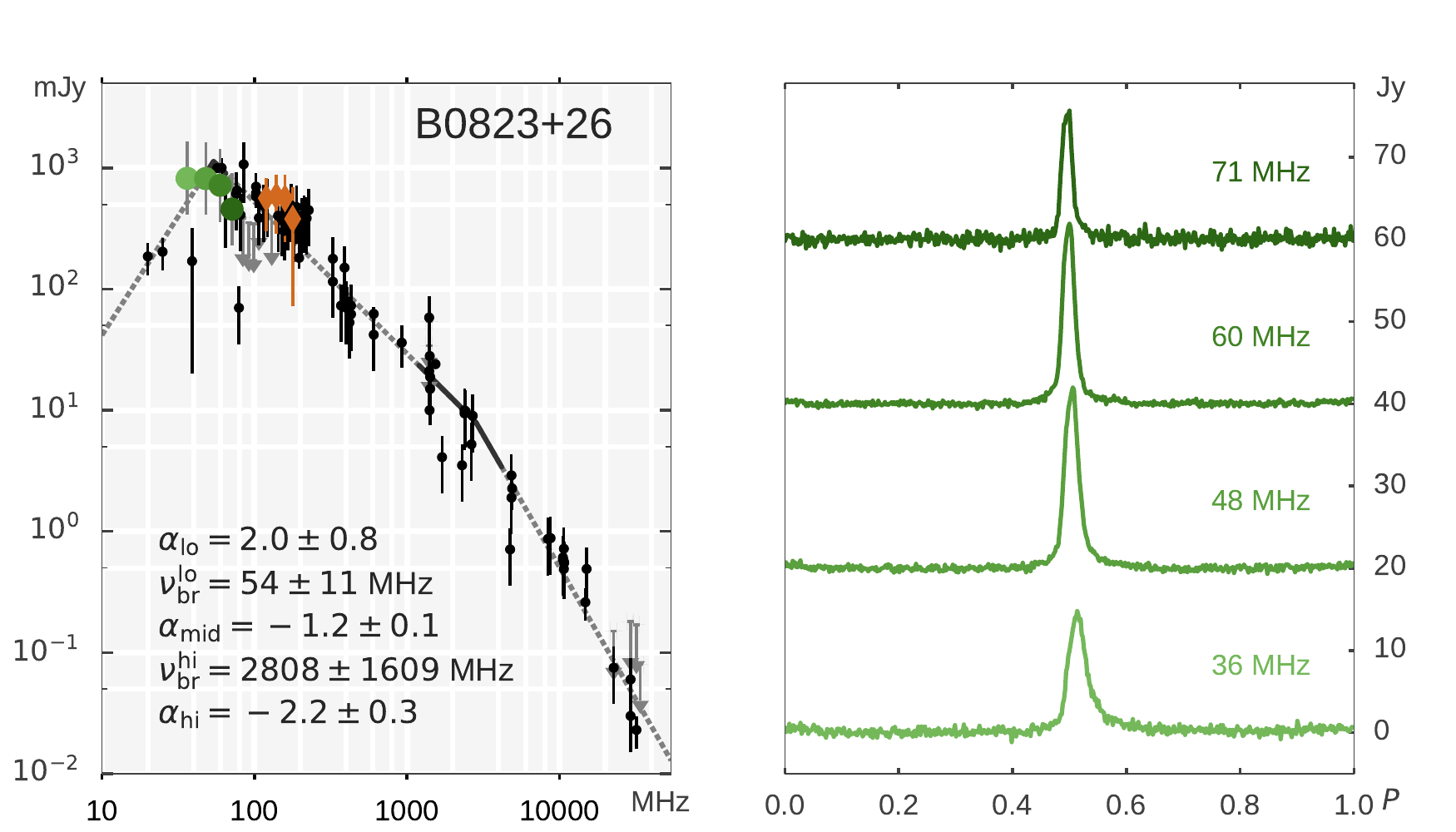}
\includegraphics[scale=0.48]{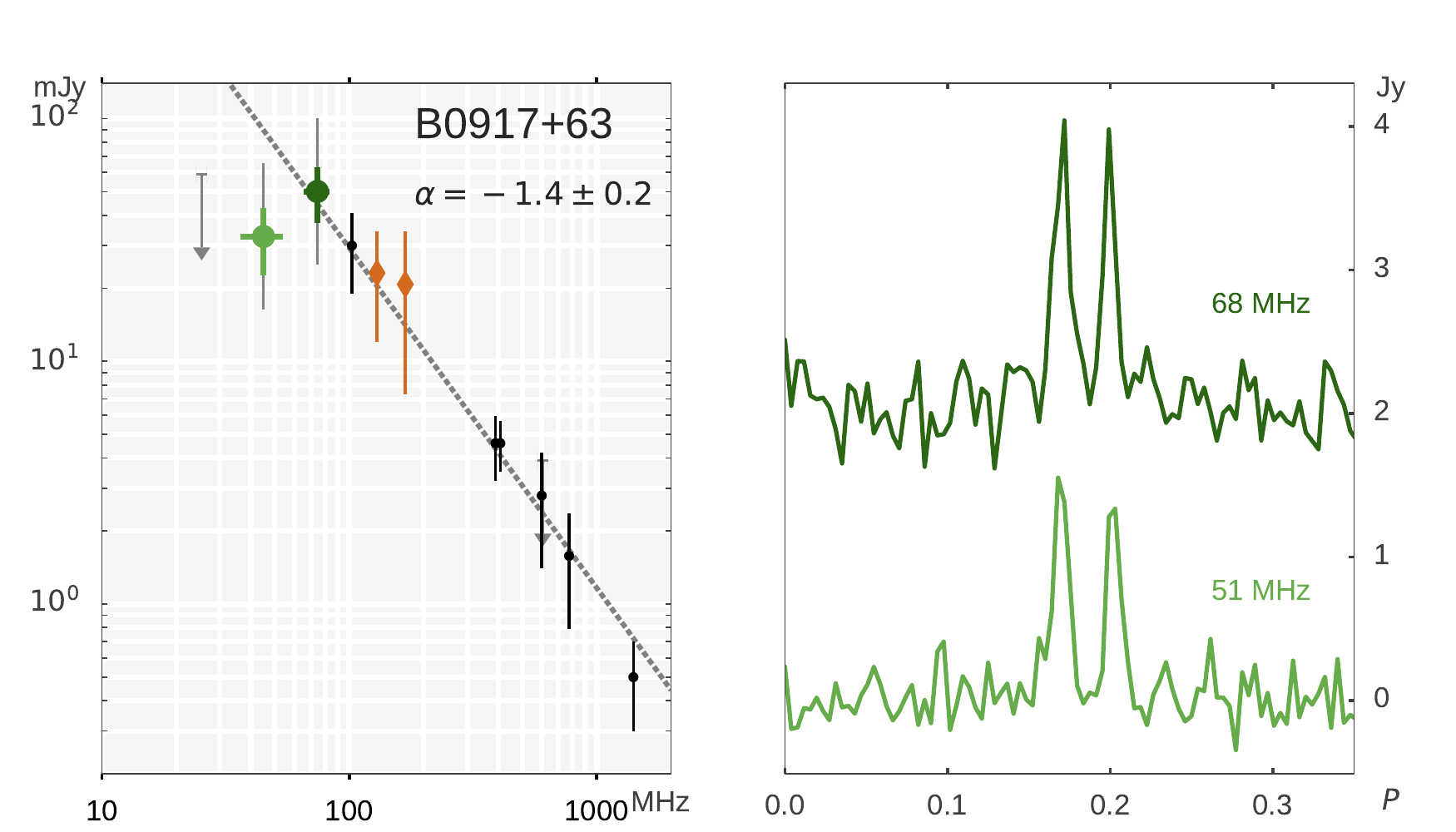}\includegraphics[scale=0.48]{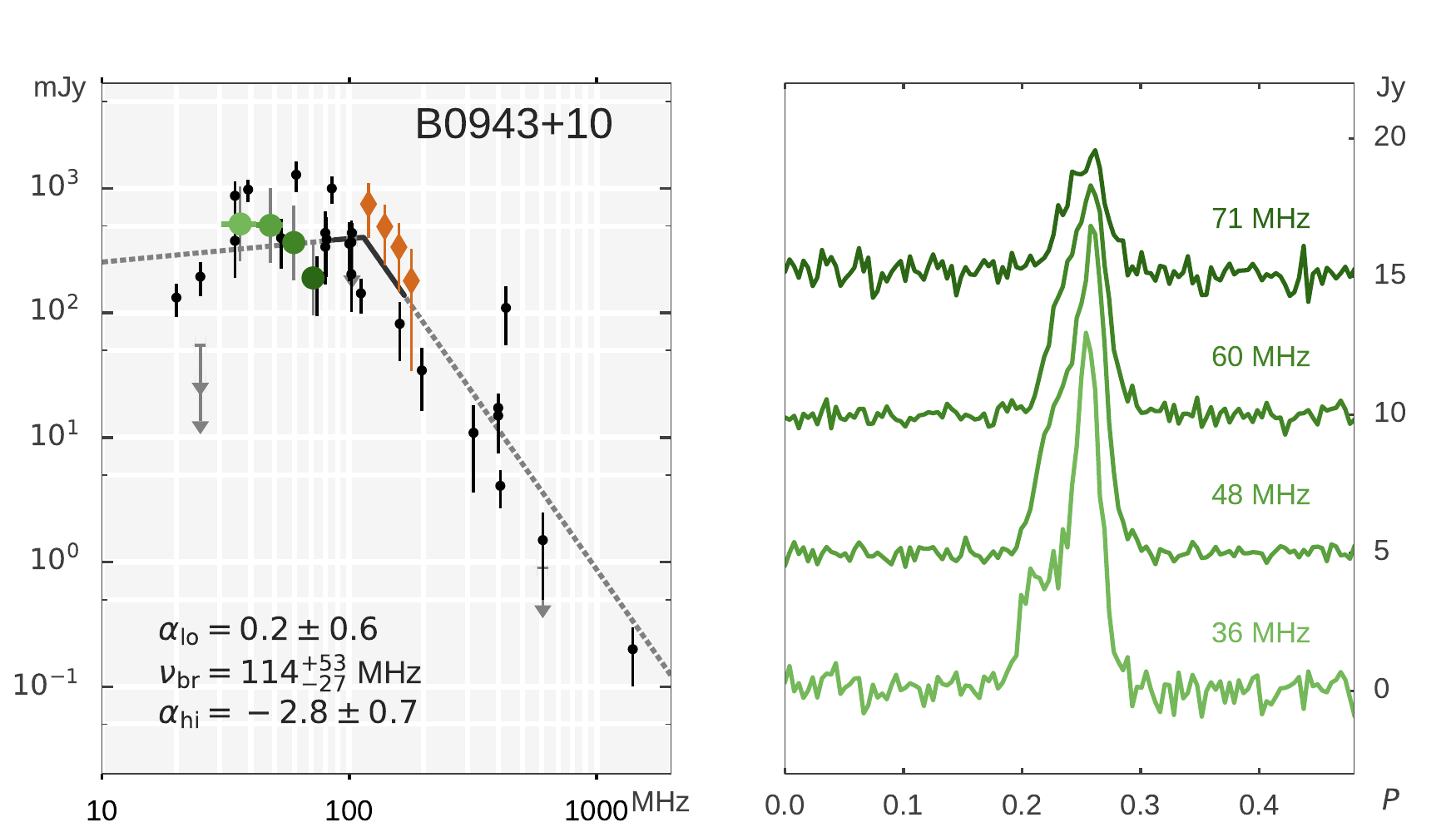}
\includegraphics[scale=0.48]{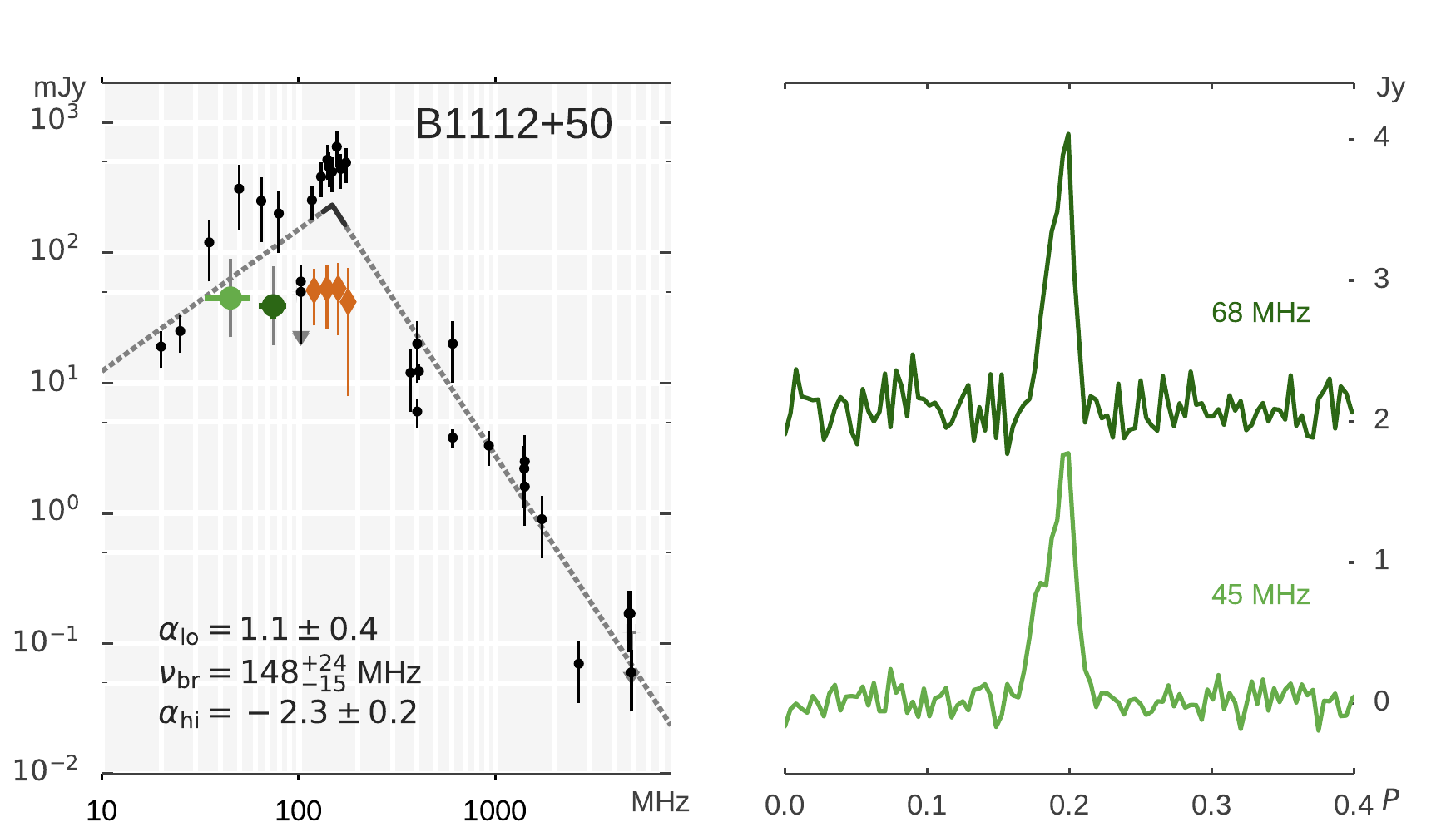}\includegraphics[scale=0.48]{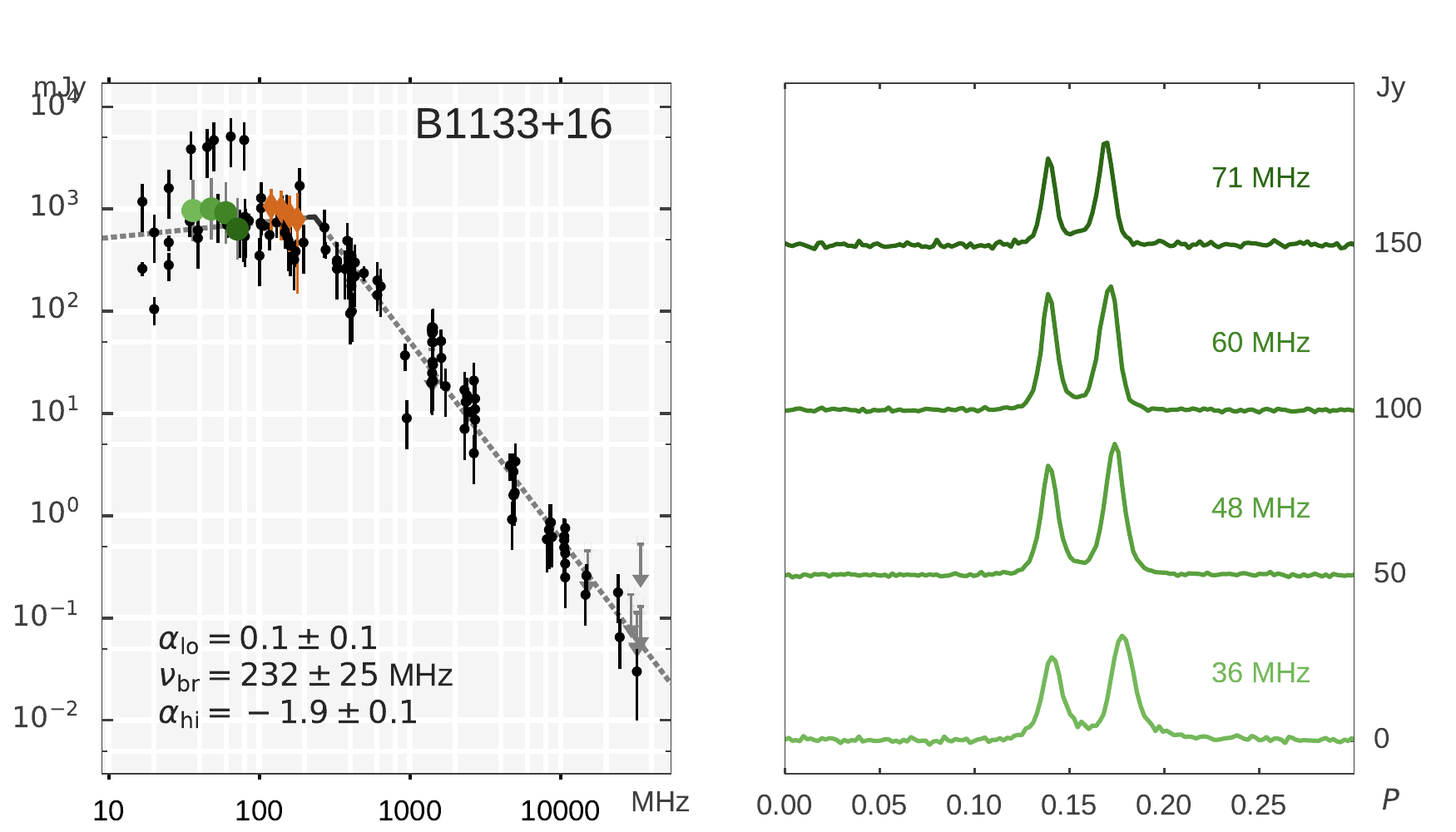}
\includegraphics[scale=0.48]{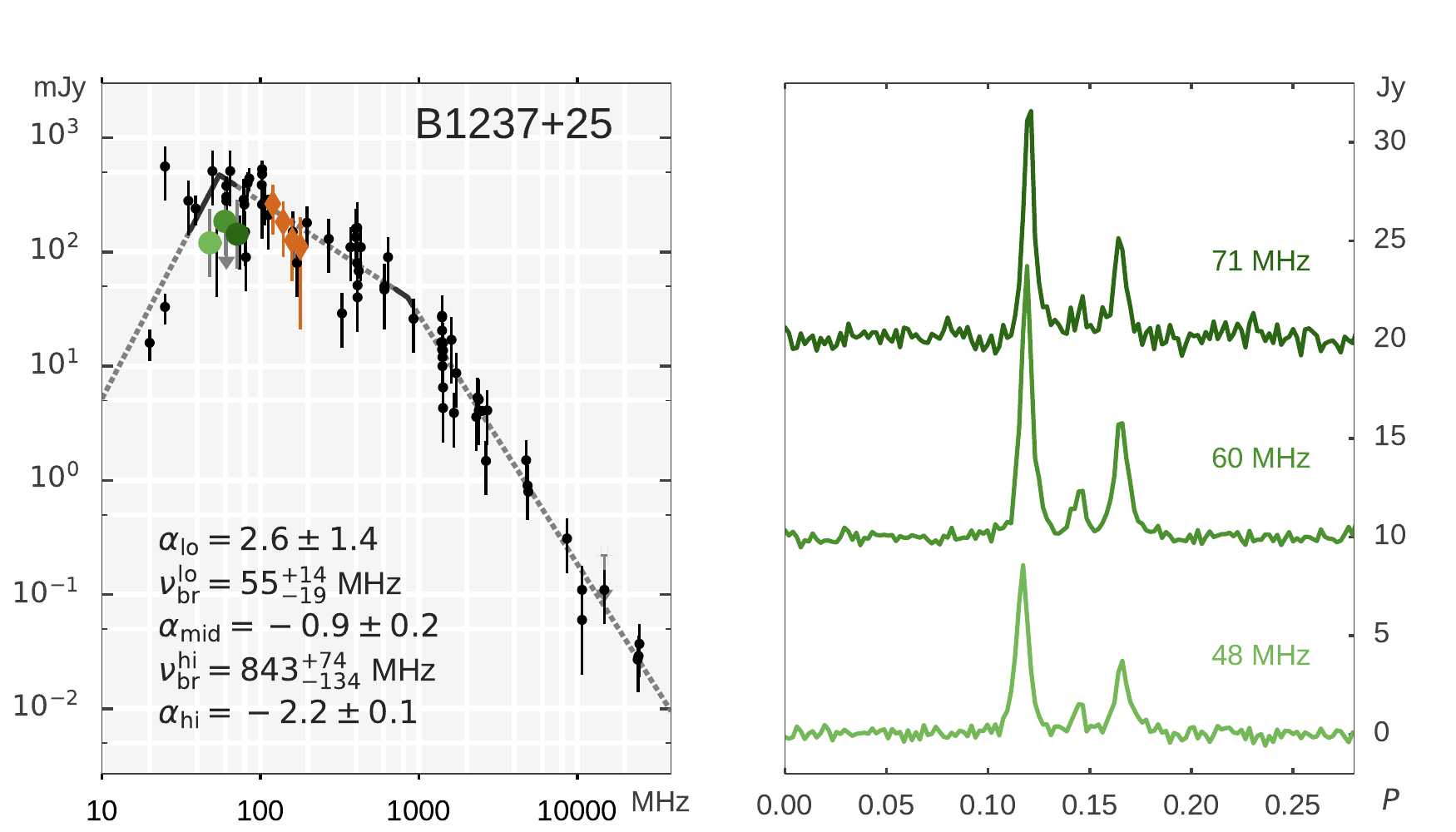}\includegraphics[scale=0.48]{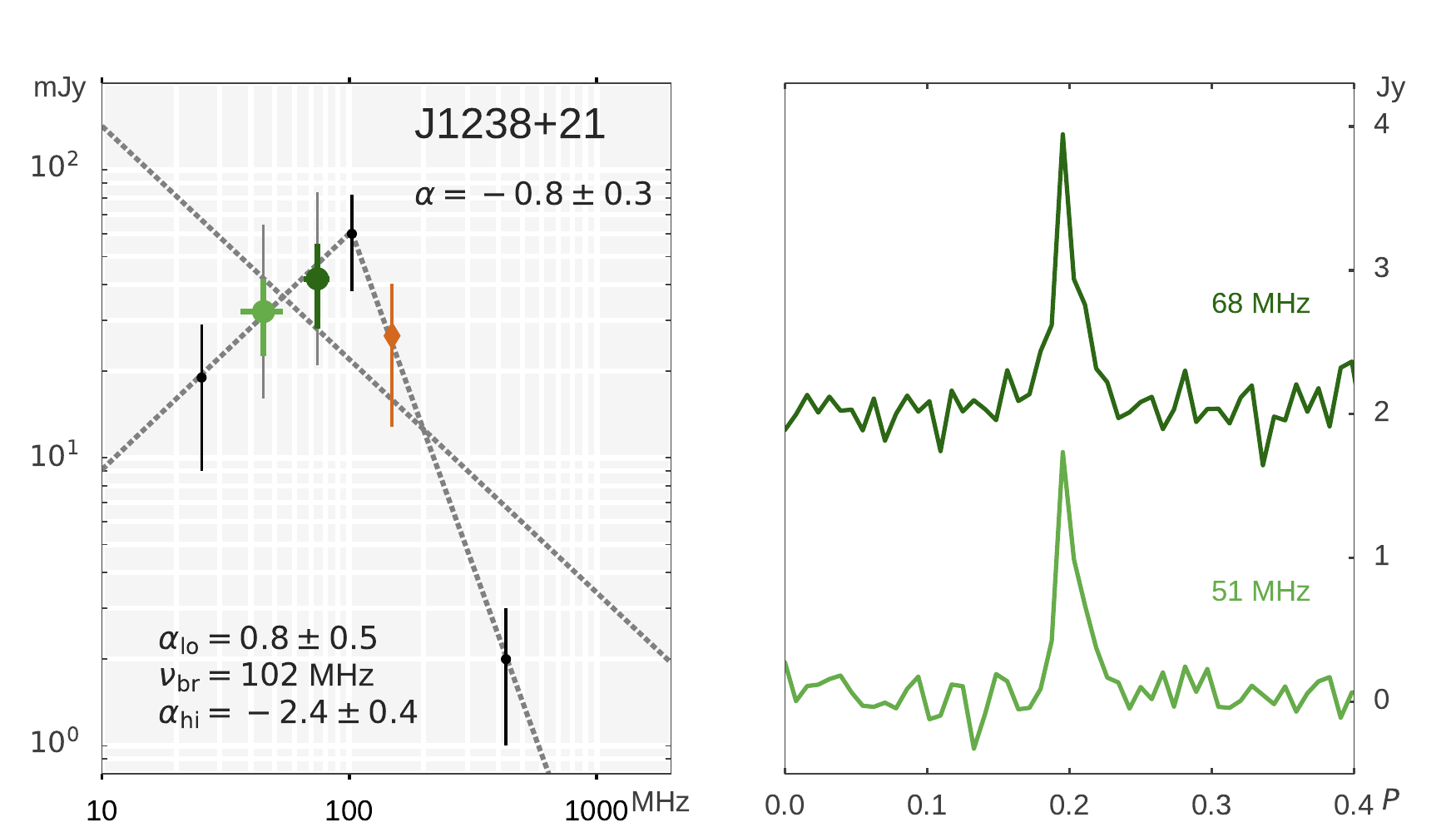}
\includegraphics[scale=0.48]{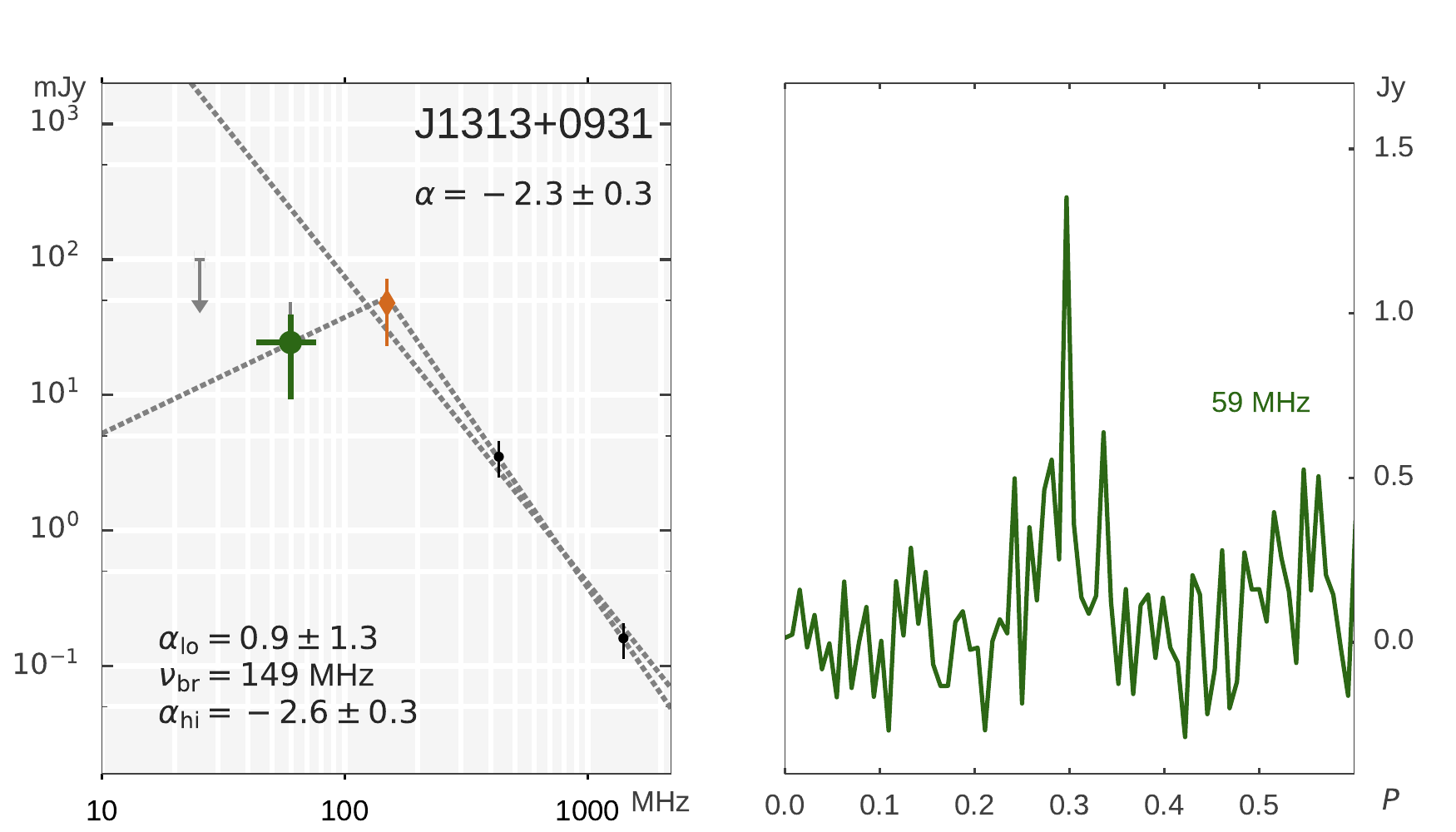}\includegraphics[scale=0.48]{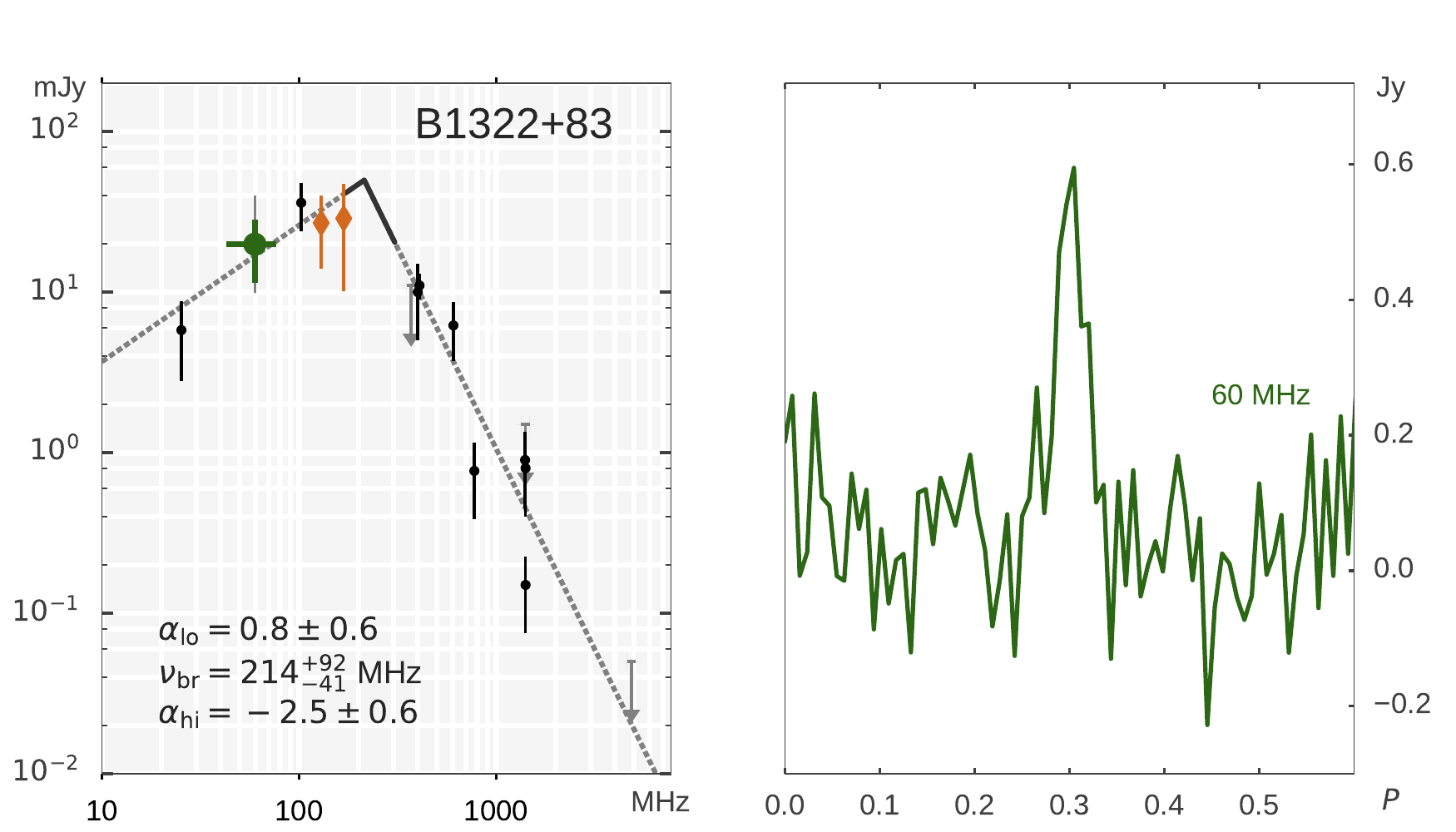}
\caption{See Figure~\ref{fig:prof_sp_1}.}
\label{fig:prof_sp_2}
\end{figure*}

\begin{figure*}
\includegraphics[scale=0.48]{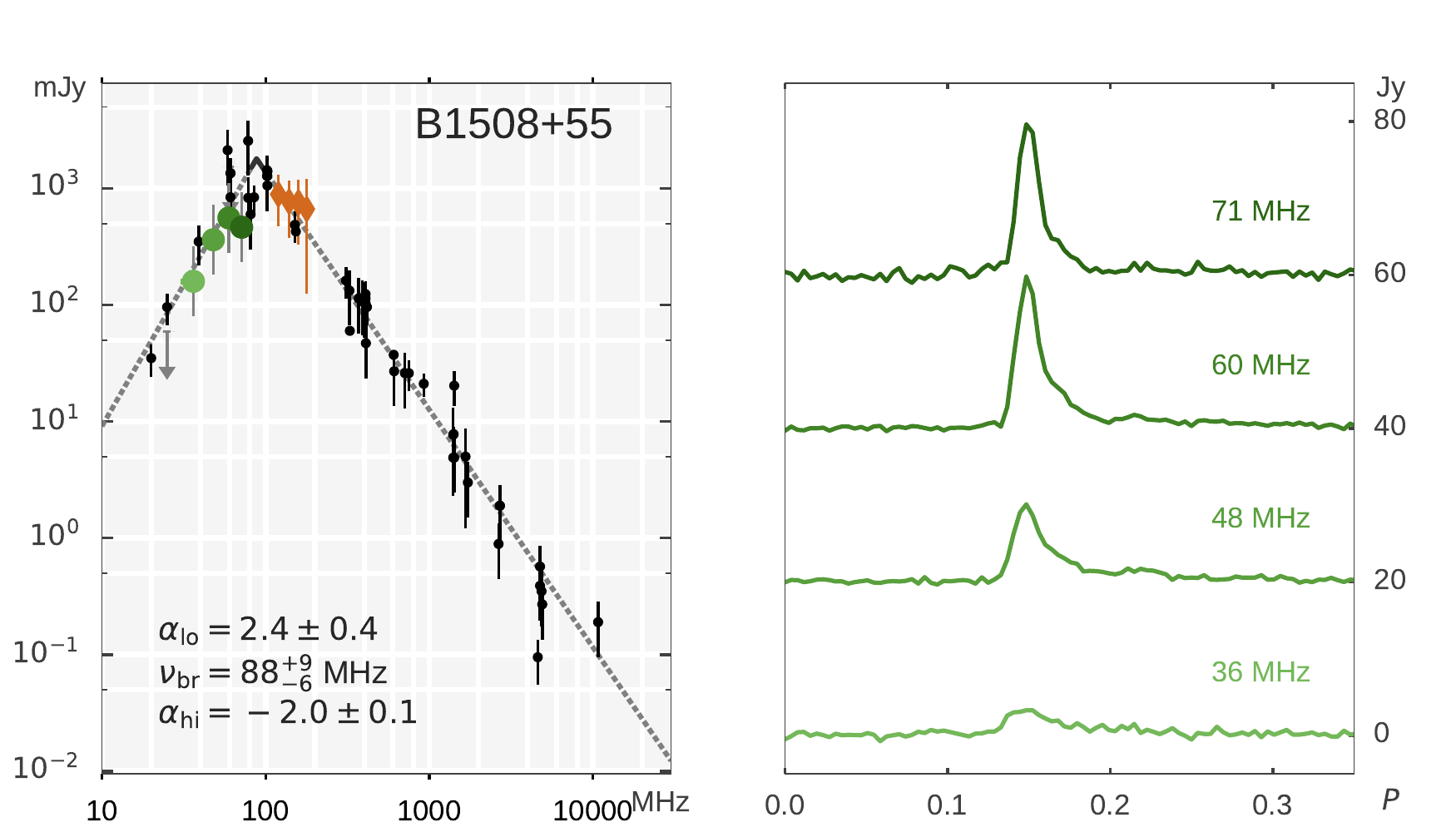}\includegraphics[scale=0.48]{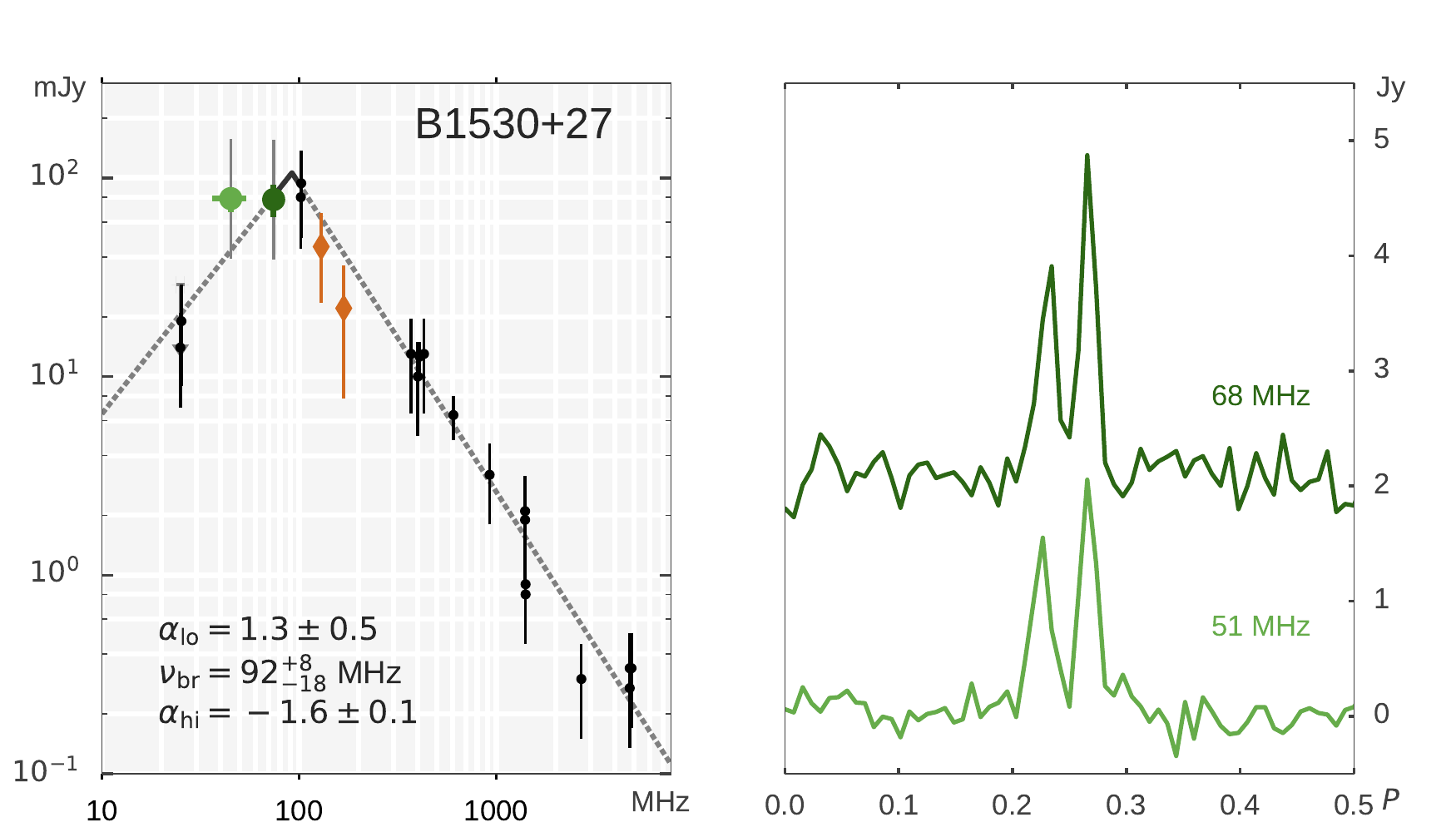}
\includegraphics[scale=0.48]{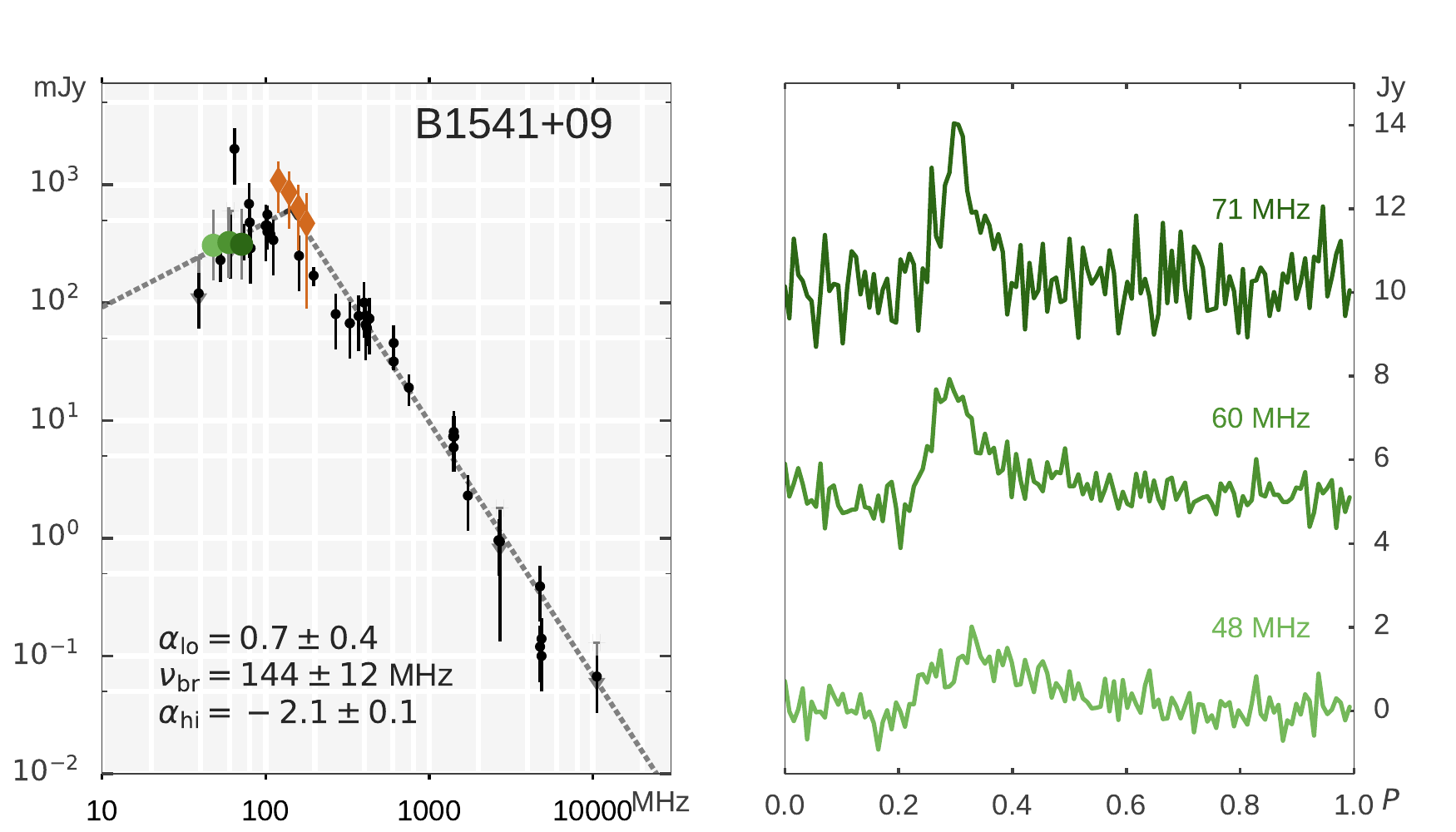}\includegraphics[scale=0.48]{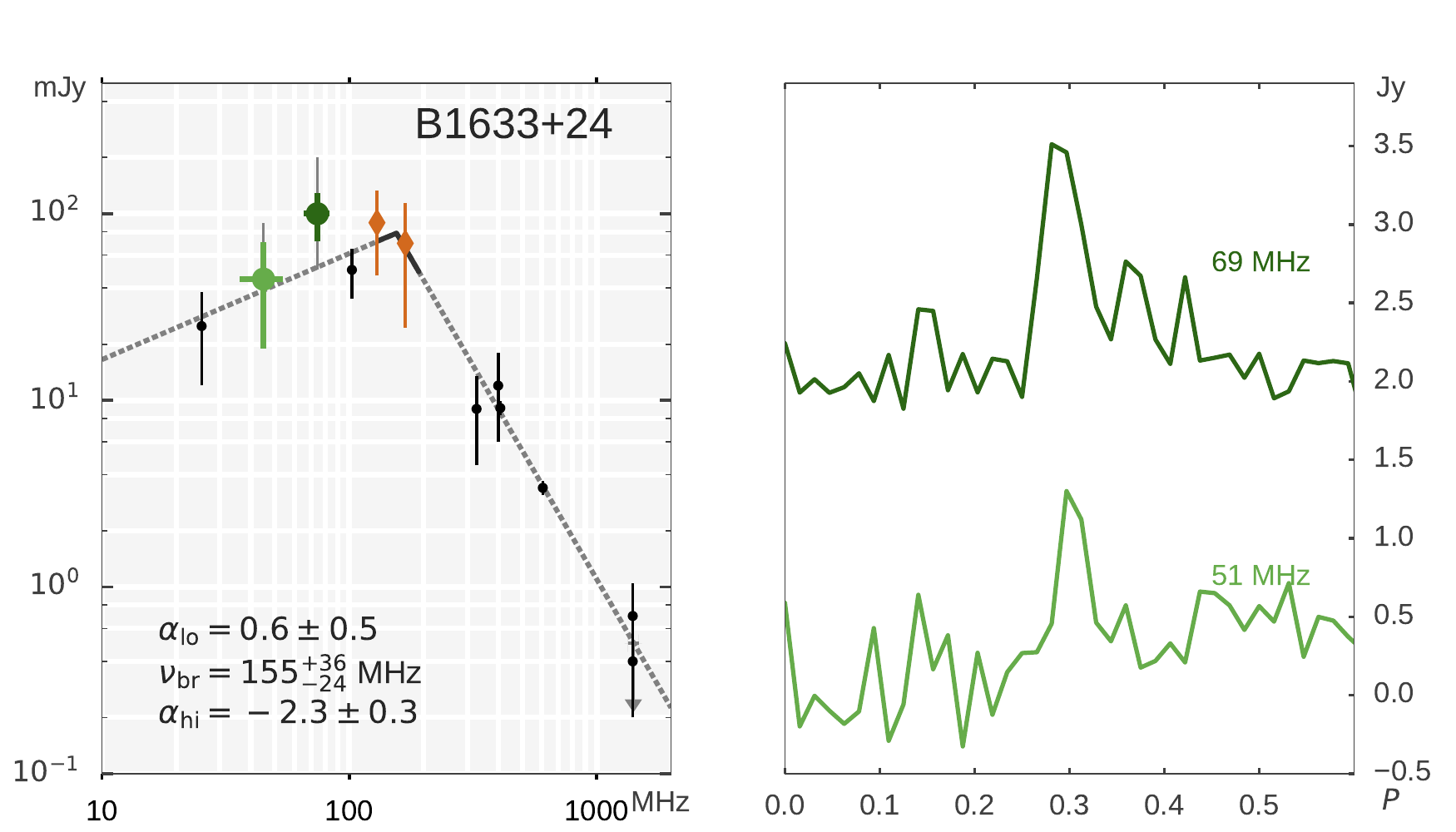}
\includegraphics[scale=0.48]{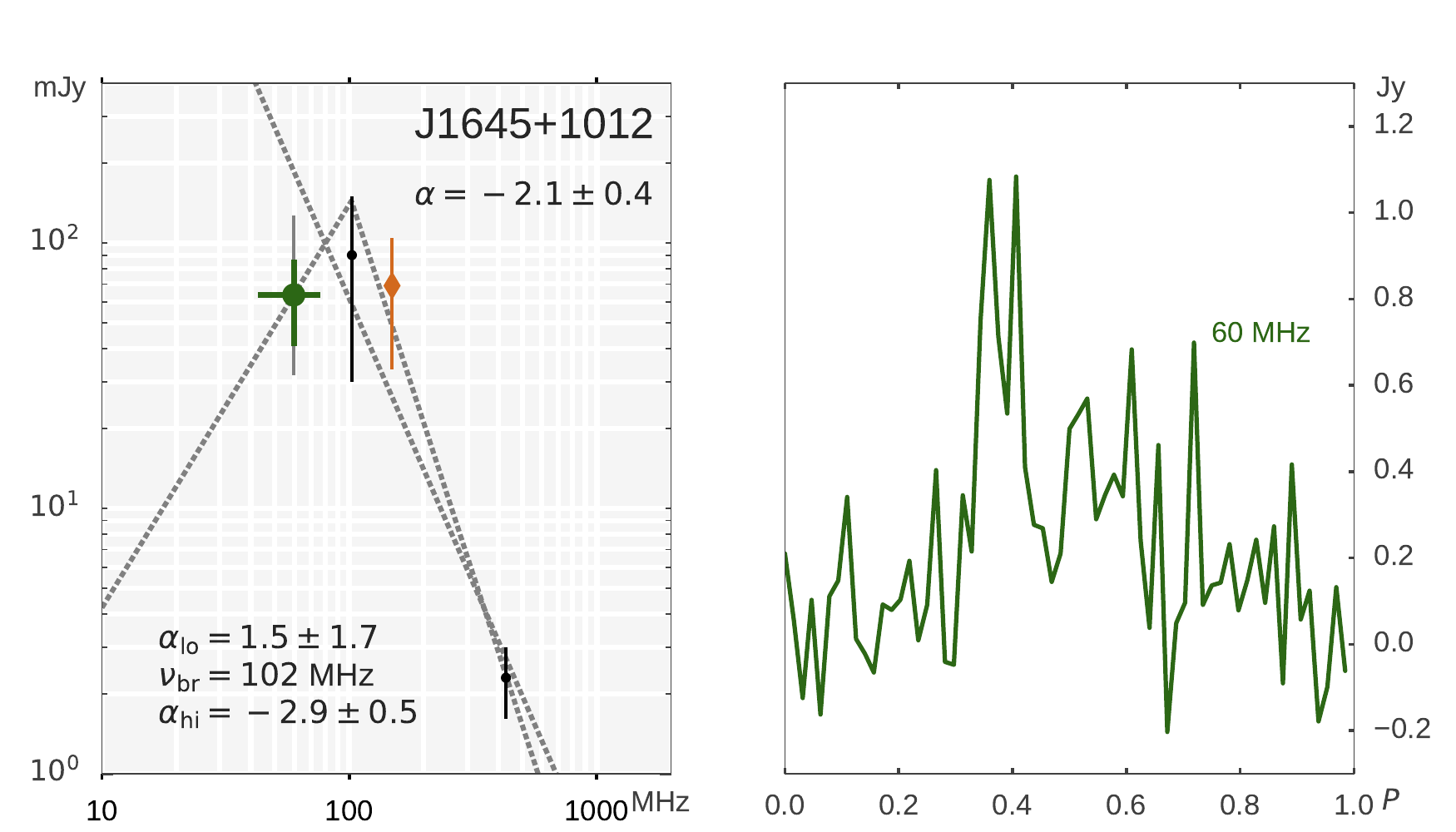}\includegraphics[scale=0.48]{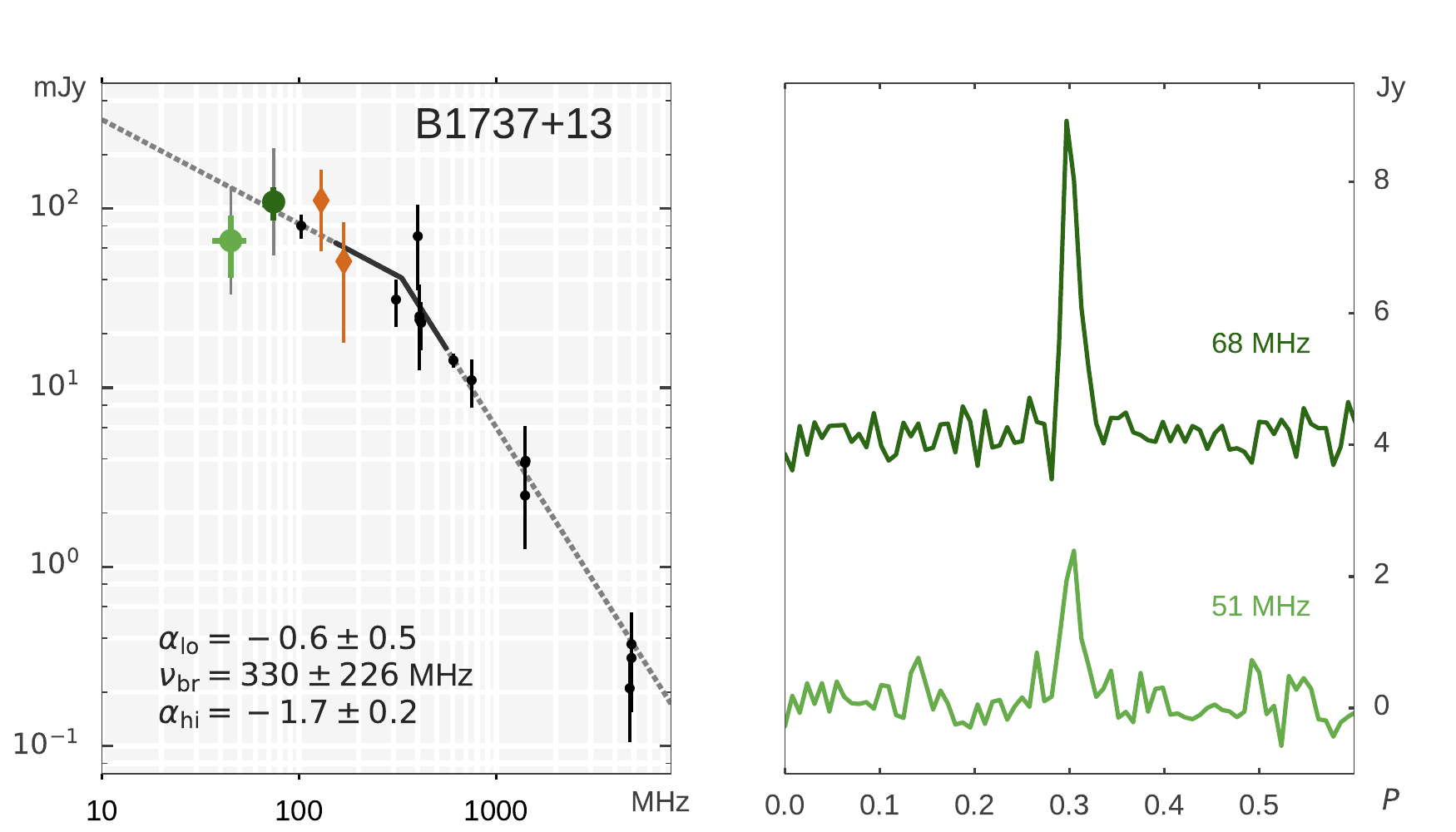}
\includegraphics[scale=0.48]{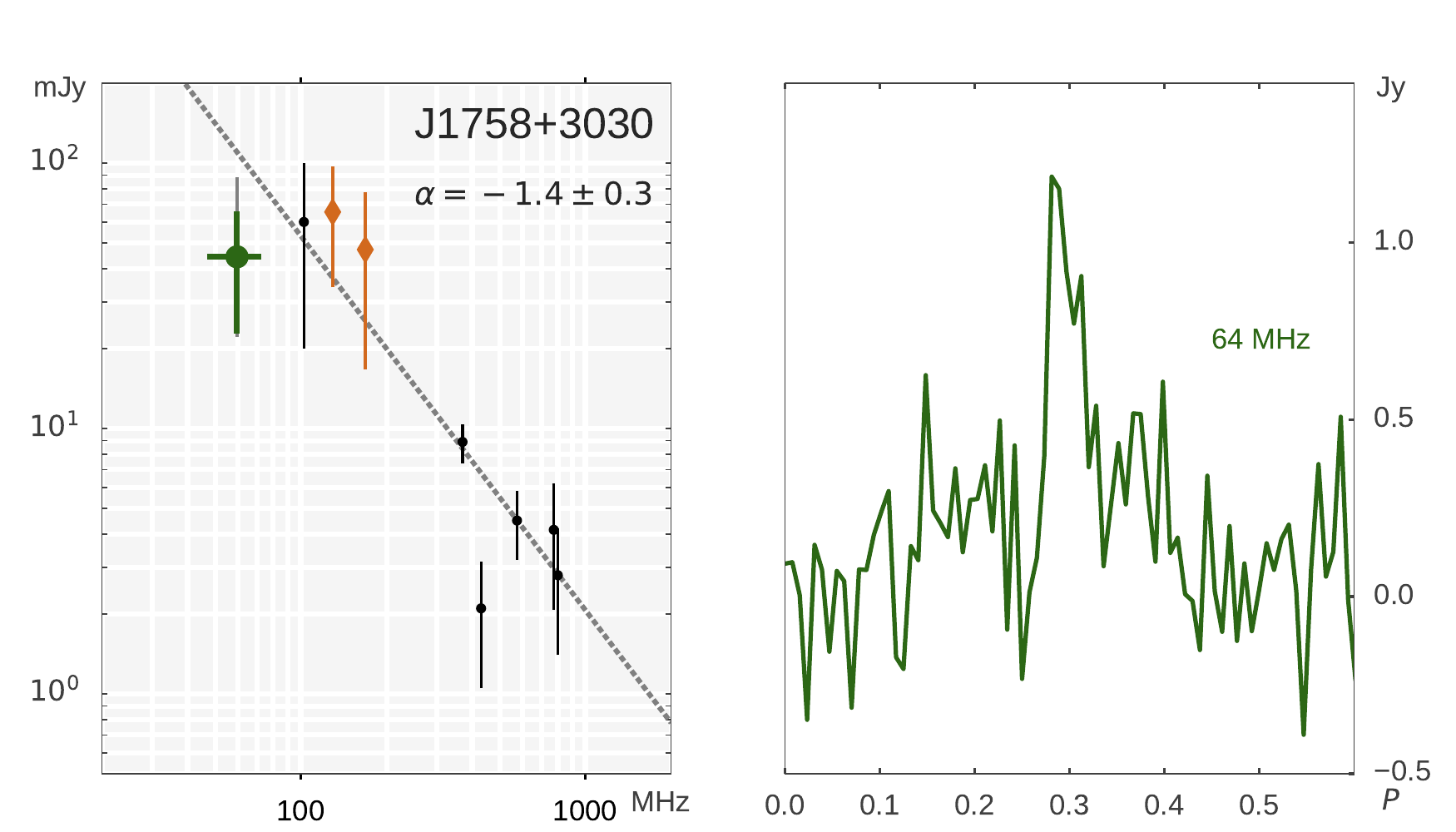}\includegraphics[scale=0.48]{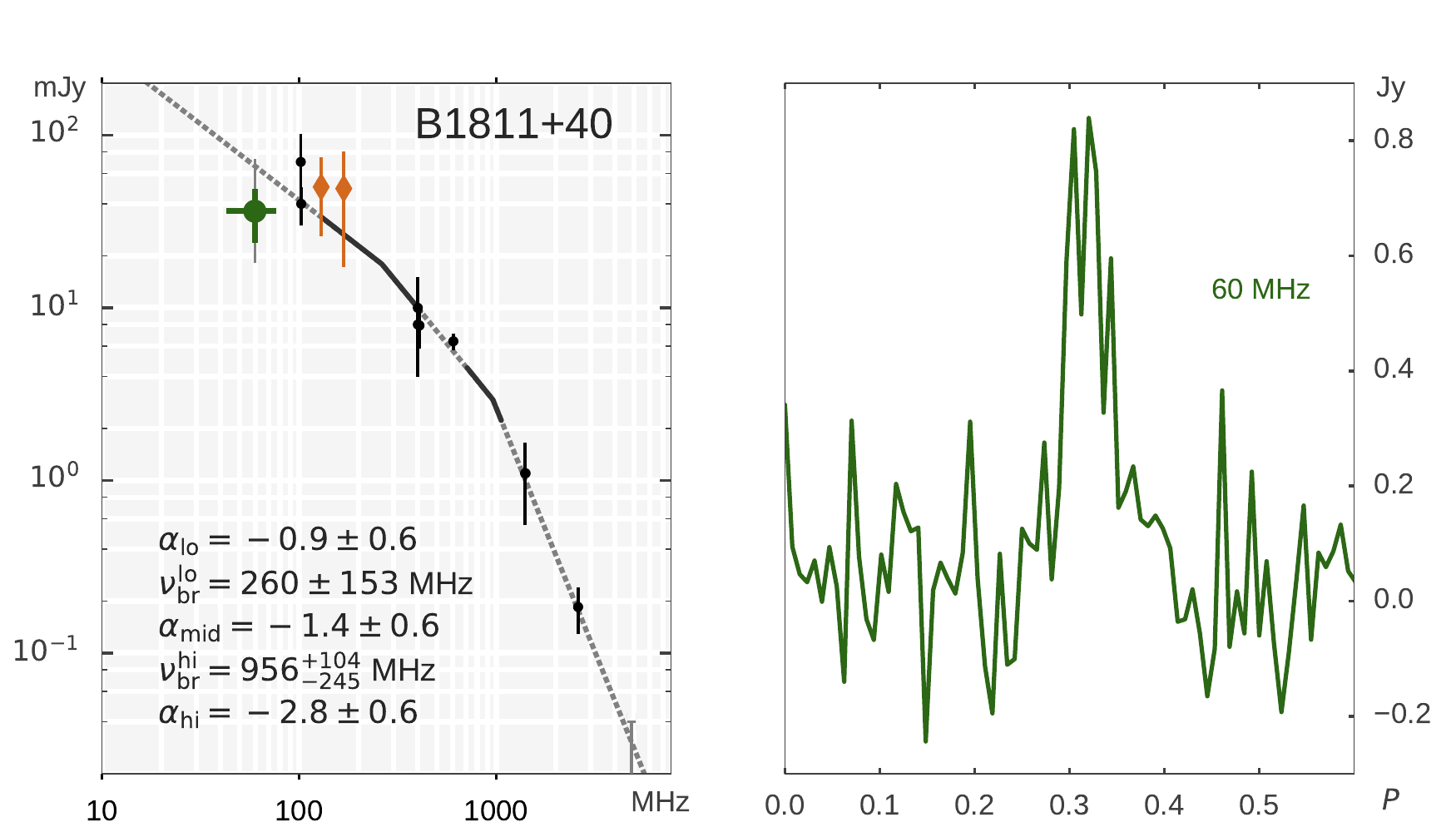}
\includegraphics[scale=0.48]{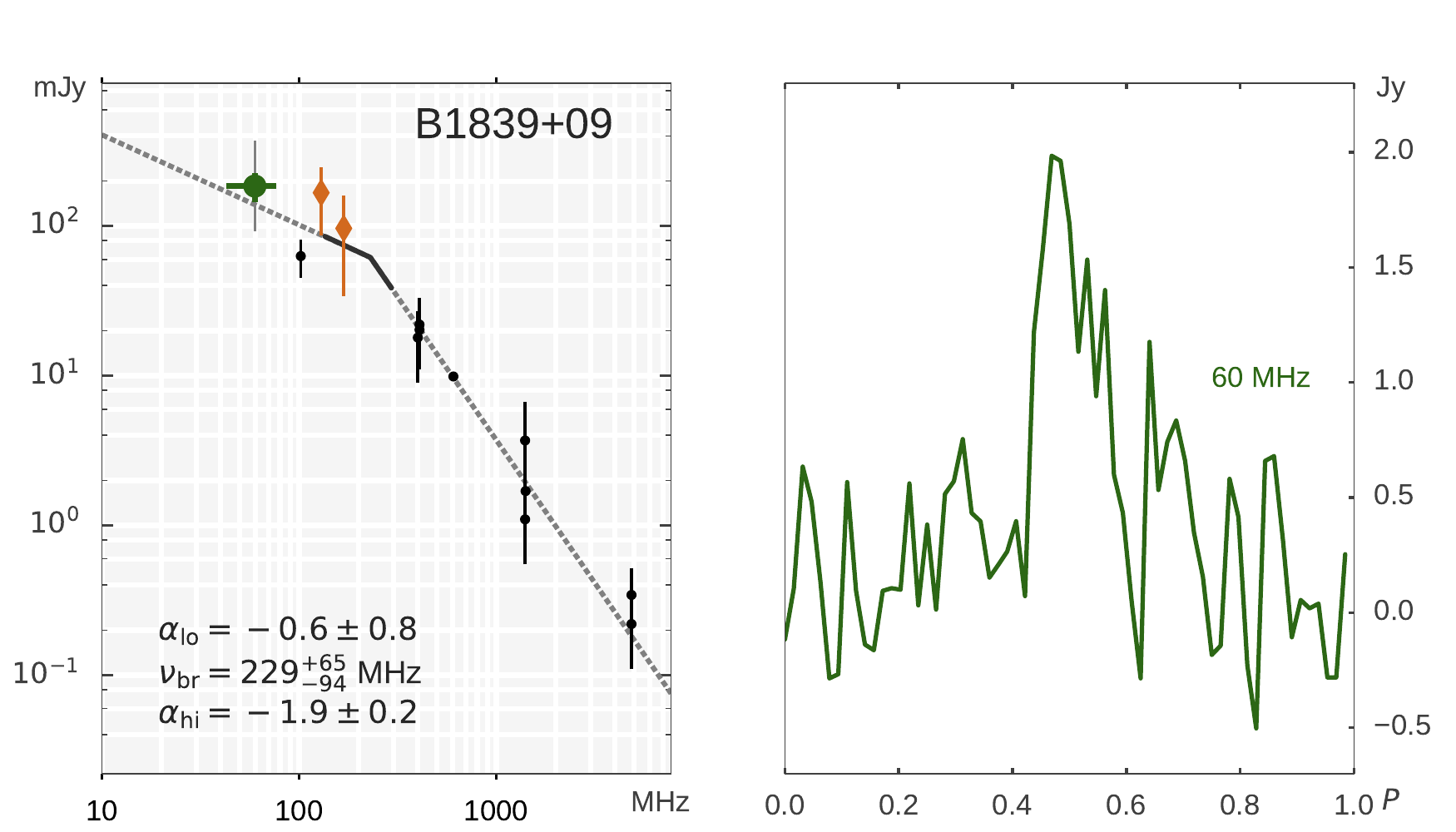}\includegraphics[scale=0.48]{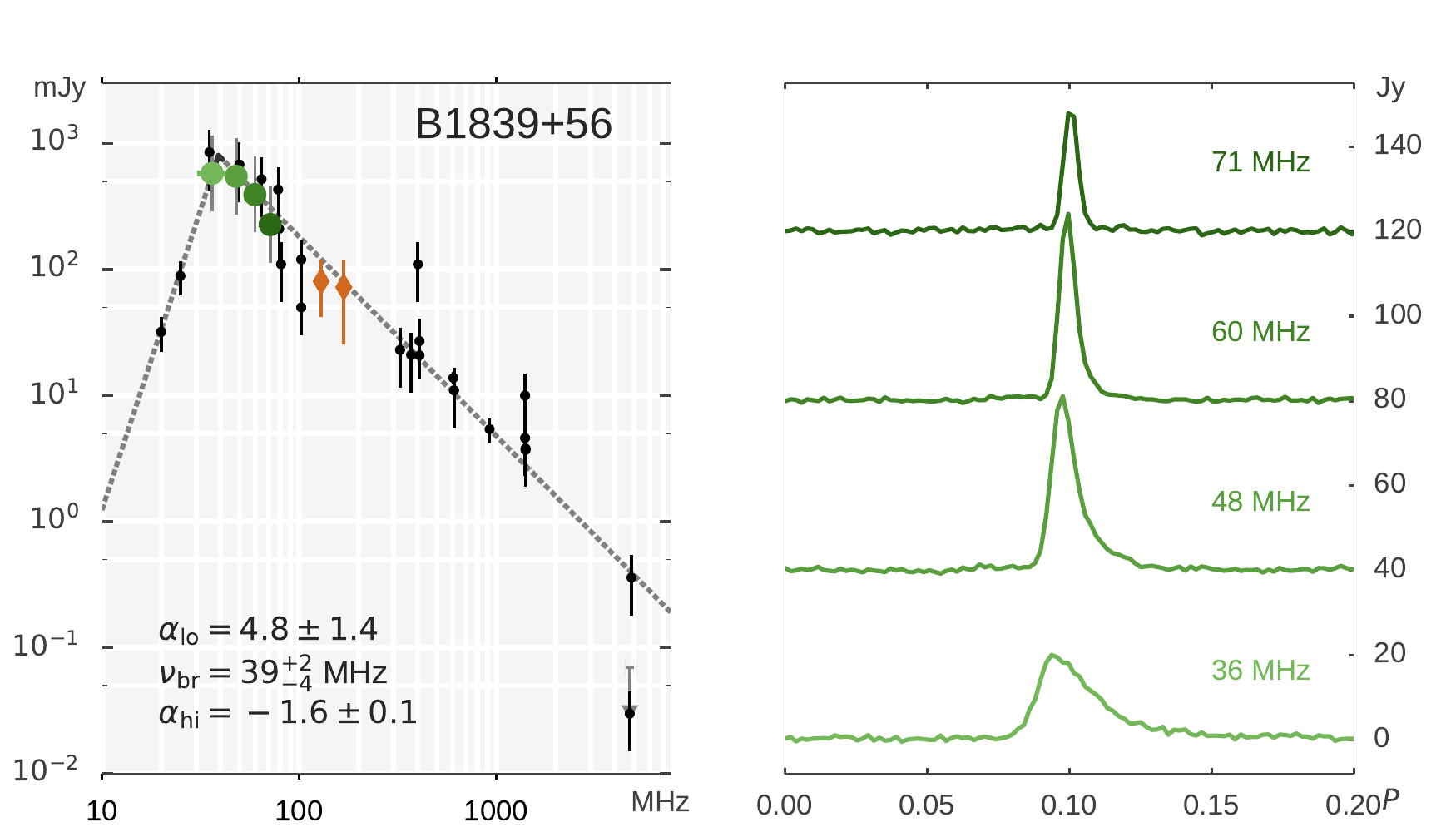}
\caption{See Figure~\ref{fig:prof_sp_1}.}
\label{fig:prof_sp_3}
\end{figure*}

\begin{figure*}
\includegraphics[scale=0.48]{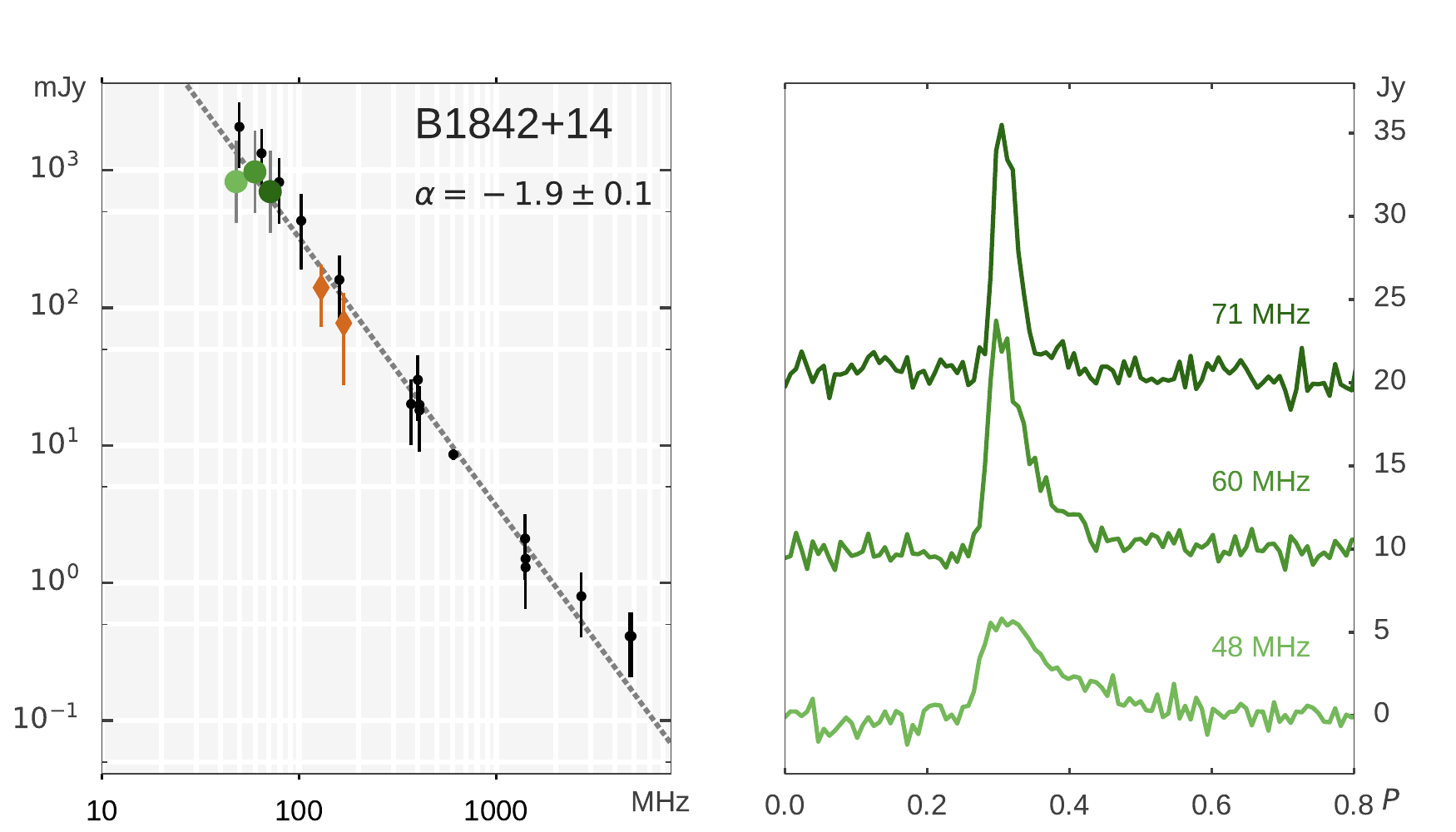}\includegraphics[scale=0.48]{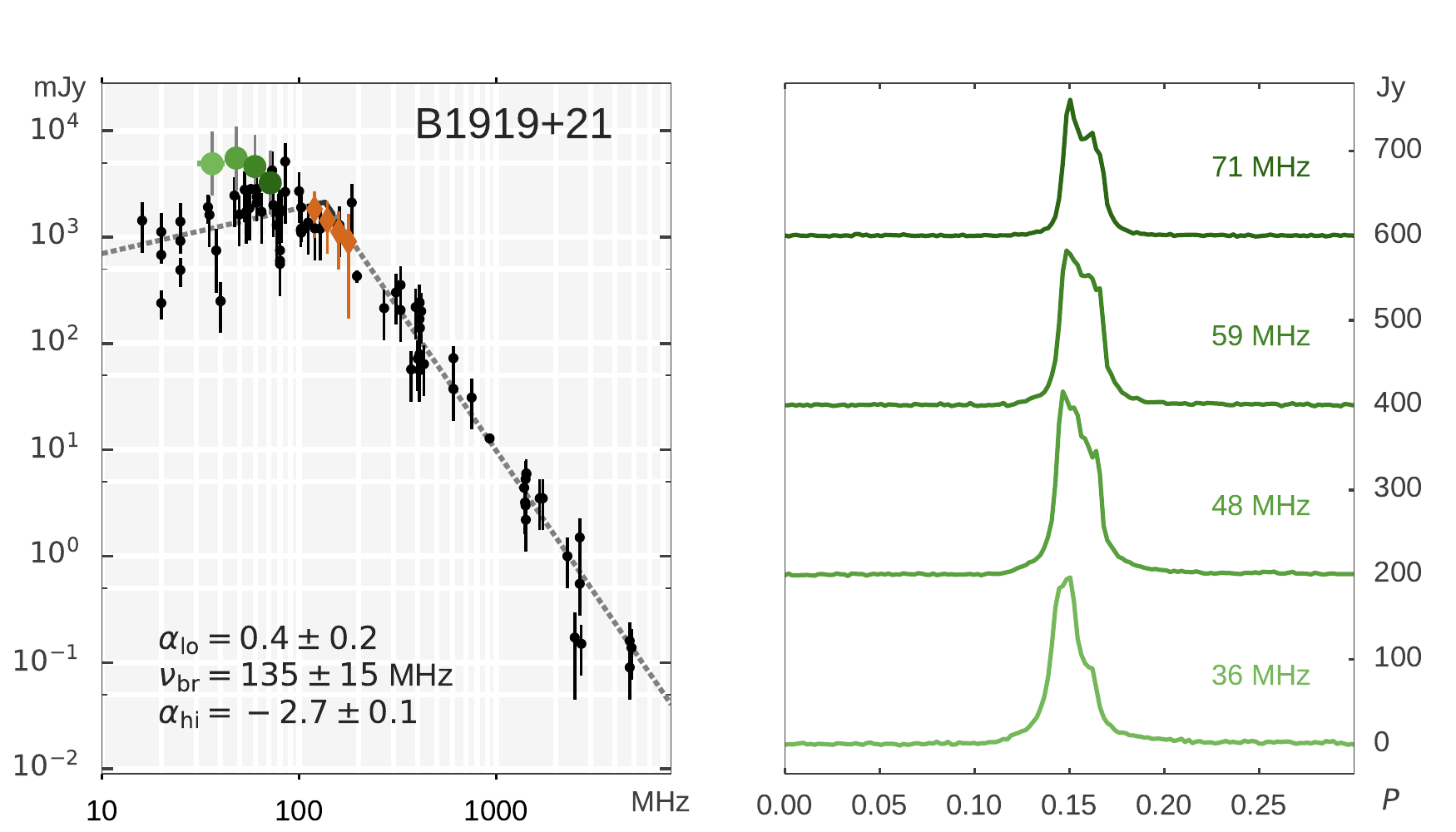}
\includegraphics[scale=0.48]{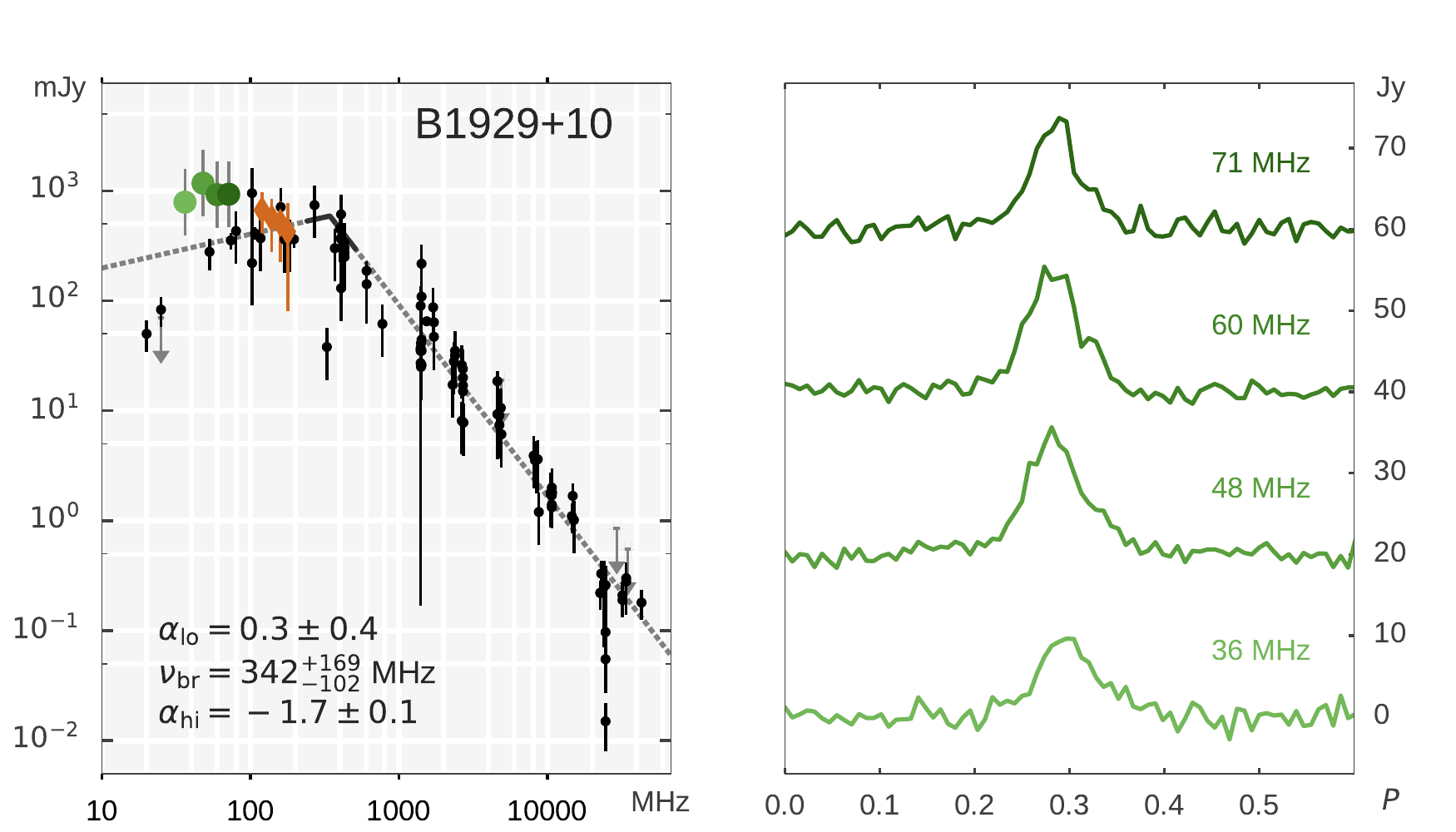}\includegraphics[scale=0.48]{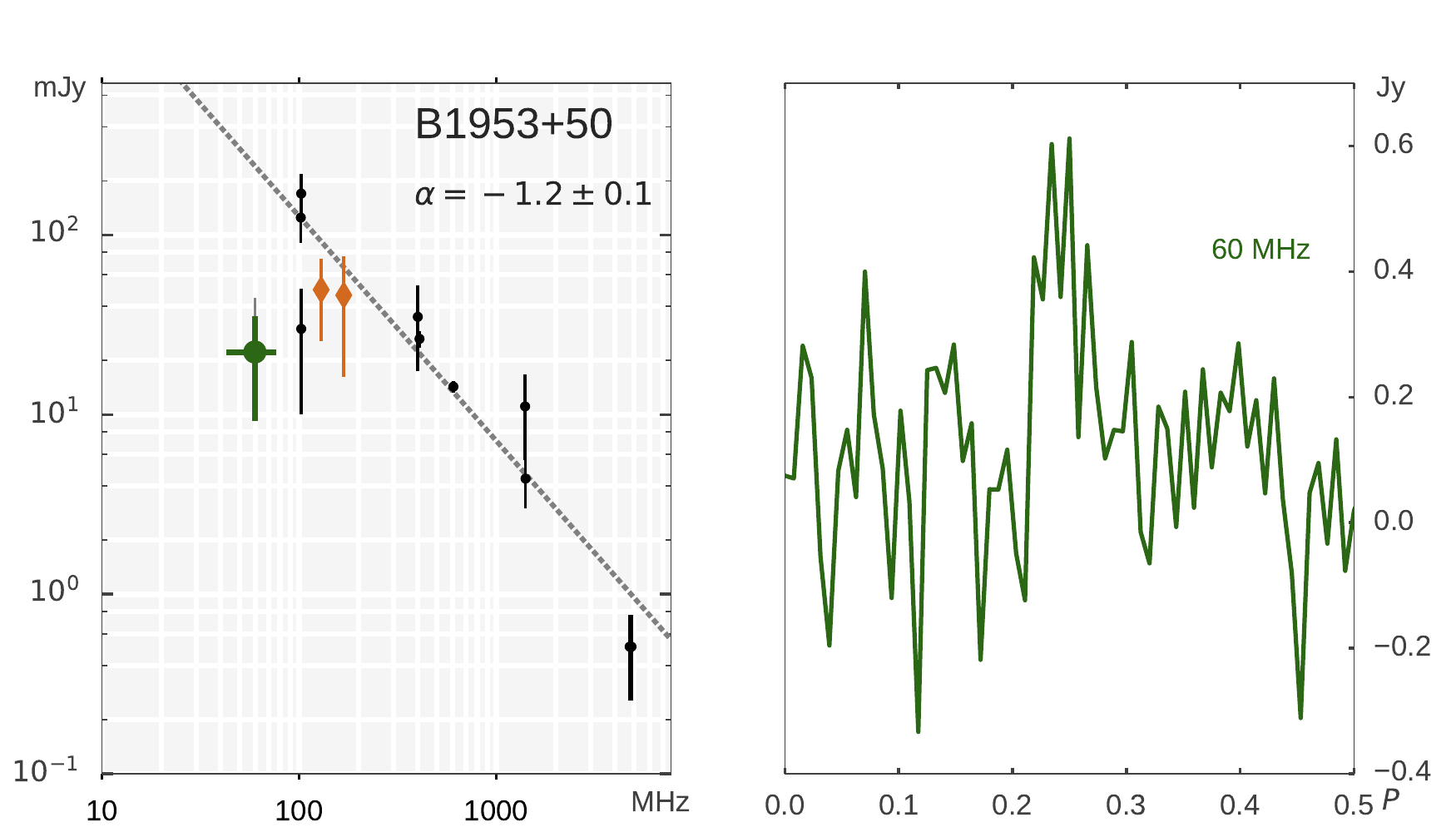}
\includegraphics[scale=0.48]{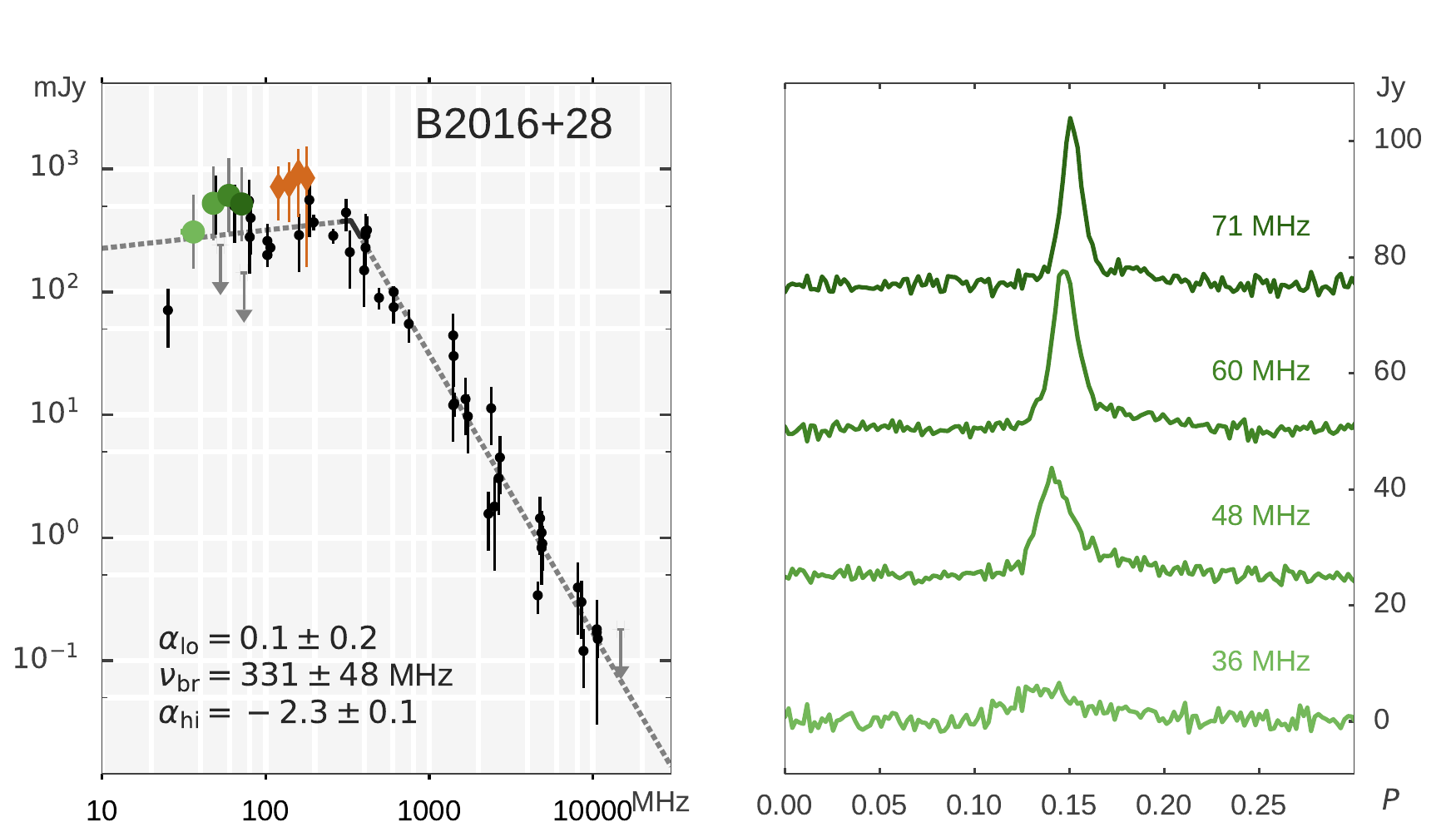}\includegraphics[scale=0.48]{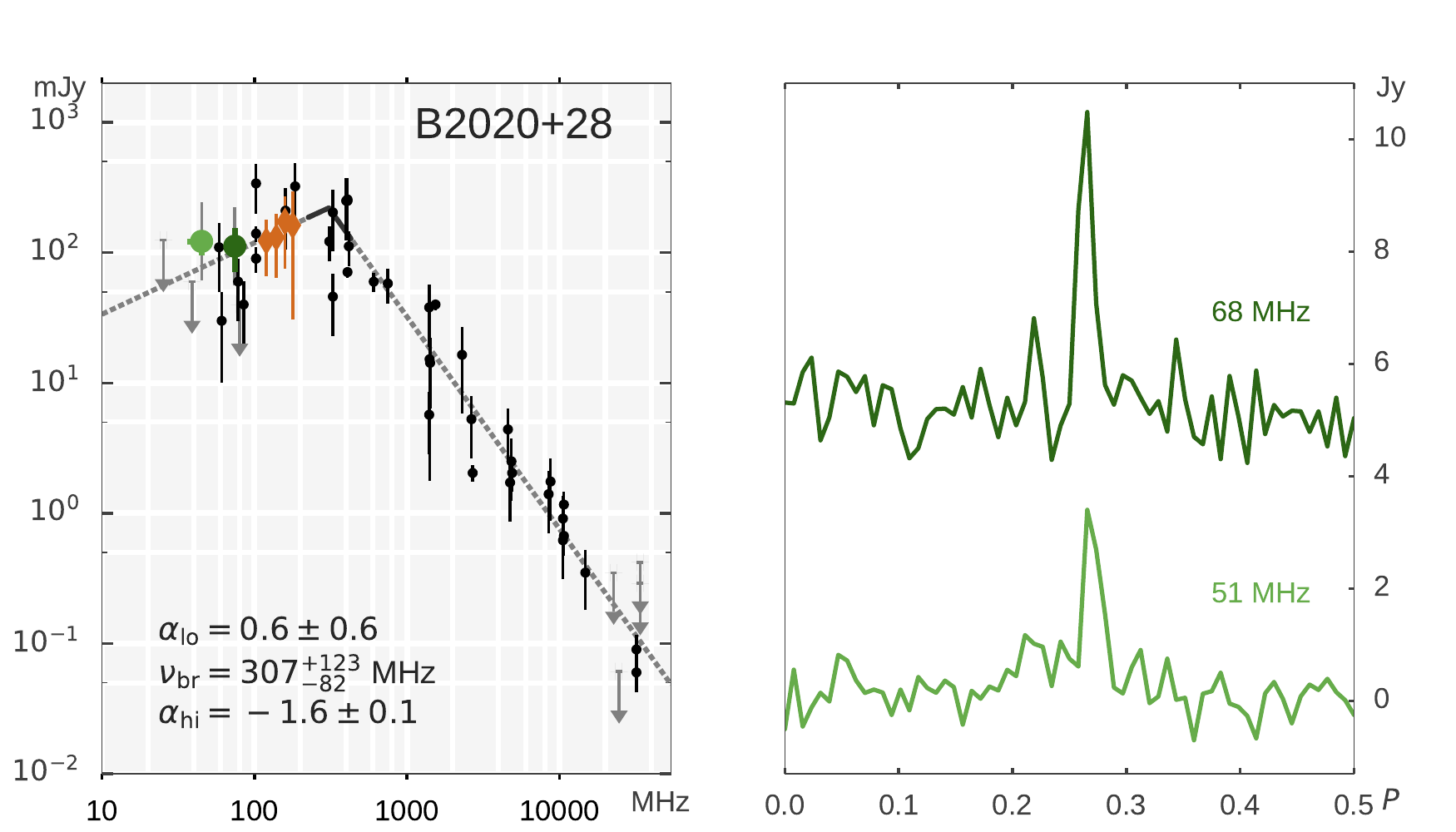}
\includegraphics[scale=0.48]{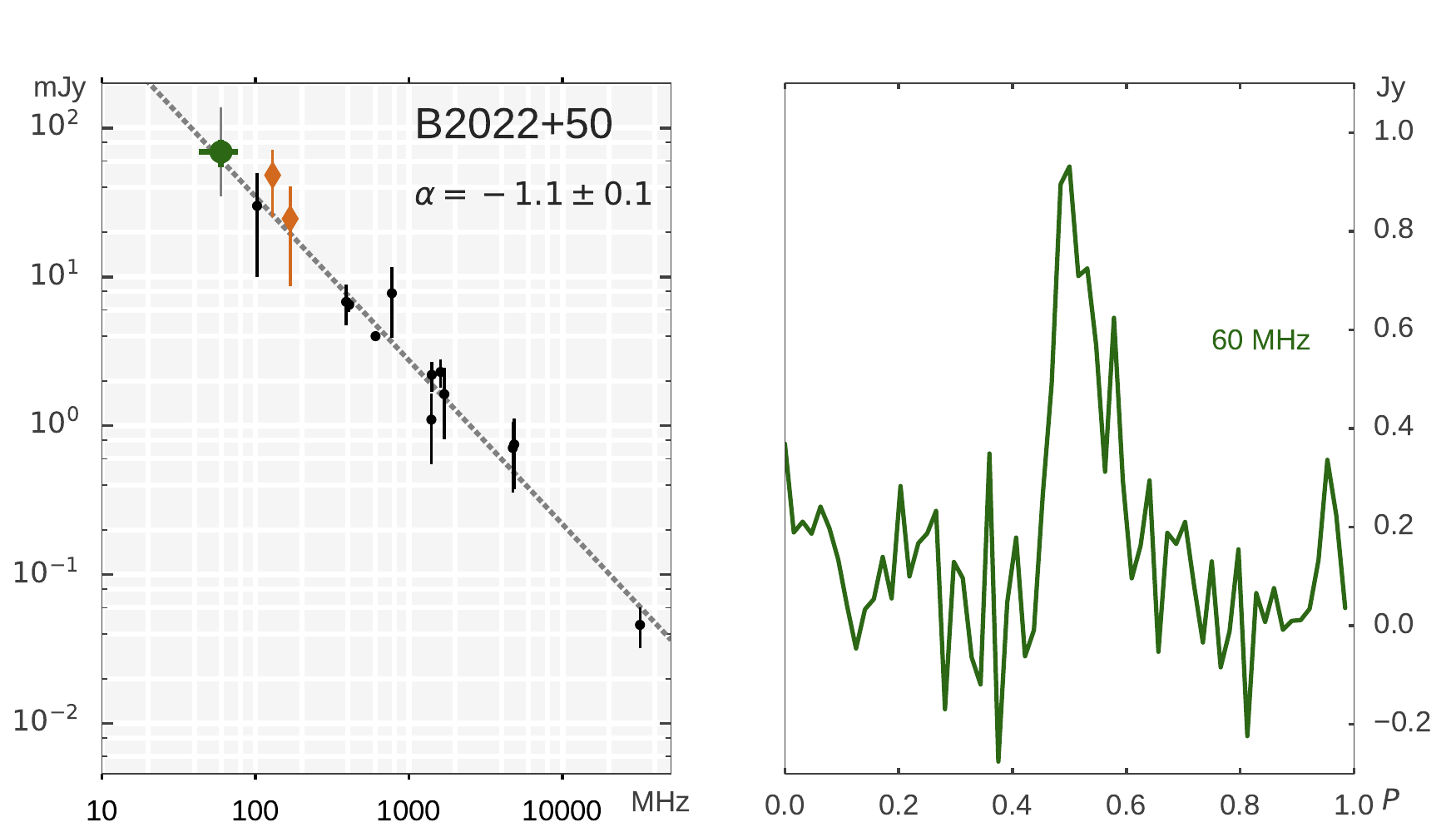}\includegraphics[scale=0.48]{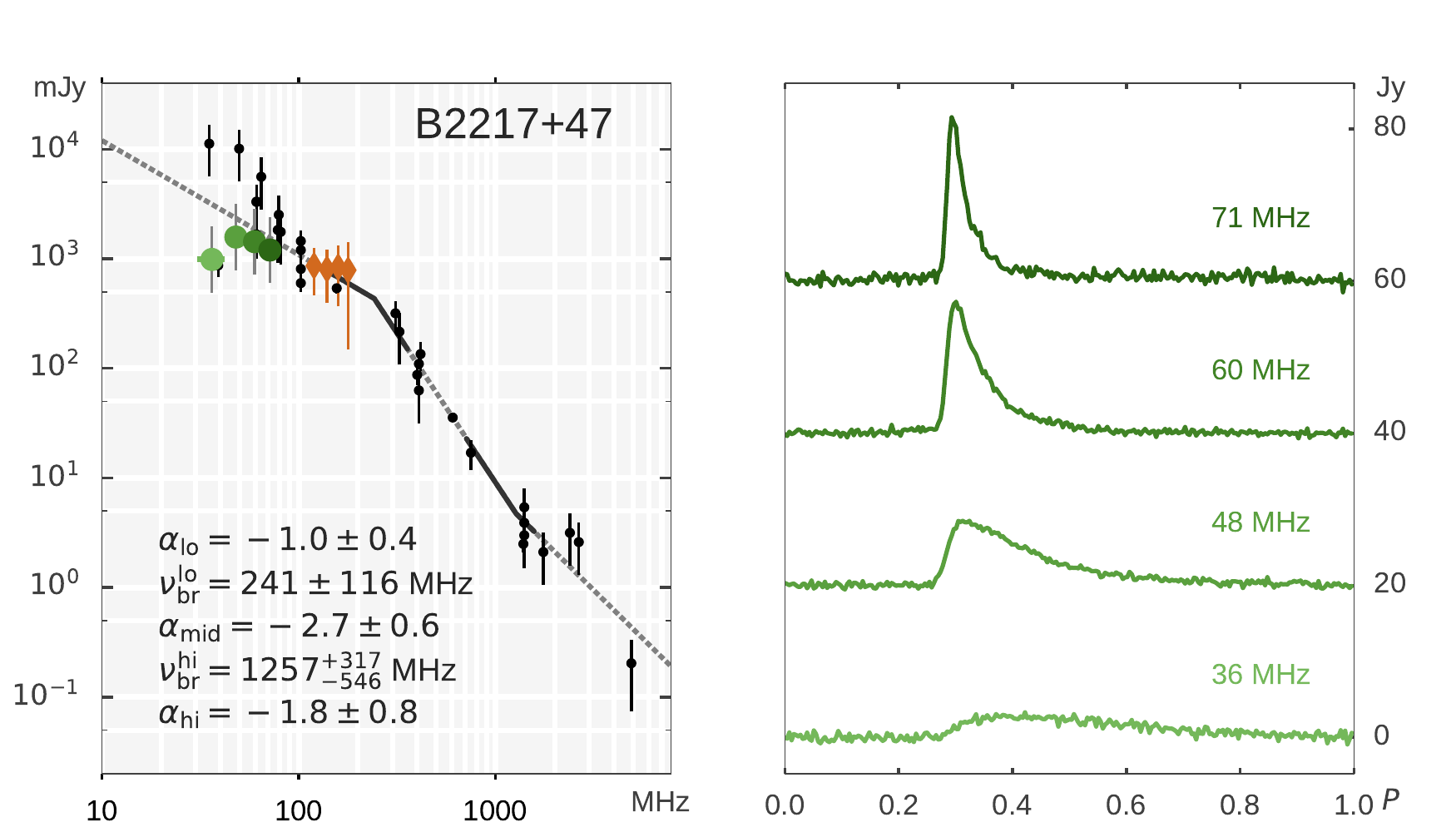}
\includegraphics[scale=0.48]{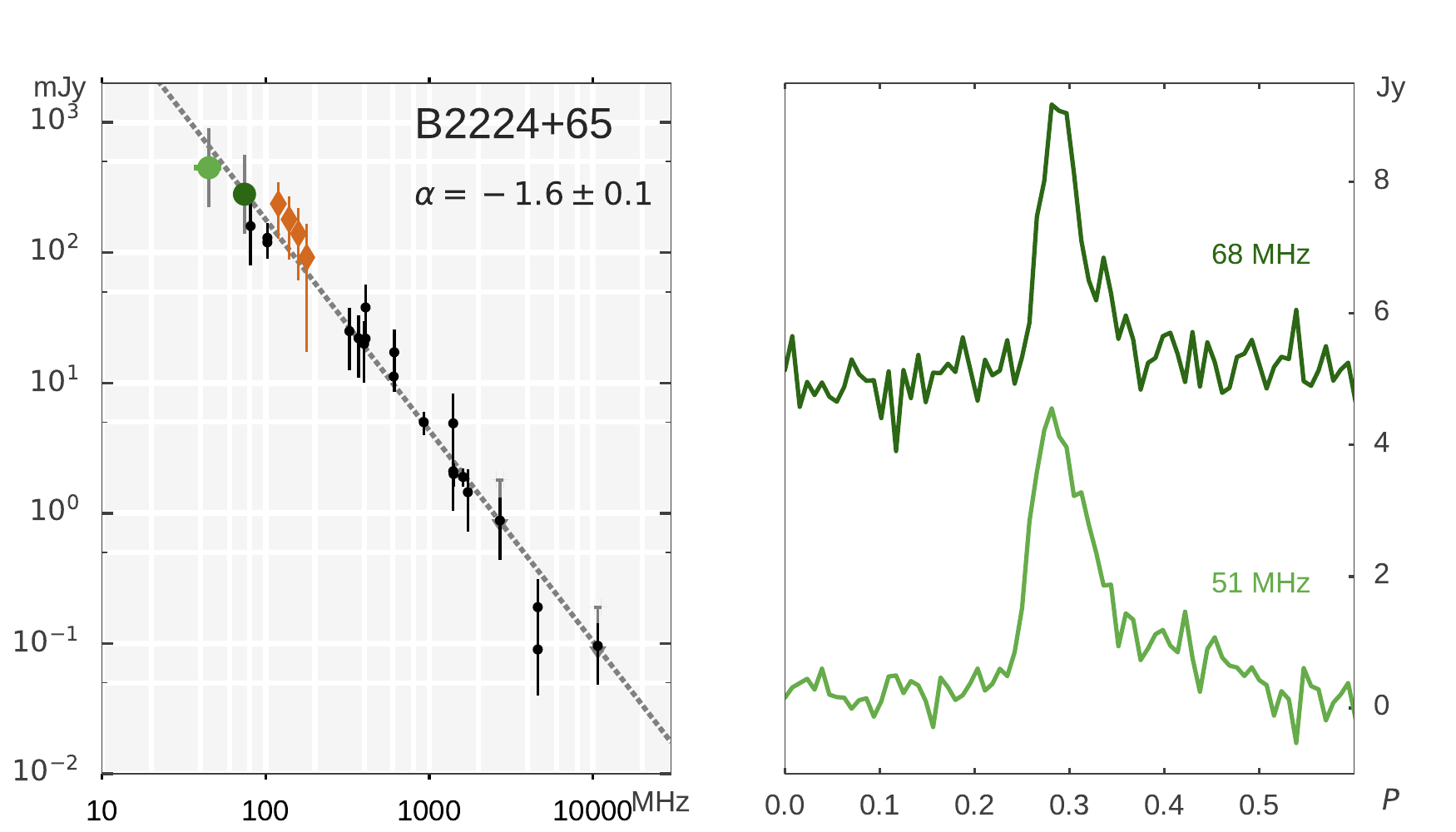}\includegraphics[scale=0.48]{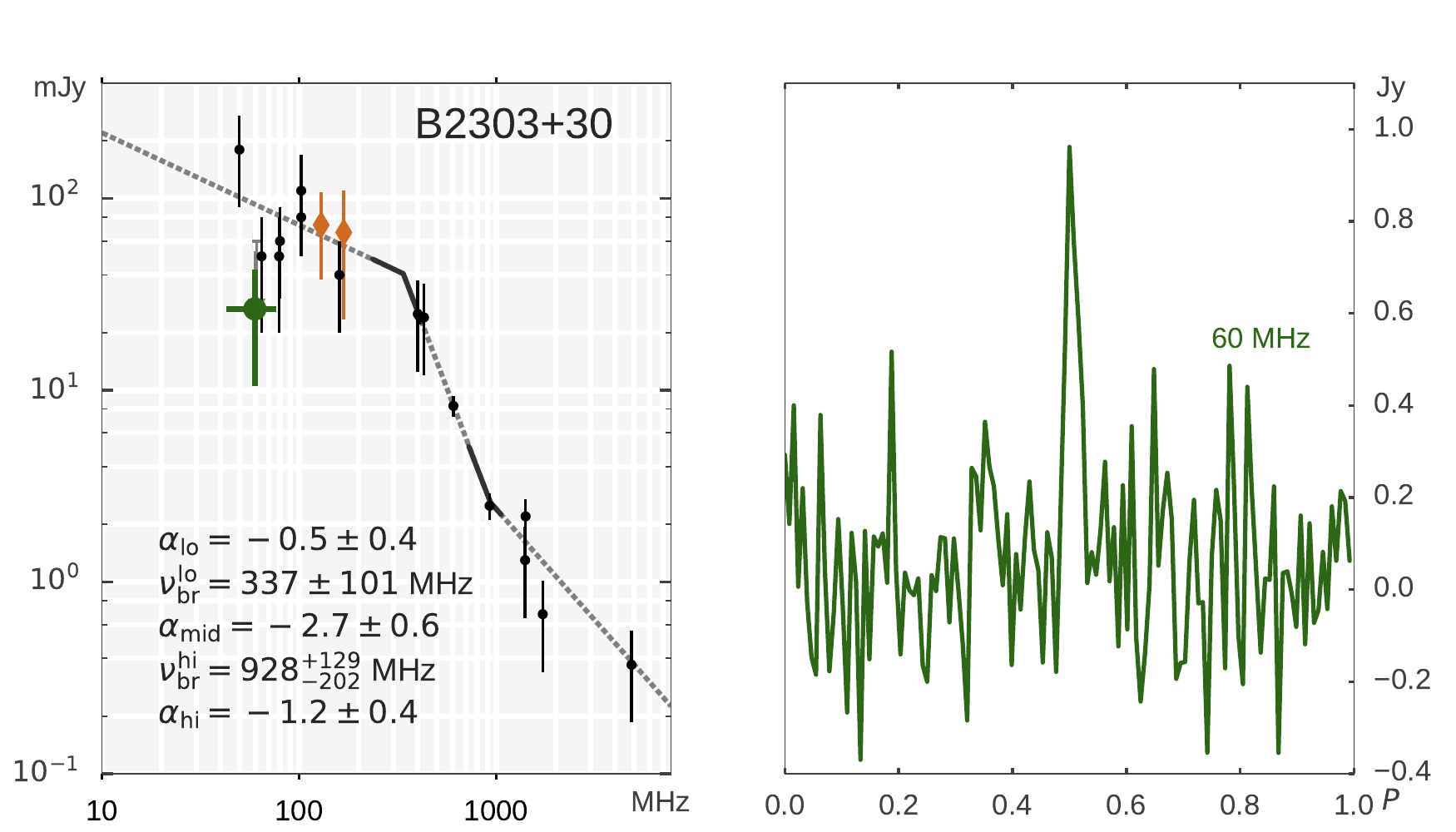}
\caption{See Figure~\ref{fig:prof_sp_1}.}
\label{fig:prof_sp_4}
\end{figure*}

\begin{figure*}
\includegraphics[scale=0.48]{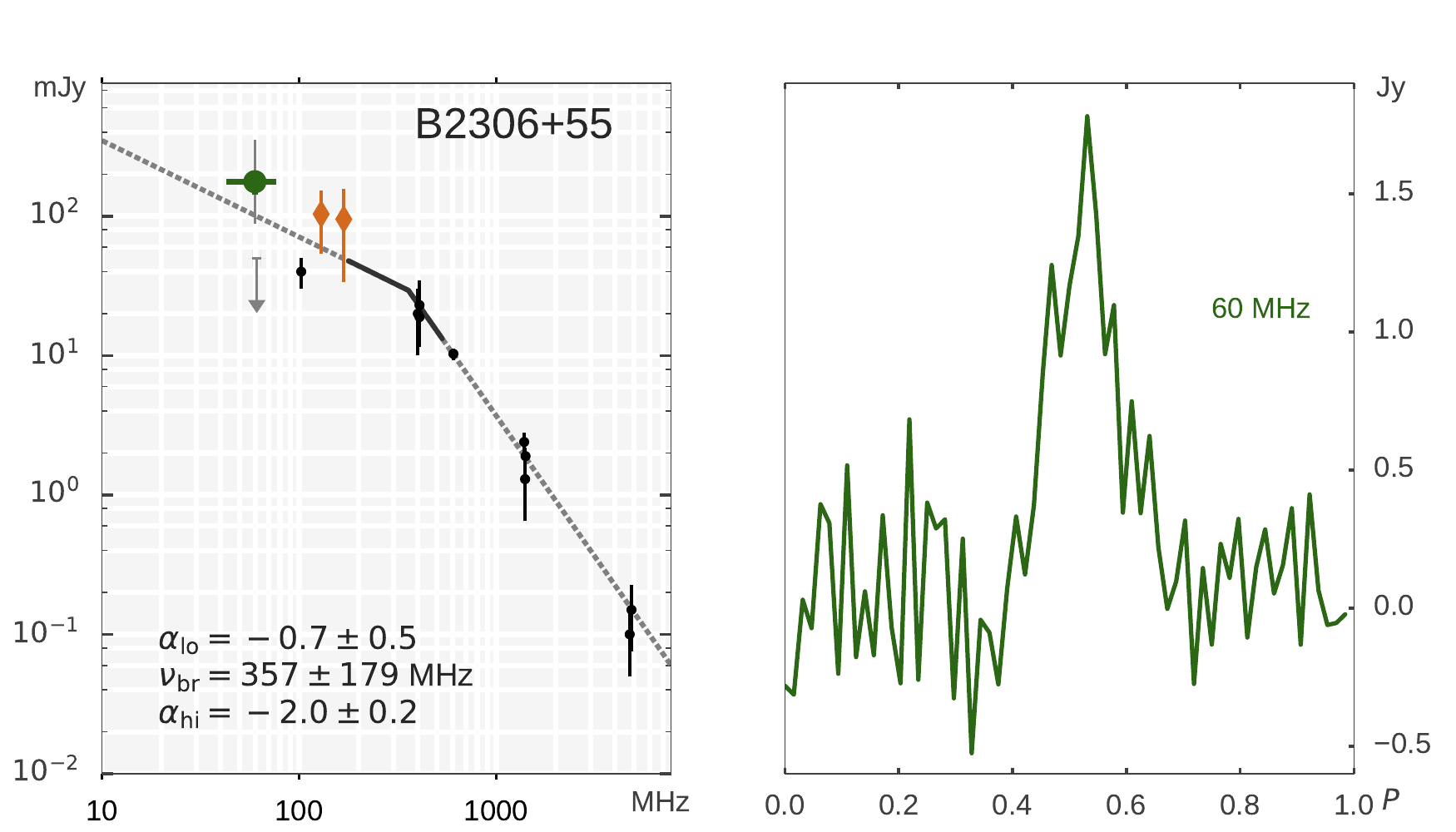}\includegraphics[scale=0.48]{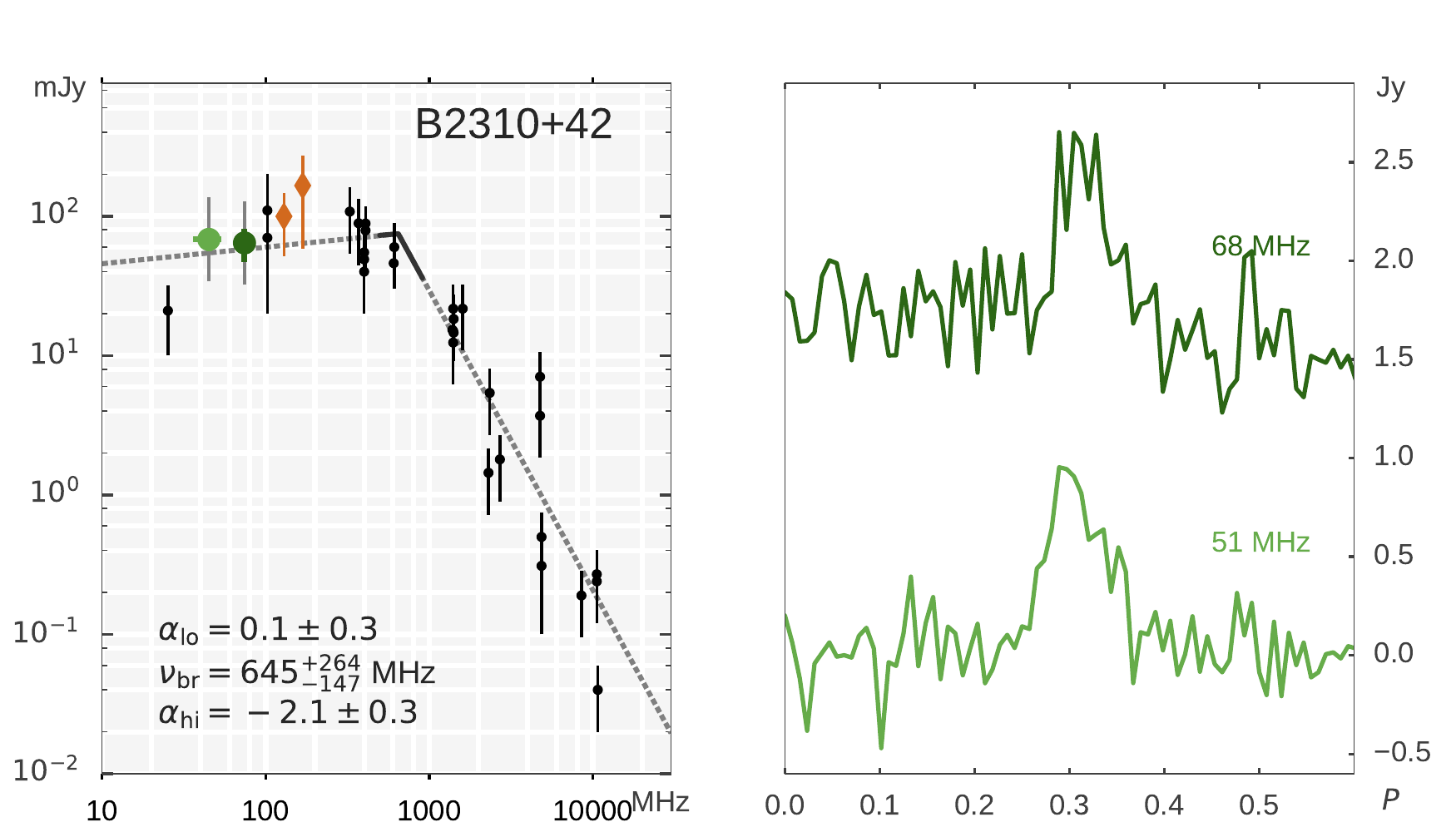}
\includegraphics[scale=0.48]{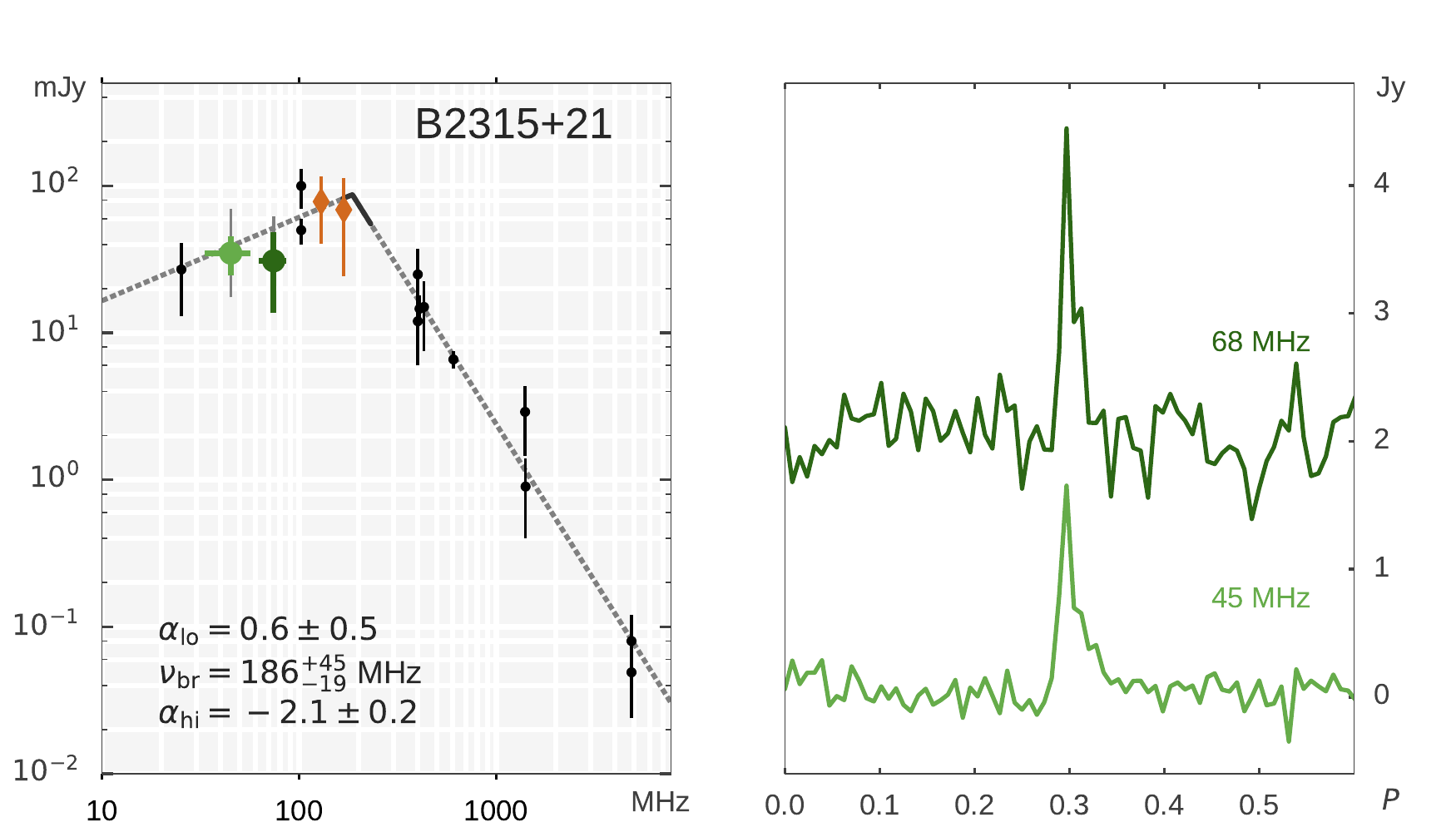}
\caption{See Figure~\ref{fig:prof_sp_1}.}
\label{fig:prof_sp_5}
\end{figure*}

\clearpage

\bibliographystyle{aa} 
\bibliography{census_bibliography}

\end{document}